\documentclass[iop]{emulateapj}

\usepackage{times}
\input{epsf}
\usepackage{graphicx}
\usepackage{color}

\shorttitle{The optical counterparts of the nearest Ultraluminous X-ray sources}
\shortauthors{Gladstone et al. }

\begin{document}


\title{Optical counterparts of the nearest ultraluminous X-ray sources}


\author{Jeanette C. Gladstone$^{*1}$, Chris Copperwheat$^{2}$, Craig O. Heinke$^{1}$, Timothy P. Roberts$^{3}$, Taylor F. Cartwright$^{1,4}$, Andrew J. Levan$^{5}$, Mike R. Goad$^6$}


\altaffiltext{*}{Avadh Bhatia Fellow, email: j.c.gladstone@ualberta.ca}
\altaffiltext{1}{Department of Physics, University of Alberta, 11322 - 89 Avenue, Edmonton, Alberta, T6G 2G7, Canada}
\altaffiltext{2}{Department of Physics, Liverpool John Moores University, Wirral, CH41 1LD, UK}
\altaffiltext{3}{Department of Physics, University of Durham, South Road, Durham DH1 3LE, UK}
\altaffiltext{4}{International Space University, 1 rue Jean-Dominique Cassini, 67400 Illkirch-Graffenstaden, France}
\altaffiltext{5}{Department of Physics, University of Warwick, Coventry, CV4 7VL, UK}
\altaffiltext{6}{Department of Physics \& Astronomy, University of Leicester, University Road, Leicester, LE1 7AL, UK}


\begin{abstract}

We present a photometric survey of the optical counterparts of ultraluminous X-ray sources (ULXs) observed with the \emph{Hubble Space Telescope} in nearby ($\la$ 5 Mpc) galaxies.  Of the 33 ULXs with \emph{Hubble} \& \emph{Chandra} data, 9 have no visible counterpart, placing limits on their $M_V$ of  $\sim$ -4 to -9, enabling us to rule out O-type companions in 4 cases. The refined positions of two ULXs place them in the nucleus of their host galaxy. They are removed from our sample. Of the 22 remaining ULXs, 13 have one possible optical counterpart, while multiple are visible within the error regions of other ULXs. By calculating the number of chance coincidences, we estimate that 13$\pm$5 are the true counterparts. We attempt to constrain the nature of the companions by fitting the SED and $M_V$ to obtain candidate spectral types. We can rule out O-type companions in 20 cases, while we find that one ULX (NGC 253 ULX2) excludes all OB-type companions. Fitting with X-ray irradiated models provides constraints on the donor star mass and radius.  For 7 ULXs, we are able to impose inclination-dependent upper and/or lower limits on the black holes mass, if the extinction to the assumed companion star is not larger than the Galactic column. These are NGC 55 ULX1, NGC 253 ULX1, NGC 253 ULX2, NGC 253 XMM6, Ho IX X-1, IC342 X-1 \& NGC 5204 X-1. This suggests that 10 ULXs do not have O companions, while none of the 18 fitted rule out B-type companions. 

\end{abstract}





\section{Introduction}

In the late 1970s and early 1980s the \emph{Einstein} telescope was used to perform studies of normal galaxies, which revealed the presence of X-ray luminous non-nuclear objects that were brighter than those in our own galaxy (Fabbiano 1989). These objects were later termed ultraluminous X-ray sources (ULXs; e.g. Makishima et al. 2000). Although studies of these luminous sources have continued for more than 30 years, their nature is still unclear (e.g. Roberts 2007, Gladstone 2010, Feng \& Soria 2011). It has been confirmed that many of these sources contain accreting black holes (Kubota et al. 2001), but currently the masses of the compact objects are still unknown. Their luminosities ($L_X \ga 10^{39}$ erg s$^{-1}$) preclude the possibility that we are observing isotropic, sub-Eddington, accretion onto a stellar mass black hole (sMBH; comparable to the Galactic sMBHs,  $3 \la M_{BH} \la 20 M_\odot$), which has led to the idea that we may be observing intermediate mass black holes (IMBHs; 10$^2$ -- 10$^{4}$ $M_\odot$; Colbert \& Mushotzky 1999). An alternative is that we may instead be observing massive stellar remnant black holes (MsBHs, Feng \& Soria 2011; defined as the end product from the death of a single, current generation star; $M_{BH} \la 100 M_\odot$; e.g. Fryer \& Kalogera 2001; Heger et al. 2003; Belczynski, Sadowski \& Rasio 2004) that are either breaking or circumventing the Eddington limit. 

X-ray analysis has been exhaustive, with early \emph{XMM-Newton} observations providing both spectral and timing evidence viewed as supporting IMBHs. Analysis of their X-ray spectra revealed the presence of cool disc emission, in combination with a power-law, a combination used effectively to study Galactic sMBH systems. However, here the disc temperature appeared much cooler than those of sMBHs, suggesting IMBHs (e.g. Miller et al. 2003; Kaaret et al. 2003; Miller, Fabian \& Miller 2004). Meanwhile, X-ray timing analysis revealed the presence of quasi-periodic oscillations (QPOs) in M82 X-1, M82 X42.3+59 \& NGC 5408 X-1 (Casella et al. 2008; Feng et al. 2010; Strohmayer et al. 2007), something observed in both Galactic sMBH and SMBH systems, that seems to scale with black hole mass (e.g. McClintock \& Remillard 2006; McHardy et al. 2006; van der Klis 2006). Combining both the spectral and timing analysis of these sources indicated that ULXs were hosts to IMBHs (e.g. Dheeraj \& Strohmayer 2012). However, recent timing studies of ULXs show that many of these systems appear to have suppressed intra-observational variability (Heil et al. 2009), while spectral studies of higher quality data have indicated the presence of a break above 3 keV, a feature that would not be expected if we were viewing sub-Eddington accretion onto an IMBH (e.g. Stobbart et al. 2006; Gladstone, Roberts \& Done 2009). Instead, we have suggested that this combination of new X-ray spectral and timing features describes a new super-Eddington accretion state, the so called \emph{ultraluminous} state (Roberts 2007; Gladstone et al. 2009; Roberts et al. 2010). This is echoed in the re-analysis of NGC 5408 X-1, which shows that the spectra, power spectral density and rms spectra of this source are better matched to models of super-Eddington accretion than sub-Eddington accretion onto an IMBH (Middleton et al. 2011). In Middleton et al. (2011a), the authors reported on the analysis of \emph{ XMM} flux-binned data from M33 X-8, and suggested a similar two component spectral fit. The spectral evolution and timing properties are unlike those of the standard sub-Eddington accretion states, leading the authors to invoke the onset of an extended photosphere and a wind to explain the observed data (photosphere and/or outflow dominated, as predicted in models of super-Eddington accretion; e.g Begelman et al. 2006, Poutanen et al. 2007). 

Although recent X-ray analysis seems to be pointing  towards the presence of MsBHs, more direct evidence is required. An attractive method is that used to confirm the first known Galactic black hole, Cygnus X-1.  Following Murdin \& Webster (1971), we must  first find the potential optical counterpart photometrically. Optical spectroscopic follow-up can then be performed to gain dynamical mass estimates for the system (Webster \& Murdin 1972;  Bolton 1972; Paczynski 1974). Such techniques have been used repeatedly with success in our Galaxy, e.g. van Paradijs \& McClintock (1995) and Charles \& Coe (2006), and more recently on the extra-galactic source IC 10 X-1 (Prestwich et al. 2007; Silverman \& Filippenko 2009). If we could apply this method to ULXs, we could settle the debate over the mass of the black holes contained within these systems. 

Here, we use optical observations to identify and classify the optical counterparts to ULXs. The identification of unique counterparts is not trivial, as many of these objects reside in crowded stellar fields (Liu et al. 2009), which is unsurprising given their apparent association with star-forming regions (Fabbiano, Zezas \& Murray 2001; Lira et al. 2002; Gao et al 2003; Swartz, Tennant \& Soria 2009). This suggests that we may be looking for young companions, a theory supported by recent results finding a prevalence for blue companions to these ULXs indicative of OB-type stars (e.g. Liu et al. 2004; Gris{\'e}  et al. 2005; Roberts et al. 2008), but it should be noted that such a blue colour may be partly due to contamination by reprocessed X-rays from the accretion disc or stellar surface (e.g. Copperwheat et al. 2005; 2007; Madhusudhan et al. 2008; Patruno \& Zampieri 2010; Gris{\'e} et al. 2012). 

To date much of the analysis of potential counterparts has focused on individual source and/or their host galaxies (e.g. Ho II X-1, Kaaret, Ward \& Zezas 2004; NGC 1313 X-2 Mucciarelli et al. 2005; Liu et al. 2007; Gris{\'e} et al. 2008; Impiombato et al. 2011; NGC 5408 X-1, Lang et al. 2007; M51 population, Terashima, Inoue \&Wilson 2006; Antennae galaxy, Zezas et al. 2002; Cartwheel galaxy, Gao et al. 2003), whilst only a small number of larger surveys have taken place (e.g. Ptak et al 2006, Swarz, Tennant \& Soria 2009; Tao et al. 2011).  

Spectroscopic follow-up has begun for some of these sources, although only a small number have been published to date. Roberts et al. (2001) studied the counterpart of NGC 5204 X-1, finding a blue, almost featureless spectrum, with similar featureless spectra found for other sources (e.g. NGC 1313 X-2, Zampieri et al. 2004; Roberts et al. 2011; Ho IX X-1, Gris{\'e} et al. 2011; Roberts et al. 2011; NGC 5408 X-1, Kaaret \& Corbel 2009; Cseh et al. 2011; Gris{\'e} et al. 2012). This suggests the light is non-stellar in origin (see Figure 1 in Roberts et al. 2011; Gladstone et al. \emph{in prep}), indicating that the light may be dominated by emission from the accretion disc. Nevertheless, the search for dynamical mass constraints has continued, as a number of spectra contain the He \textsc{ii} 4686 {\AA} high excitation line. This line has been associated with accretion discs in Galactic sources, and has been used successfully to gain such mass constraints in the past (e.g. GRO J1655-40; Soria et al. 1998). Initial results from the optical analysis of multi-epoch spectra of two ULXs (NGC 1313 X-2 \& Ho IX X-1) have detected radial velocity variations; however, they may not be sinusoidal (Roberts et al. 2011; Gladstone et al. \emph{in prep}). Studies have also been performed on the optical counterpart to ULX P13, in NGC 7793 (Motch et al. 2011). Here variations in the He II line are also present, but  superimposed on a photospheric spectrum. This reveals the possible presence of a late B-type supergiant companion of between 10 -- 20 $M_\odot$ (Motch et al. 2011). Again, radial velocity variations were detected for this source, with further data and analysis required to confirm its period and nature. 

To date the confirmation of black hole mass has proved elusive for known counterparts. This paper seeks to find more potential counterparts for further study. We focus our search on nearby galaxies, in order to have the best chance of finding unique optical counterparts, while maximising the potential for photometric and spectroscopic follow-up of these systems.

The paper is arranged as follows.  First we outline the sample selection, and the data reduction processes (Sections~\ref{section:optical sample} \& \ref{section:optical data}). We then go on to combine optical and X-ray imaging data to identify all possible counterparts in Section~\ref{section:identify}. Section~\ref{section:c/ps} applies multiple techniques to classify these candidate counterparts, while Section~\ref{section:summary} presents a discussion of our results, implications of the analysis, and routes for further study.

\section{Sample selection}
\label{section:optical sample} 

We compiled a complete list of the known ULXs, drawing from a number of ULX catalogues available in the public domain, including Roberts \& Warwick (2000), Swartz et al. (2004), Liu \& Mirabel (2005), Liu \& Bregman (2005), Ptak et al. (2006) and  Winter, Mushotzky, \& Reynolds (2006). A concise primary list was formed by merging duplicate identifications and removing any sources for which subsequent research has indicated that a ULX was not present (based on luminosity criteria, or the object later being identified as a non-ULX, e.g. a foreground star or background quasar). 

Many of the optical counterparts identified to date are faint ($\ga$ 24 mag; Roberts, Levan \& Goad 2008). Therefore, we place an additional distance constraint on our sample of 5 Mpc, to allow for potential photometric \& spectroscopic follow-up. At this distance, a B0 V star would have an apparent magnitude of 24.4 ($M_V$ = -4.1, Zombeck 1990), so more distant objects would be impractical for spectroscopic studies with current international ground based facilities. 

We retain the ULXs residing within NGC 3034 (M82) in our sample as some distance estimates have indicated that this galaxy may be located within 5 Mpc (e.g. $\sim$ 3.6 Mpc, Freedman et al. 1994). This provides a sample of 45 nearby ULXs that we list in Table \ref{tab:full list}, along with their published luminosities, distances, Galactic absorption and extinction columns for both the optical and X-ray bands.  We also include extra-galactic or total $N_H$ columns for each source, as found via literature search (where available).  It is not clear where the additional absorbing material is located, as it could be gas clouds in the host galaxy, or associated with the ULX itself (e.g. photosphere/wind). 

Table \ref{tab:full list} also indicates that the sources residing within 5 Mpc cover the majority of the standard ULX X-ray luminosity range ($L_{\rm X} \sim 10^{39}$ -- $\sim 10^{41}$ erg s$^{-1}$; only excluding the new hyperluminous X-ray sources e.g. Sutton et al. 2012). The X-ray luminosities listed within this table are taken from the references denoted in superscript. As the data is collated from published results we highlight inconsistencies in their calculation. Many are the observed X-ray luminosities, although some are intrinsic/de-absorbed (identified by `I'). The luminosities listed in Table \ref{tab:full list} include values derived from observations using three separate X-ray telescopes (\emph{ROSAT} -- R, \emph{Chandra} -- C and \emph{XMM-Newton} -- X), with detectors that are sensitive to differing energy ranges. 

Liu \& Bregman (2005) used the \emph{ROSAT} archive in combination with the online tool \textsc{webpimms}\footnote{{\tt http://heasarc.gsfc.nasa.gov/Tools/w3pimms.html}} to extrapolate a luminosity over the 0.3 -- 8.0 keV energy range. In each case they used a photon index of 1.7 and Galactic $N_H$. Luminosities calculated using \textsc{webpimms} are marked with a `P'. The authors Swartz et al. (2004), Humphrey et al. (2003) and Bauer et al. (2001) each used data from the \emph{Chandra} X-ray telescope to provide luminosities over the ranges 0.5 -- 8.0, 0.3 -- 7.0 and 0.5 -- 10.0 keV ranges respectively. Some authors provide 0.3 -- 10 keV luminosities calculated using observations from the \emph{XMM-Newton} telescope including Winter et al. (2006), Stobbart et al. (2006) and Feng \& Kaaret (2005). Trudolyubov (2008) also used \emph{XMM-Newton}, but considered only the 0.3 -- 7.0 keV band pass, whilst Strohmayer \& Mushotzky (2003) calculated the bolometric luminosity for each source. Finally, Liu \& Mirabel (2005) collated information from published works and so provide details of the observed peak luminosity over an identified range specific to each source (all luminosities taken from this work were calculated over the 0.5 -- 10.0 keV energy range). 

We searched the \emph{Chandra} archive and \emph{Hubble Legacy Archive} (HLA) for publicly available observations of each of the 45 sources (using data available November 2011). Twelve objects which lack either \emph{Chandra} or \emph{Hubble} data are marked with a  `$^*$' in Column 1 of Table \ref{tab:full list}, and we do not discuss them further in this paper. We are left with a sample of 33 ULXs residing within 5 Mpc.

\section{Data collation \& reduction}
\label{section:optical data}

With the inception of the HLA\footnote{{\tt http://hla.stsci.edu}}, designed to optimise the science from \emph{HST}, we are able to collate pre-processed data. These data sets are produced using the standard \emph{HST} pipeline products, which combine the individual exposures using the  \textsc{iraf} task \emph{MULTIDRIZZLE}\footnote{{\tt http://www.stsci.edu/$\sim$koekemoe/multidrizzle}}. Each field is astrometrically corrected (whenever possible), by matching sources in the field to one or more of three catalogues; Sloan Digital Sky Survey (SDSS), Guide Star Catalogue 2 (GSC2) and  2MASS, in order of preference. Information provided on the HLA pages states that this is only possible in $\sim$ 80 \% of ACS - WFC fields, due to crowding or lack of matching sources, or sources that are unresolved from the ground (and therefore not present in the catalogues). As a result, we check and improve on these to produce the best possible astrometry. 

To astrometrically correct these fields, it is important to have a number of sources in the field. We therefore opt to use observations with large fields of view wherever possible, and so select the \emph{HST} instrument and detector based on these criteria. In order of preference, we select observations in available bands from the Advanced Camera for Surveys (ACS) - Wide Field Camera (WFC), the Wide Field Planetary Camera 2 (WFPC2) and ACS - High Resolution Camera (HRC).   

Another consideration is the variability that occurs within these systems, which would affect the emission observed in the optical and UV bands. From the X-ray spectra of these systems, we see variability on longer inter-observational time-scales of days to years (e.g. Fabbiano 2004). Such variability is likely to affect the optical and UV emission, as seen in Galactic X-ray binary systems (e.g. Charles \& Coe 2006). However, the X-ray  variability appears to be suppressed on shorter, intra-observational, time-scales (Heil et al. 2009). Thus, we expect that near-simultaneous ($\la$ 24 hours) observations in multiple bands are unlikely to be significantly impacted by variability.

We fold these considerations into our observation selection criteria, awarding observations of multiple bands, over short time-scales, higher priority. Where these were not available, different bands were selected from multiple observations, in order to give a fuller view of the source. In these cases we must seriously consider the potential impact of optical variability, as observations could have been made months, or even years apart (see Section \ref{subsection:op var}). The observation IDs, instrument, date of observations \& mode selection and filter band information and exposure times for selected observations are listed in Table \ref{tab:observations}.

We also collated and downloaded \emph{Chandra} observations for each of our ULX sample. Wherever more than one X-ray observation was available in the public archive, we chose the longest appropriate observation containing that source. In the case of transient ULXs this was not always the most recent observation of the field. The observation IDs, instrument setup and exposure times for each of these X-ray observations are also listed in Table  \ref{tab:observations}.

We attempt to improve the astrometry in order to maximise the chance of obtaining unique optical counterparts to these sources. We approach the optical data first, by checking and improving the  astrometric corrections of each field using the the \textsc{iraf} tools \emph{CCFIND}, \emph{CCMAP} and \emph{CCSETWCS}, in combination with either the 2MASS or USNO 2.0 catalogues (depending on number of sources available in the field). This process also provides the average astrometric error (3 $\sigma$) across the field. 

We find that 14 of the 98 \emph{HST} observations used in our analysis do not contain enough catalogued objects in the optical/ultraviolet fields of view to allow for accurate astrometry corrections. In these cases we compare the field to an alternative corrected observation in a similar waveband. We match sources in these observations and perform relative astrometry corrections using the \textsc{iraf} tools \emph{IMEXAM} and \emph{GEOMAP}. The tool \emph{GEOXYTRAN} is then used to translate the position of the ULX to the relative field coordinates. 

In the case of NGC 4190, we find that we are unable to correct the astrometry of any field by matching to known catalogues. We therefore opt to take advantage of the increased field of view afforded us by SDSS, collecting an image of this region from their archive. The astrometry of this image is corrected using the 2MASS catalogue, and relative astrometry performed on each of the \emph{HST} images. We note that where relative astrometry is required, the additional errors arising from this are also folded into our calculations. 

The astrometry of the X-ray observations must also be checked. We chose to use the reduced primary data provided by the \emph{Chandra X-Ray Center} (\emph{CXC}).  Each observation was checked for any known aspect offset\footnote{{\tt http://cxc.harvard.edu/cal/ASPECT/fix\_offset/fix\_offset.cgi}}. There is a small intrinsic astrometric uncertainty in \emph{Chandra} observations, an error of 0.6 \arcsec for ACIS-S, 0.8 \arcsec in ACIS-I, 0.6 \arcsec for HRC-S and 0.5 \arcsec for HRC-I fields (90 \% confidence region for absolute positional accuracy\footnote{{\tt http://cxc.harvard.edu/cal/ASPECT/celmon/index.html}}). This known error is folded into the initial positional error calculation.

The X-ray astrometry can be further improved by cross-matching sources to the same catalogues used in the optical. The tool \emph{WAVDETECT}, was used to identify sources in the 0.3 -- 7.0 keV energy band, within 6' of the target (in some cases we were forced to use a smaller region, details can be found in Table \ref{tab:positions}). We cross-correlated the positions of sources with $>$20 counts with the 2MASS or USNO catalogues (choice depending on which was used for the respective \emph{HST} fields). Care was taken in the selection of sources, for example galactic centres were considered unsuitable as it can be difficult to get accurate centroiding for such sources in the optical (e.g. NGC 4736, Eracleous et al. 2002). 

Another concern was the limited number of sources available for cross-matching. In cases where no suitable sources were found, we  revert to the 90 \% confidence region for absolute positional accuracy of \emph{Chandra}. In some cases only one object was available for cross-matching in the region of the sky surrounding the ULX.  Some previous works have used this single source to perform relative astrometry, but we suggest caution in doing so, as corrections can only be made in the XY plane, with no consideration for rotational error. In these cases we show both the corrected position and the unimproved 90 \% confidence error region. All sources falling within the larger 90 \% error region are considered, but the corrected position can be used to help identify the more likely ULX counterpart candidate. 

If multiple sources are present in both the \emph{Chandra} and \emph{HST} fields, we used a weighted average to cross-correlate the positions of these sources, and find the required shift. When applying corrections with few sources, it is only possible to correct by shifting in the \emph{xy} plane, which does not account for all rotational error in the telescope, but the impact is minimised. As a result, errors may be underestimated. We still attempt such corrections but apply them with care, using them only as guidance in selection of `most likely' when multiple counterparts lie within the larger error circle. 

When considering transient sources, the object is not always visible in the deepest X-ray observations. In these cases we were able to match the position of this source to the position of other X-ray sources in another observation.

Finally the accurate source positions were found using \emph{WAVDETECT}, and any calculated shifts applied. The positions of each ULX are listed in Table \ref{tab:positions}, along with their associated errors. These errors are found by combining in quadrature the astrometry errors from both fields with the source's individual positional error to provide the resulting error regions for each individual ULX.

\section{Identification of possible counterparts}
\label{section:identify}

The derived positions and their respective error regions are applied to each field in order to search for potential counterparts to these sources. Figure \ref{fig:pictures} contains a 25 $\times$ 25 arcsec colour image (tri-colour wherever possible), and a finding chart (6 $\times$ 6 arcsecs) for each ULX. Error regions are plotted in each case, with blue representing the standard 90 \% confidence ellipse, while the error derived from relative astrometry corrections is plotted in magenta.

To construct the colour images, we select available wavebands for each part of the optical/UV spectrum. We use filter bands ranging from F656N to F814W to represent the red end of our range, filter bands F475W to F606W  for the green band, whilst F220W -- F450W are blue. In each case where more than one band is available, we opt for a band that gives the clearest view of any potential counterparts. If this is not required, we opt for the band with the smallest error region. In some cases we have no data in one or more of the colour bands, in which case the images are presented in one or two colour bands only. 

The positions of all potential counterparts are also marked in each field. One of these is NGC 598 ULX1, which was identified as the nucleus of the galaxy by Dubus et al. (2004), as is clearly seen in Figure \ref{fig:pictures} (pg. 1). The revised \emph{Chandra} position of IC 342 ULX2 shows that this source is also nuclear. As a ULX is non-nuclear by definition, we remove both of these sources from our catalogue. Nine of the remaining ULXs have no optical counterparts in their error regions. Of the 22 remaining sources, 13 have a unique candidate counterpart, and the rest have up to five objects within their error ellipse. 

\begin{figure*}
\leavevmode
\begin{center}
\includegraphics[height=64mm, angle=0]{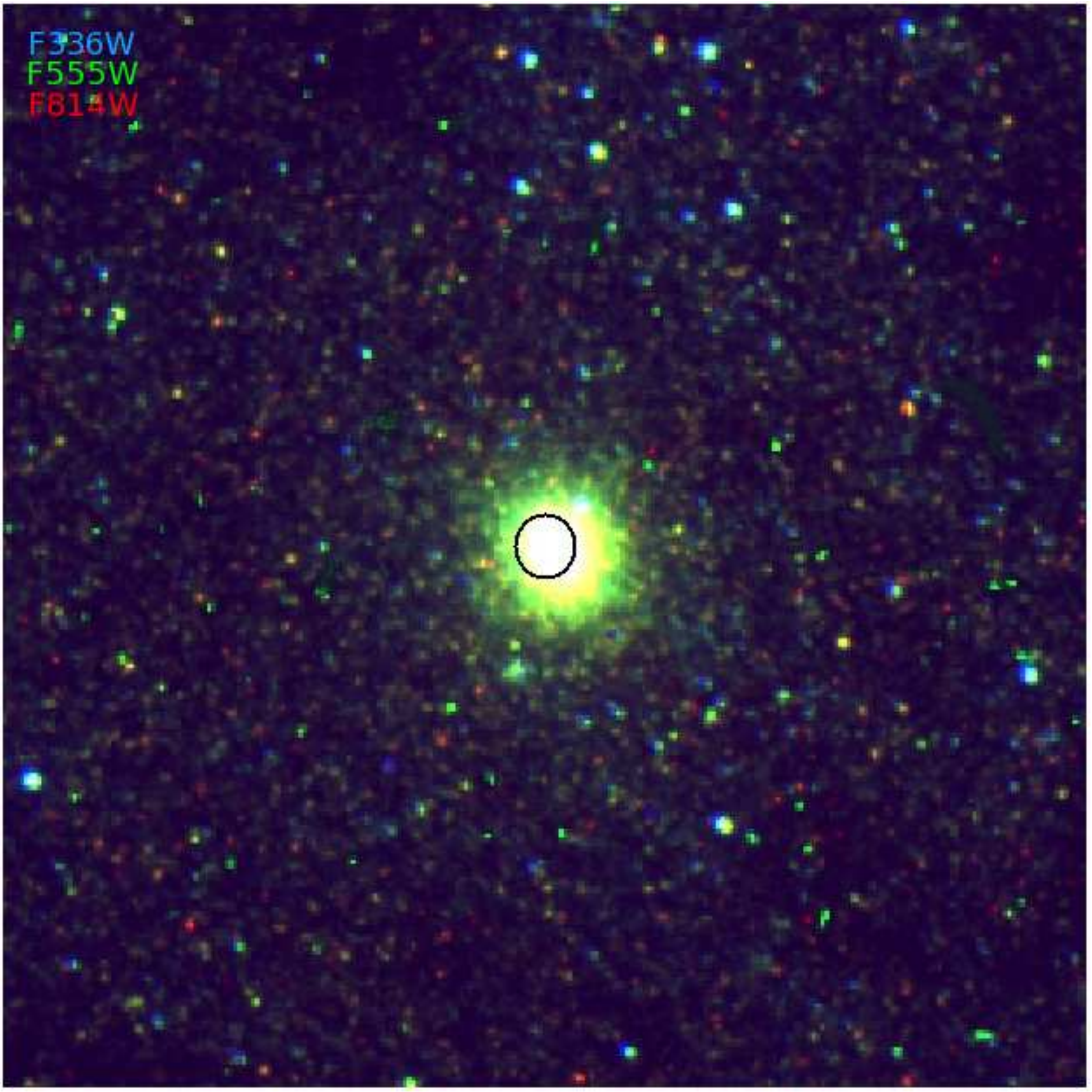} \hspace*{0.4cm}
\includegraphics[height=64mm, angle=0]{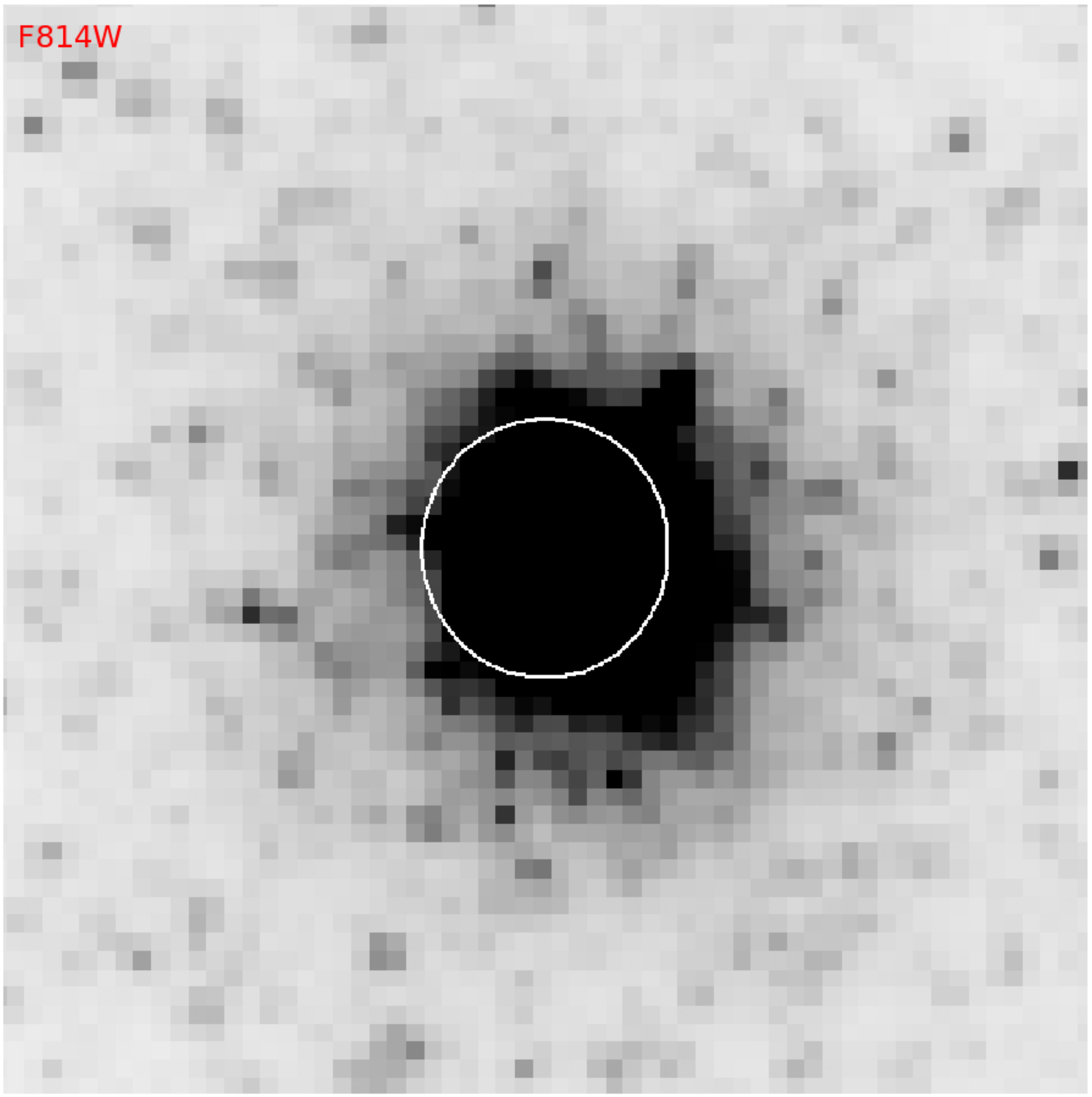} 
\vspace*{0.3cm} 

\includegraphics[height=64mm, angle=0]{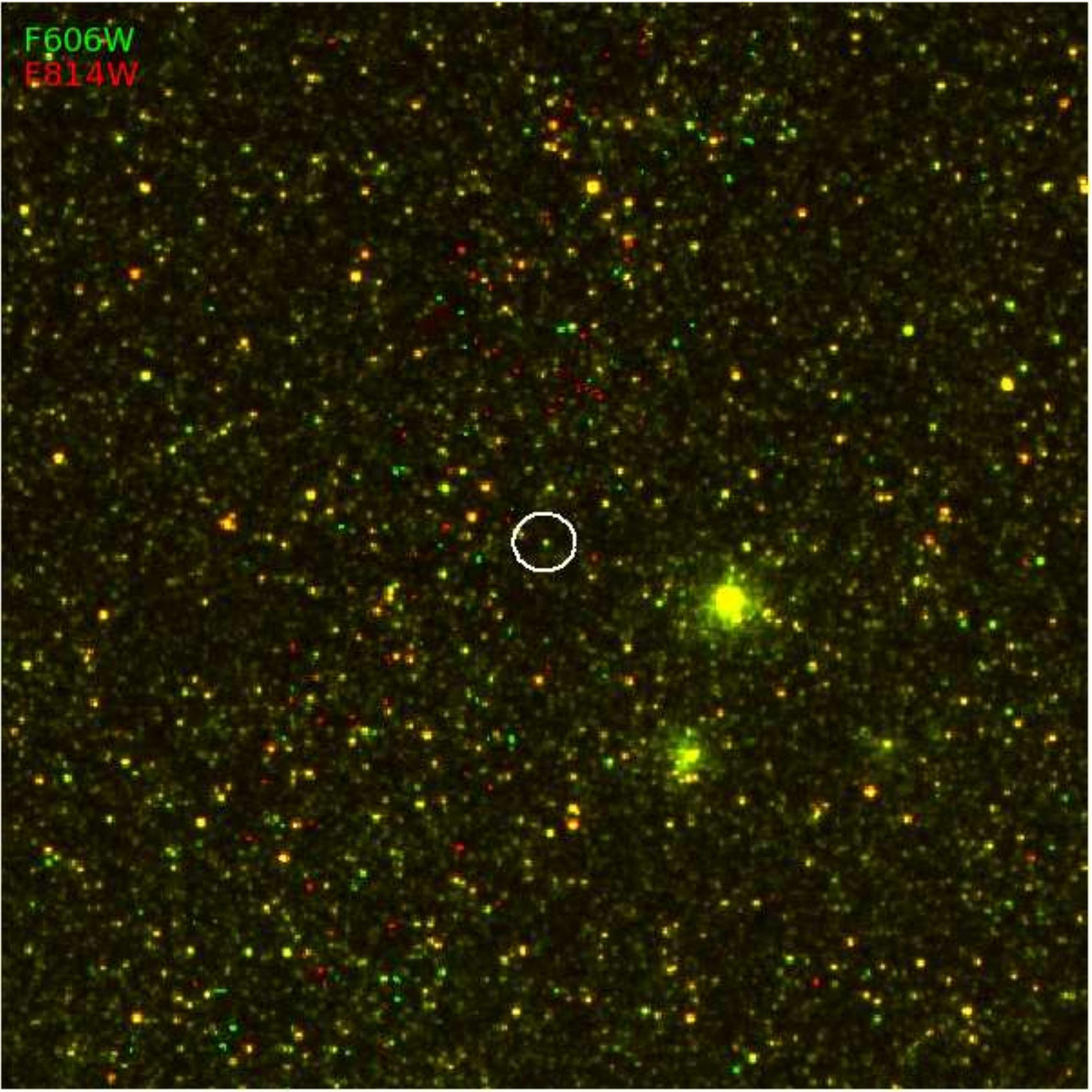} \hspace*{0.4cm}
\includegraphics[height=64mm, angle=0]{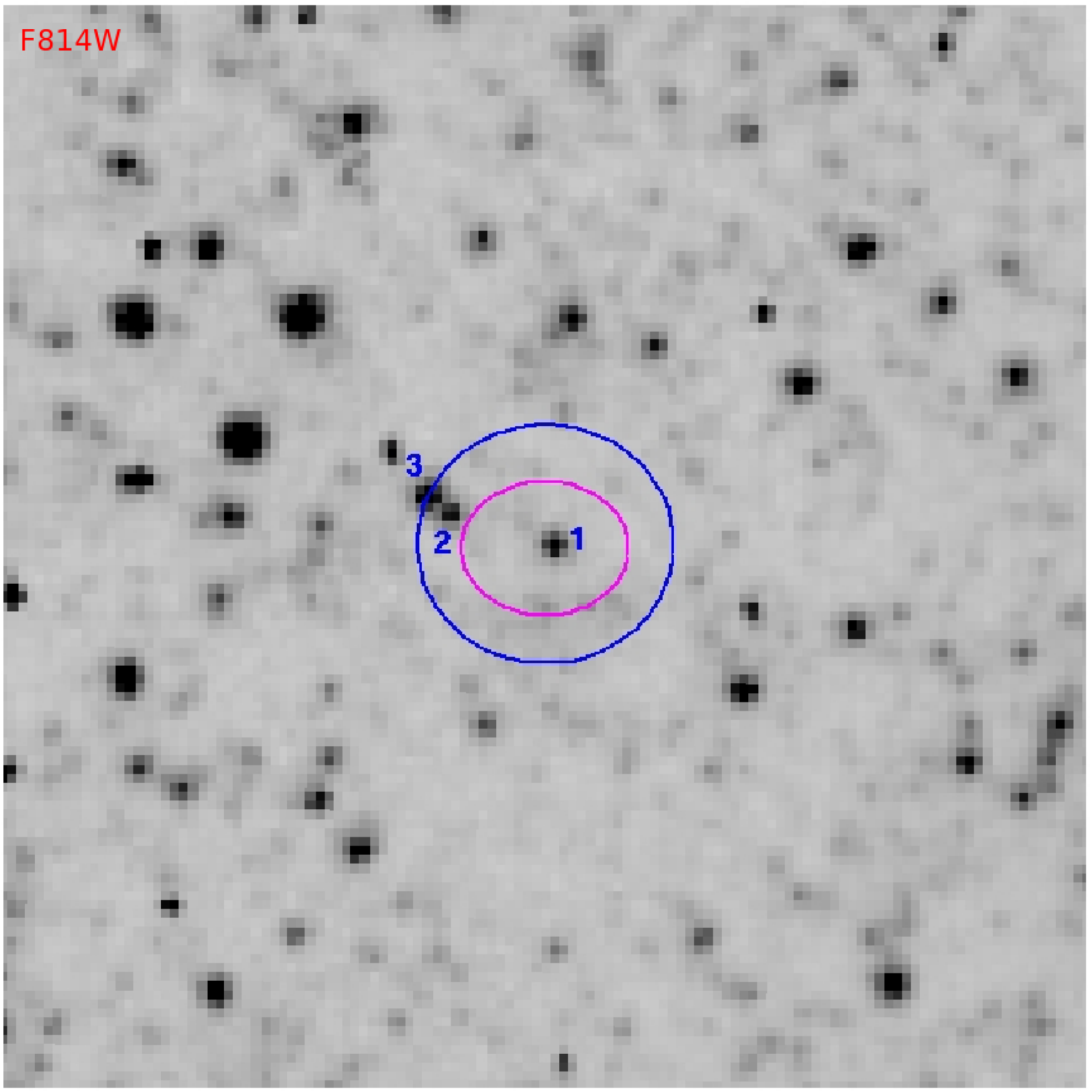} 

\end{center}
\caption{\small{\emph{HST} colour images (\emph{left}) and finding charts (\emph{right}) of the ULX locations. \emph{Left panel}: Colour images are $25 \times 25$ arcseconds in size, over-plotted with positional error ellipse and constructed using the following filter band pass where available; blue = F220W -- F450W, green = F475W -- F606W and red =  F656N -- F814W. \emph{Right panel}: Finding charts are made from only one individual band, showing a regions $6 \times 6$ arcseconds, over-plotted with the combined positional error ellipse for that band (specific wave bands given in brackets after source name, below). Potential counterparts are highlighted numerically with associated magnitudes given in Table \ref{tab:c/ps}. The remainder of Figure \ref{fig:pictures} is provided at the end of the paper.
\emph{Specific notes}: displayed ULX regions are, from top to bottom, NGC 598 ULX1 (F814W) \& NGC 55 ULX1 (F814W). NGC 598 ULX1 is contained within the nucleus of the galaxy. NGC 55 ULX1 has three potential counterparts within the error circle, however counterpart 3 is ruled out as it lies outside the error circle in band F606W. It will not be considered further in this paper.
}}
\label{fig:pictures}
\end{figure*}

In the nine error circles which lack optical counterparts, limits are obtained for the approximate V band observation of each ULX field (listed in Table \ref{tab:c/ps}), given the observed background within the positional error circle for each ULX. We compare these values to the expected V magnitude of each stellar type (Zombeck 1990) at the distance of the galaxy (converted to $m_{555}$ Vega-mag using \textsc{synphot}), with Galactic extinction corrections applied using the $E(B-V)$ values in Table \ref{tab:full list}. O stars appear to be ruled out for 5 of these ULX. However, it is possible for the companion star to be a main sequence B star in all cases (some only valid for later-type B stars). We are unable to obtain a $\sim$ V band image for M83 XMM2, as the error region for this source lies on the ACS-WFC chip gap in the F555W band.We thus compare limits for the (only) available image, a WFPC2 F336W image, to the expected $m_{336}$ values for O5 V, B0 V \& B5V stars (derived using stellar templates for the Bruzual-Person-Gunn-Stryker catalogue using the \textsc{synphot} tool \emph{CALCPHOT}). O stars are ruled out by this comparison, but B stars are acceptable.  We should note, however, that this does not take into account any extinction from the host galaxy or that is intrinsic to the system itself. If this extinction is high, as is seen for those sources in NGC 3034, this may be masking a brighter blue object. 

In those fields where potential counterparts are identified, we collate Vega magnitude zero points ($Zpt$) to allow for the derivation of \emph{HST} filter dependent Vega magnitudes of each source. For observations made using the ACS instruments, we are able to take these values directly from Sirianni et al. (2005). For WFPC2 data, the HLA pipeline converts the units contained within the science field to electrons s$^{-1}$ (like ACS) rather than DN (Data Number), and hence the tabulated zero points given in the \emph{HST} Data Handbook for WFPC2\footnote{{\tt http://www.stsci.edu/hst/wfpc2/Wfpc2\_dhb/wfpc2\_ch 52.html\#1933986}} must also be converted, by applying a correction for the gain\footnote{{\tt http://www.stsci.edu/hst/wfpc2/Wfpc2\_hand\_current/ ch4\_ccd14.html\#440723}}, using $Zpt =$ tabulated zero point $+ 2.5 \times log(\rm{gain})$. 

Aperture photometry is performed on all potential counterparts using \textsc{gaia}\footnote{{\tt http://astro.dur.ac.uk/$\sim$pdraper/gaia/gaia.html}}. Aperture corrections are applied to all fields, irrespective of instrument and detector, following the procedures laid down by Sirianni et al. (2005), with values of corrections for WFPC2 observations taken from the \emph{HST} WFPC2 cookbook\footnote{{\tt http://www.stsci.edu/hst/wfpc2/Wfpc2\_dhb/wfpc2\_ch 52.html\#1933986}}).

Galactic extinction corrections are also applied, using $E(B - V)$ values listed in Table \ref{tab:full list}. These are used in combination with the filter specific extinction ratios depending on the instrument. Extinction ratios for WFPC2 data were taken from Schlegel, Finkbeiner \& Davis (1998) where available, or calculated using \textsc{synphot} when this was not possible. The calculated extinction ratios are found by folding a template spectrum through the instrumental response allowing for foreground extinction using Cardelli laws (chosen for consistency with Schlegel et al. 1998). Although this correction is spectrum-dependent, we find that this dependence is small, and we choose a 10,000 K blackbody as a first order estimate for these corrections, since the observed candidate counterparts are of unknown type. For magnitudes calculated using ACS fields, filter-dependent extinction ratios are given by Sirianni et al. (2005). Again, since the extinction ratios are also dependent on stellar type (and no blackbody is available), we choose to use the corrections for an O5 V star, following the example of Roberts et al. (2008). Although this choice will affect the calculated magnitudes of our sources, the impact will be minimal in most bands, with a larger (although still marginal) impact in the bluest bands (F435W and bluer).  The aperture and Galactic extinction-corrected Vega magnitudes are given in Table \ref{tab:c/ps}.

If we wish to get a clear view of the binary system, we must obtain intrinsic magnitudes for these sources. To do this we must also take account of any absorption from either the host galaxy or that is intrinsic to the ULX itself. One method that has been used previously to correct for this extinction is to use the measured absorption from X-ray spectral fitting (e.g. Roberts et al. 2008), which will give an upper limit to the extinction of the optical light from the binary. 

To calculate the maximal optical extinction column, we use the relation for X-ray-to-optical dust-to-gas ratios published by G{\"u}ver \& {\"O}zel (2009; $N_H = (2.21 \pm 0.18) \times 10^{20} A_V$, with 2$\sigma$ errors). However, X-ray spectral fitting can be degenerate, and the $N_{H}$ columns derived can vary substantially depending on the author's model choice.  We minimise the impact of model choices by using the highest quality X-ray spectra available, and fitting these spectra with current physical models used to describe ULXs.  We begin by  collating $N_H$ values from published results along with their respective errors, preferring long observations, physical models and statistically good fits. When physically motivated models have not been applied to an object, we use values from publications of the deepest X-ray observation of these sources which show statistically good fits based on phenomenological models. Our adopted $N_H$ values can be found in Table \ref{tab:full list}. Using the standard Galactic extinction curve ($R_V = A_V \/ E(B - V) = 3.1$; Cardelli et al. 1989), we  estimate the intrinsic optical magnitudes of these sources. The relevant errors are calculated by combining the error on $N_H$, obtained from literature, with that of G{\"u}ver \& {\"O}zel (2009). The intrinsic Vega magnitudes for each potential counterpart are listed in Table \ref{tab:c/ps}.

We also compile details of previous identifications of potential counterparts, in order to compare our results and search for the most likely candidates. Hereafter in the text, bracketed values refer to the candidate counterpart ID (e.g. IC 342 X-1 (1) for candidate counterpart 1). Details of each potential counterpart are listed in Table \ref{tab:c/ps}, along with any previous identifications of potential counterparts. We compare our findings to previous work in Section \ref{subsection:previous}. 

\section{Counterparts}
\label{section:c/ps}

In our sample, we find 40 potential counterparts to the 22 ULXs. Thirteen of these have previously been reported in the literature (ignoring NGC 598 ULX1), with the remaining 27 potential counterparts identified here for the first time. Up to 22 of these potential counterparts may be the true counterparts, but it is possible that all the potential counterparts in some error circles are chance coincidences. Therefore, we calculate the likelihood of chance coincidences given the density of the local stellar population of each ULX. We search for stellar objects within an annulus around each object with inner radius of 1'' and outer radius of 3''.  For Circinus ULX1, we use a rectangle of size 3'' x 8'' instead, due to the chip geometry. We expect an average of 27$\pm5$ objects to be present within the positional error regions of our ULX sample, yet we have observed a total of 40 potential counterparts. This indicates an overpopulation of 13$\pm5$, which is our best estimate of the number of true counterparts identified in all ULX fields. There is a greater likelihood of foreground object contamination when considering those ULXs in Circinus, as it lies in the Galactic plane, so we temporarily remove these sources from our calculations. If we also removed NGC 3034 (M82) from our likelihood calculations, because the sources lie in an extremely obscured region of the galaxy, such that it is very unlikely that any optical emission would be detected from any real counterparts, this results in an excess of 15 $\pm$ 4 for 36 ULXs, with 19 of these having detected candidate counterparts. Thus we believe that for $\sim$ 83 \% of ULXs with candidate counterparts, the true counterpart lies among our candidate counterparts.  However, we note the caveat that if the ULX is located in a star-forming region with higher stellar density than areas $\sim$ 1\arcsec  away, our number of true counterparts could be overestimated.

In this section, we compare each candidate to stellar models to constrain the nature of the donor star in these ULX binary systems. In Section \ref{subsection:SED} we consider the spectral energy distribution (SED) and the apparent magnitude of each candidate to characterise the star (assuming the light is stellar in origin).  In Section  \ref{subsection:op var}, we check whether variability may impact our characterisation, which is relevant when we are incorporating optical data from multiple epochs. In Section \ref{subsection:previous} we compare our findings to previous works. In Section \ref{subsection:disc} we introduce an additional component in the form of an accretion disc, and also consider the effects of irradiative heating of the star and disc.

\subsection{Spectral typing from stellar templates and magnitudes}
\label{subsection:SED}

We initially consider the case where the donor star contributes 100 \% of the optical light, and it has a luminosity and spectrum which is consistent with a single star. For these preliminary fits, we use only Galactic corrected magnitudes, ignoring any intrinsic extinction. We attempt to classify the donor stars by comparing the candidates with template SEDs. Previously, authors have done this by converting the filter band magnitudes to UBVRI magnitudes to compare with typical values for different stellar types (e.g. Soria et al. 2005; Ramsey et al. 2006; Roberts et al. 2008; Tao et al 2011). The \emph{HST} filters are not an exact match to other photometric systems, which can lead to large errors in typing stars using the \emph{HST} filter bands (see Sirianni et al. 2005 for more detailed discussion). 

\begin{figure}
\begin{center}
\leavevmode
\includegraphics[height=80mm, angle=0]{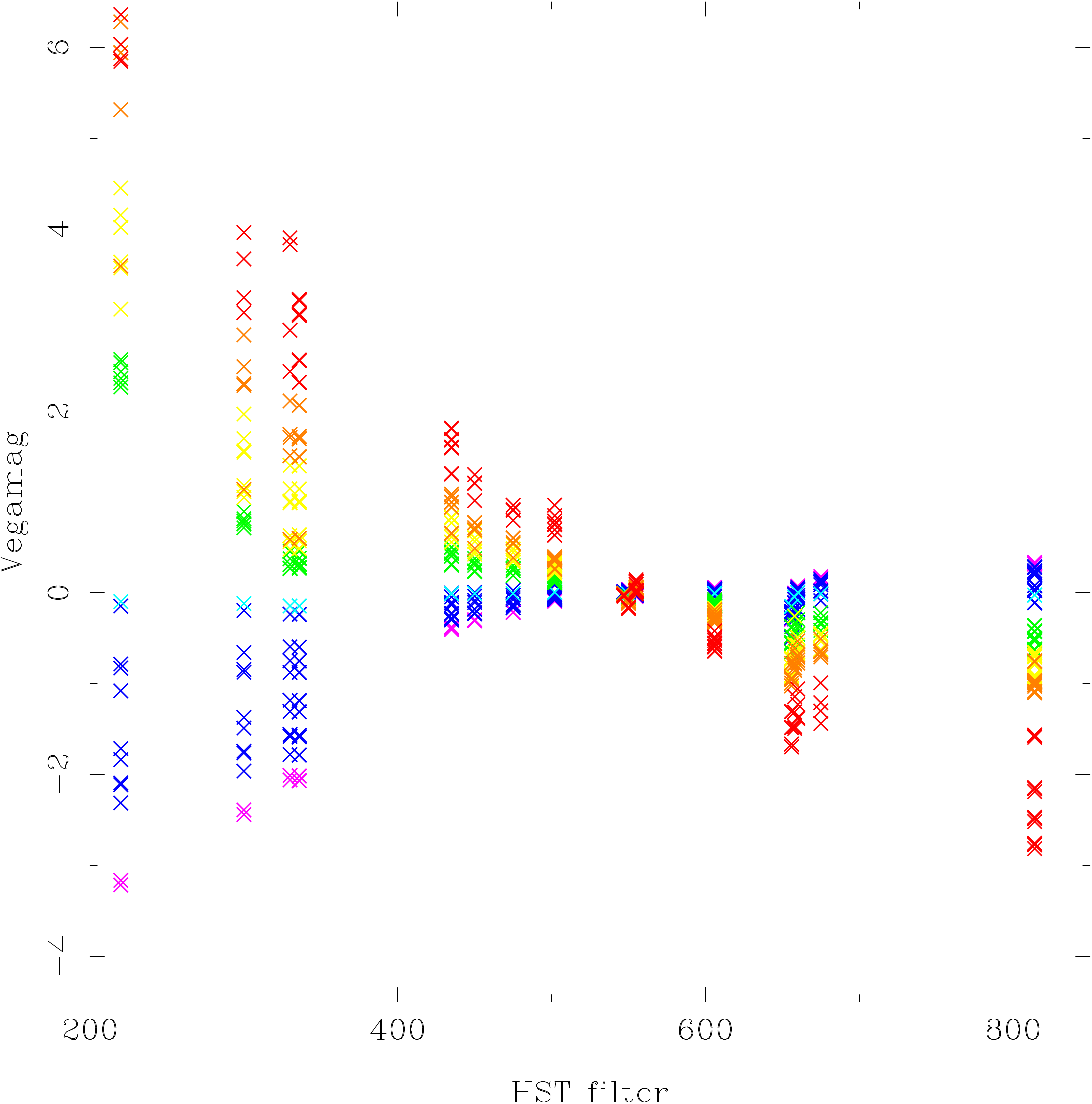} 
\end{center}
\caption{Photometric stellar templates for \emph{HST} filter bands, constructed using the \textsc{iraf} package \textsc{synphot}. The templates normalised to a V band Vega magnitude of zero, and are grouped according to type as follows; magenta~-~O stars, blue~-~B-type, cyan~-~A type stars, green~-~F stars, yellow~-~G type, orange~-~K stars and red~-~M.}
\label{fig:star types}
\end{figure}

Here, we perform typing by folding standard stellar spectra through the \textsc{synphot} tool \emph{CALCPHOT}, a package that allows the user to calculate the photometric magnitudes observed for a given stellar type. In order to simplify our comparisons, we choose to normalise all spectra to a V-band magnitude of zero. We use the Bruzual-Persson-Gunn-Stryker (BPGS) standard stellar templates associated with the \textsc{synphot} package. Although the atlas has the broad band coverage required for our analysis, it contains few giants/bright-giants/ supergiants, which could affect our results.

\begin{figure*}
\begin{center}
\leavevmode
\includegraphics[height=56mm, angle=0]{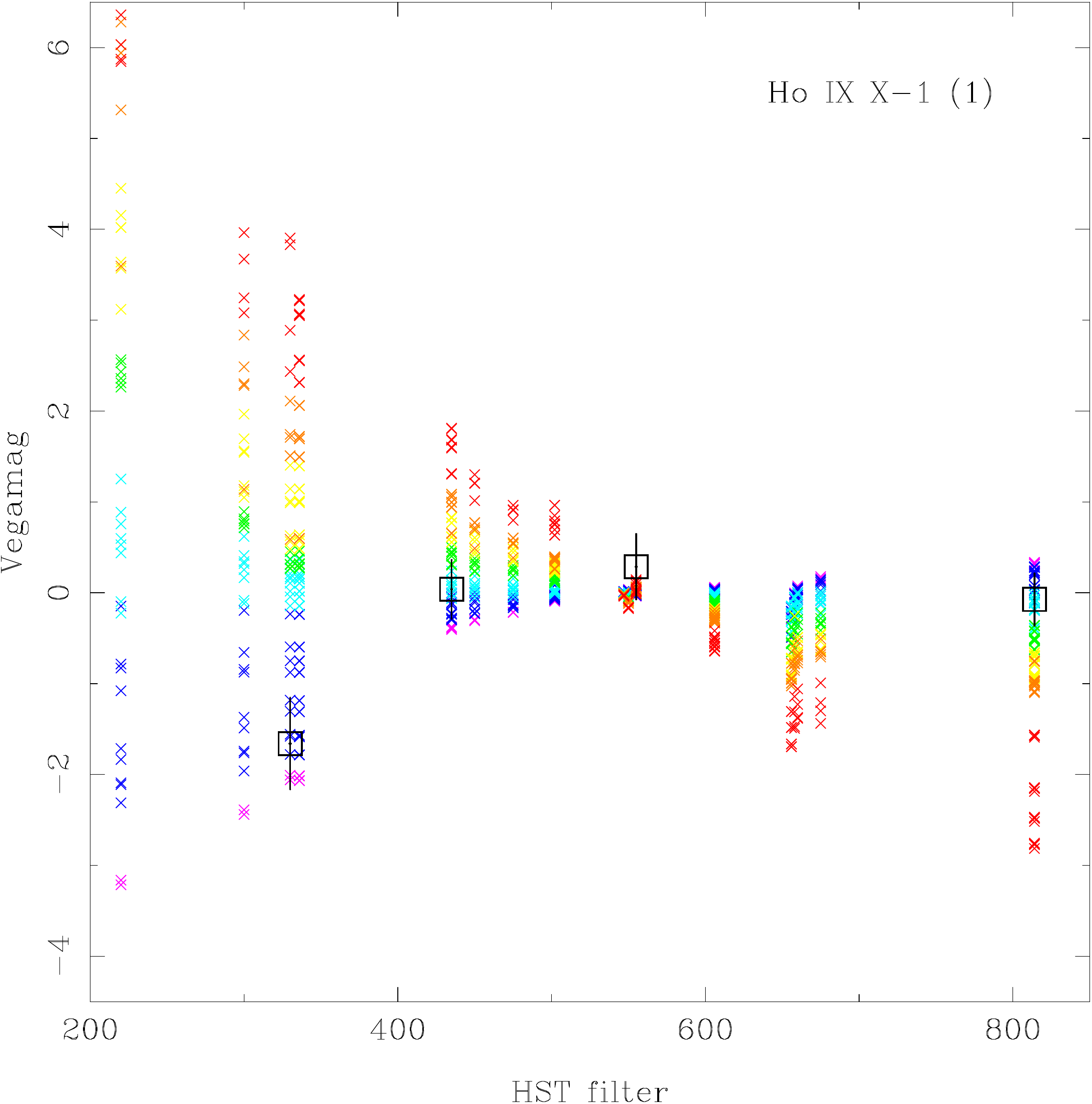} \hspace*{0.3cm}
\includegraphics[height=56mm, angle=0]{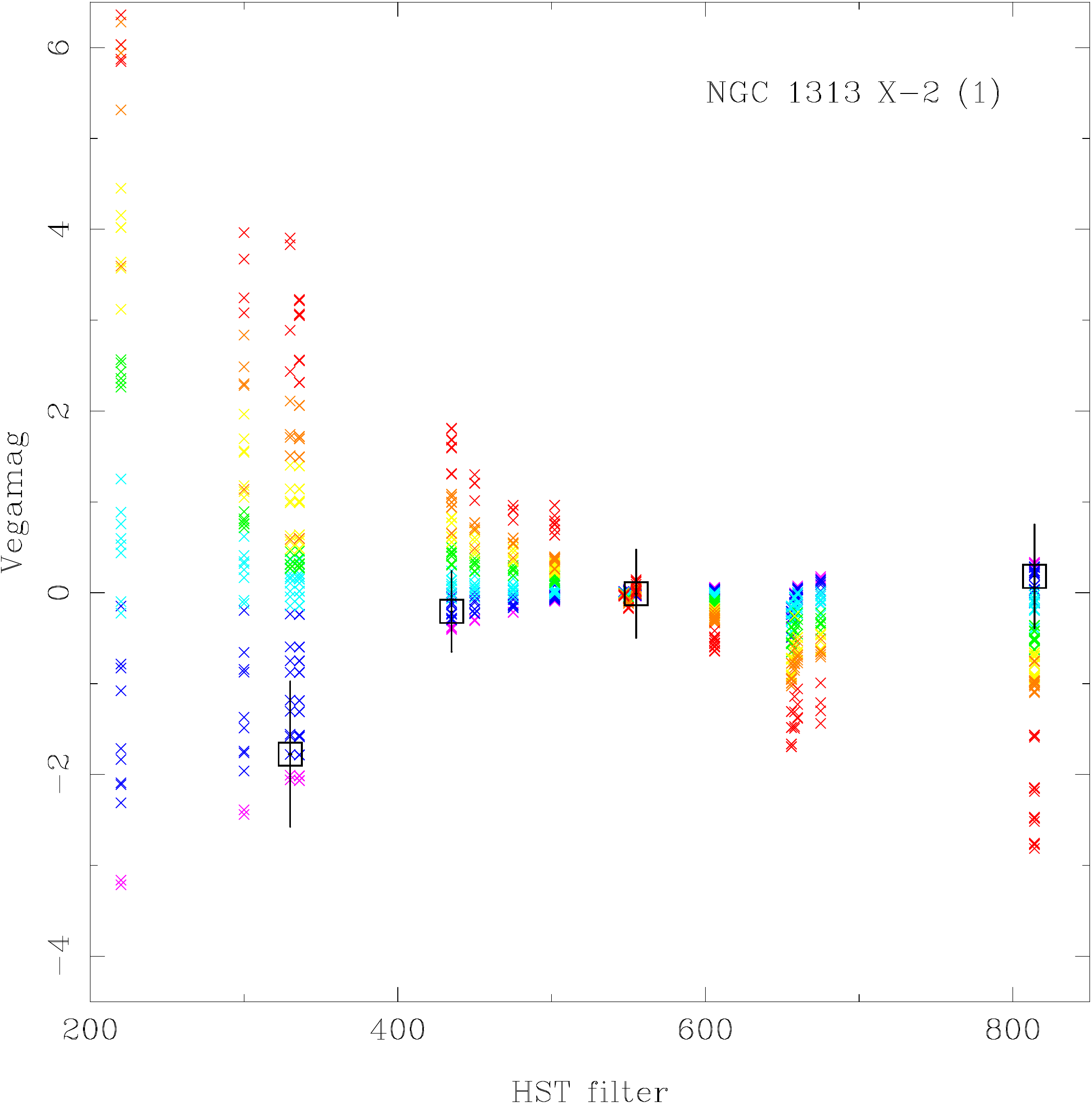} \hspace*{0.3cm}
\includegraphics[height=56mm, angle=0]{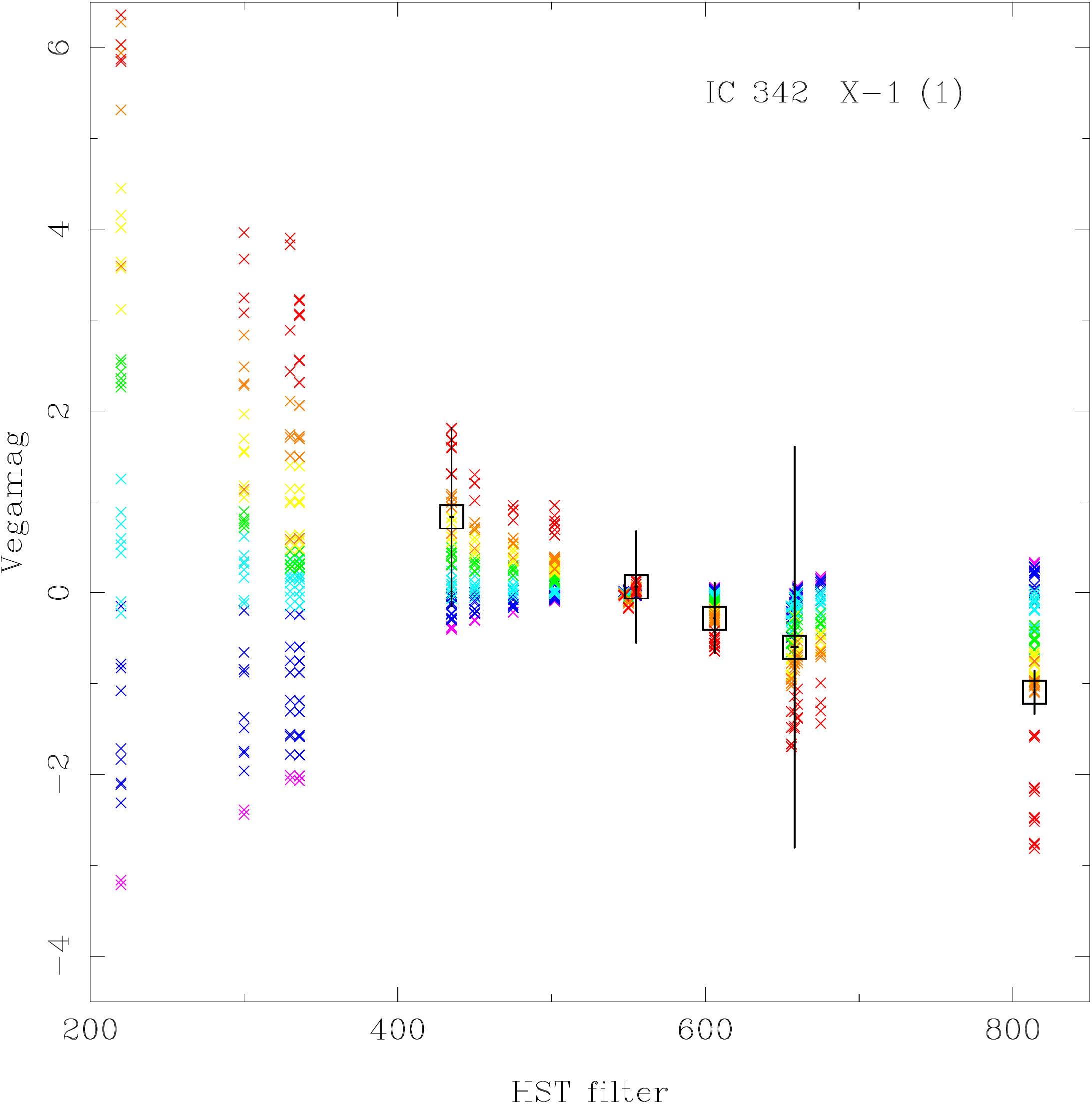}
\includegraphics[height=56mm, angle=0]{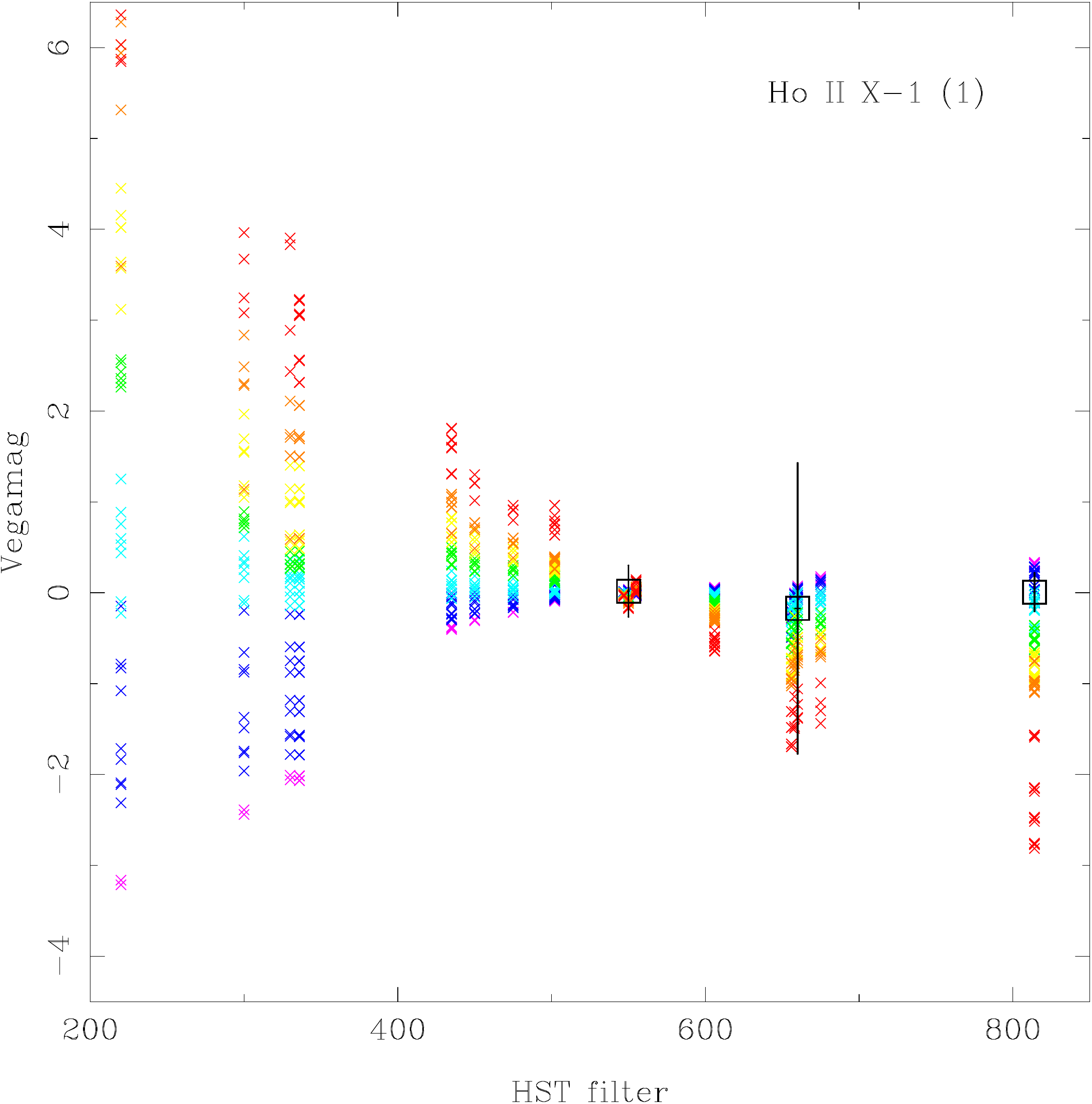} \hspace*{0.3cm}
\includegraphics[height=56mm, angle=0]{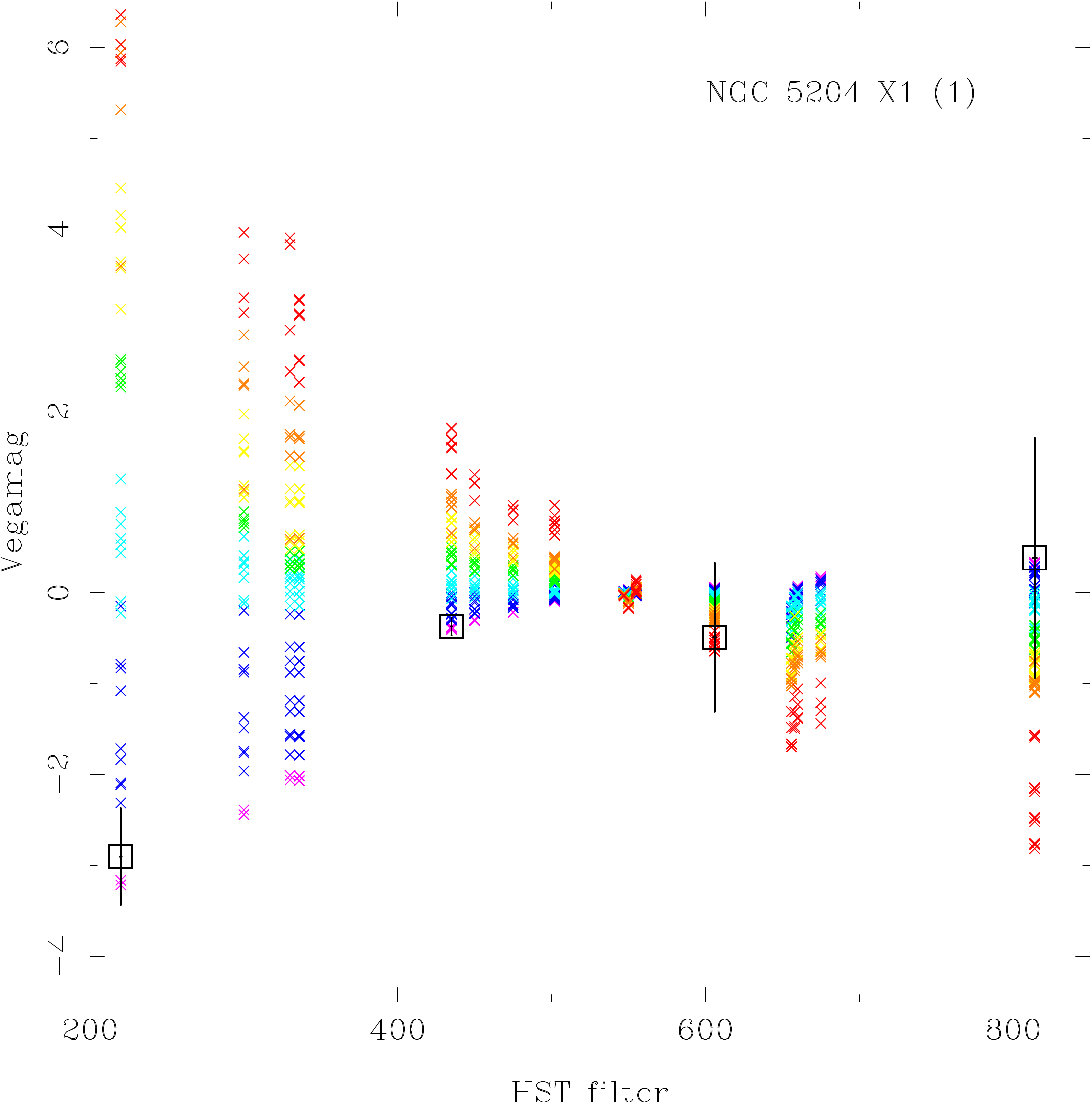} \hspace*{0.3cm}
\includegraphics[height=56mm, angle=0]{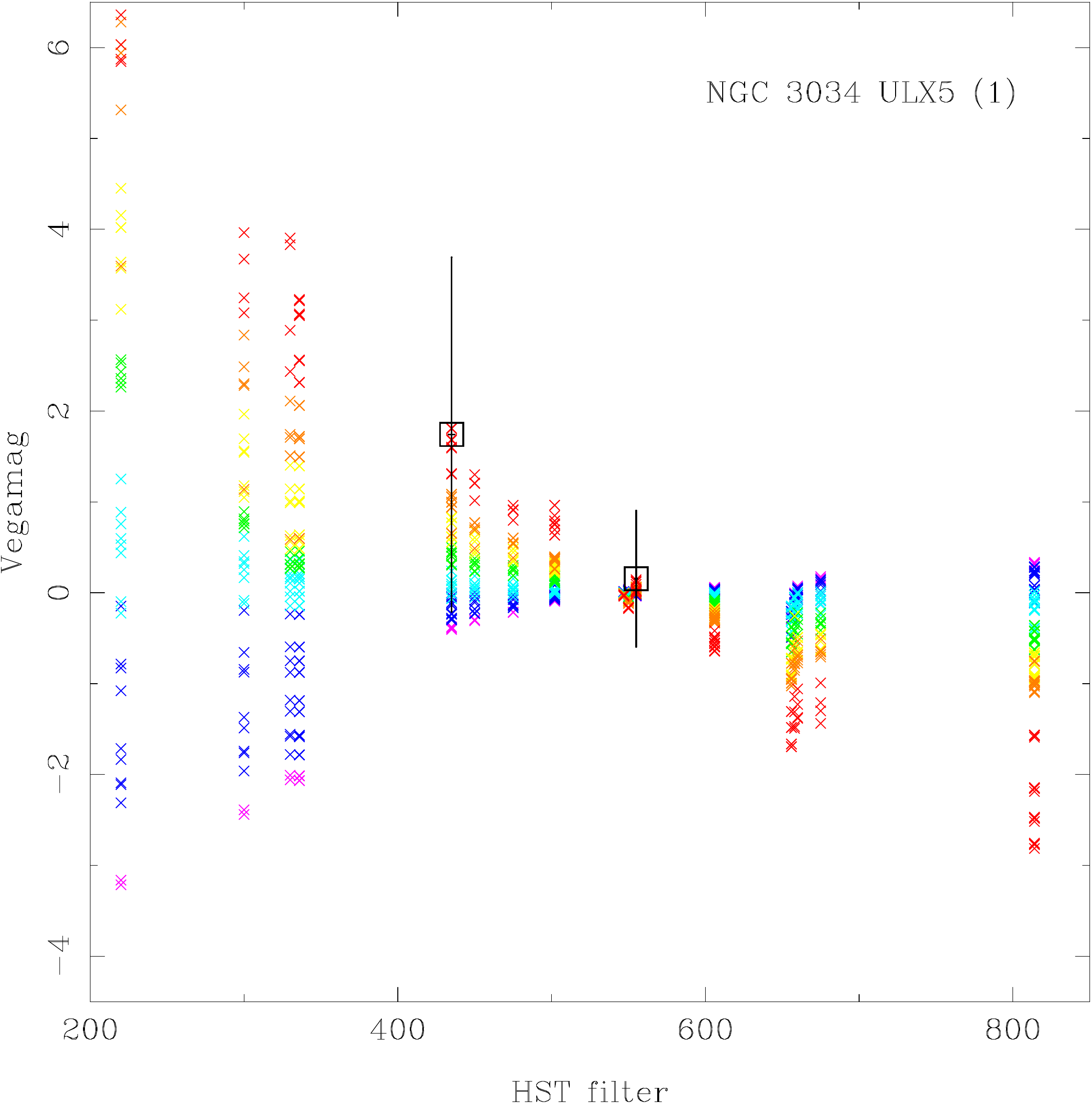} 
\end{center}
\caption{Vega magnitudes of six of our sample, plotted against the stellar templates taken from the Bruzual-Person-Gun-Stryker atlas. From here it is evident that a range of types and data quality is available within our sample. Some are well constrained, such as Holmberg IX X-1 (1), whilst others, such as NGC 3034 ULX5 (1), have such large errors as to cover the whole range of possible stellar templates.}
\label{fig:match types}
\end{figure*}

The conversion is performed for all instrument/detector/filter combinations used in the analysis of our ULX fields, with the resulting values plotted in Figure \ref{fig:star types}. Our templates range from O to M, varying in size from main sequence to supergiant. The templates are grouped by colour, according to type (listed in figure caption). 

Simple $\chi^2$ minimisation is performed to determine the best fitting stellar type, whenever more than one filter band is available, with some examples of resulting fits shown in Figure \ref{fig:match types}. This fitting requires that an offset be calculated in each instance (the shift required between its current magnitude in F555W and zero, for the best fitting model). This offset can also be considered to be the $m_{555}$ of the source (which will be similar in value to $m_V$). 

SED fitting is only possible when more than one filter band is available, however, of the 40 candidate counterparts identified, 15 are observed in only a single filter band (due to chip gaps, or depth of exposures in other filters; see Ptak et al. 2006). Of the 25 remaining potential counterparts, we note that in ten instances fitting is performed where only two bands are available; five of our sample have three data points available for fitting, seven contain four \emph{HST} bands and three contain 5 data bins for comparison to standard stellar types. The resultant types and \emph{offsets} are displayed in Table \ref{tab:mags}, with sample fits shown in Figure \ref{fig:match types}. 

The best fits achieved by this process, given in column 3 of Table \ref{tab:mags}, suggest that the majority of these 25 objects are not best fit by OB stars, as was previously suggested (e.g. Liu et al. 2007, Copperwheat et al. 2007). Only two are best fit with O-type stars, while eight prefer B-type stars, and six are best fit by M stars. The best fit luminosity classes are typically not supergiants (only two), but main-sequence (14) or giants (9). 

We list all types that can explain the observed SEDs within their errors in Column 9 of Table \ref{tab:mags}. These allow for more blue companions, with 20 now consistent with OB stars. Five of our sample appear to require later-type sources (NGC 55 ULX1 (2), NGC 253 ULX2 (1), NGC 253 XMM6 (2), IC 342 X-1 (1) \& NGC 5204 X-1 (2)). However, the redder nature of these candidates could also be a function of reddening, as a result of the host galaxy or their local environment. For example, IC 342 X-1 \& NGC 3034 ULX5 have been shown to be heavily absorbed in X-rays on many occasions, which is evident from the high values of $N_H$ listed in Table \ref{tab:full list}. Holmberg II X-1 may not show as high an absorption column, but this source resides in an excited He \textsc{ii} region (Pakull \& Mirioni 2002; Kaaret et a. 2004). Although X-rays appear not to be heavily obscured by this nebula, it may be affecting the optical emission from the star. Nebulae have also been associated with other ULXs, including NGC 5204 X-1 (Roberts et al. 2001) and IC 342 X-1 (Pakull \& Mirioni 2002). Thus, it's possible that the typing of these objects is incorrect, but this evidence for possible later-type companions to some ULXs is intriguing, and worth further study.

We also consider the absolute magnitude that would be observed from each of these stellar types (Zombeck 1990; Wegner 2006). These are combined with the distance modulus for each source (in Table \ref{tab:full list}) to derive the apparent magnitude for a star of that class at the required distance\footnote{Exact absolute magnitudes for all star types are not given in Zombeck (1990) and Wegner (2006), so to compensate for this we use absolute magnitudes for the most similar star type available within the text. This will lead to discrepancies in magnitudes of at most 1 -- 2, depending on type.}. We use the absolute magnitude for the V band to allow for easy comparison to our fits, and fold this through {\it CALCPHOT}  to derive the value for the \emph{HST} band F555W (i.e. $m_{555}$ for that stellar type). As all templates are normalised to a V band magnitude of zero, this means that the choice of instrument \& detector will have minimal impact on the derived apparent magnitude in this \emph{HST} filter band. Since more than half of our observations were taken by ACS using the WFC, we choose this combination to derive the apparent magnitudes that would be observed. The absolute magnitude, distance modulus and derived apparent F555W magnitudes are given in Table \ref{tab:mags}.

As we do not have a F555W observation for every source, we use the offset value to derive the observed absolute F555W band magnitude. To calculate the value of $M_{555}$ for each candidate counterparts, we combine the offset from fitting with the distance modulus for each potential counterpart, listing the resultant value in column 10 of Table \ref{tab:mags}. These calculations provide us with absolute magnitudes over the range $-1.4 < M_{555} < -8.2$. We then used this information in combination with the absolute magnitudes listed in Zombeck (1990) to list all possible stellar types that can be observed at approximately this absolute magnitude (see Column 11 of Table \ref{tab:mags}). As Zombeck (1990) only lists those classified with subtypes 0 or 5 for each stellar type (e.g. O5, B0, B5, etc.), we classify some stars as being early ($\la$ 5) or late ($\ga$ 5) within a stellar type. 

We compare the observed apparent magnitudes of candidate counterparts with the calculated magnitudes (using the {\it offsets} gained from $\chi^2$ fitting). To do this we group our sources into four categories for ease of comparison; those that differ by $\ga 10$ magnitudes (extreme difference), $5 \la \Delta m_{555} \la 10$ magnitudes (large difference), those that differ by $\la 2$ magnitudes (comparable), and those that lie in the range of  $2 \la \Delta m_{555} \la 5$ magnitudes (other). The ranges are designed to allow for magnitude and stellar template fitting errors, and for slight variations on tabulated absolute magnitudes while clearly identifying those that show striking differences. We find that 5 exhibit `extreme' differences, 4 display  `large' differences, 9 show `comparable' values, and 7 are classified as `other'.

The sources with an `extreme' value of $\Delta m_{555}$ are generally best fit by later-type main sequence stars. However, in most cases we would be unable to see such a object, as it would be too faint to be observed at that distance. The first is NGC 5204 X-1 (2) and is discussed in depth in Section \ref{subsection:op var}, so we will not consider it further here. The SED of NGC 253 XMM6  (1) allows for any spectral type, while $M_{555}$ limits the range to OB or later II. We will also return to this source in Section \ref{subsection:previous}.  The SEDs of the three remaining candidate counterparts only allow later-types within their errors, while in each case the  $M_{555}$ value indicates a need for bright-giants or supergiants.  This potential mis-classification is probably due to a lack of bright-giants/supergiants in the chosen catalogue.

Those sources that have a `large' value of $\Delta m_{555}$ also appear to be late-type sources. These sources are IC 342 X-1 (1) \& (2), Ho II X-1 (1) and NGC 3034 ULX5 (1). If we again consider the types allowed within errors, we find that two appear fairly well constrained (IC 342 X-1 (1) and Ho II X-1 (1)). The difference in observed and derived $m_{555}$ may be largely due mistyping due to the catalogue used. It may also be the case that this mistyping is due to reddening, which must be considered as we are using Galactic corrected magnitudes, but given the increase in errors, no fits would be achievable.

Of the remaining sixteen sources, nine display similar apparent magnitudes to their {\it offsets} ($\Delta m_{555} \la $ 2) and so are `comparable', whilst the seven remaining sources display a greater divergence. In four of these cases we were unable to collate the exact value of $M_V$ for the specified source type, which could induce 1 -- 2 magnitudes of errors. so these candidate counterparts can also be considered as `comparable' within the increased errors. Thus, we have consistent stellar typing and magnitudes for thirteen of our sources. Of these 13 sources, we find that eight are likely OB-type stars. The other five sources are a mixture of mid to late-type stars. The two potential counterparts classified as A-like stars are less well constrained, although in both cases any errors veer more towards the blue. Of the three later-type classifications (G to M), NGC 253 XMM6 (2) seems well constrained. This is very interesting, as it is unlike many previous classifications of ULX counterparts and as such, deserves further study. Such a find is also supported by the recent discovery of two near-by low-mass X-ray binary (LMXB) ULXs (e.g. Middleton et al. 2012; Soria et al. 2012). Two of our ULXs (NGC 253 ULX1 and XMM6) had multiple \emph{XMM-Newton} observations in early 2006, 8 months before the \emph{HST} observations we used, which did not show X-ray activity.  If X-ray activity did not restart during these 8 months, then these \emph{HST} observations may be given an unprecedented view of the ULX companion stars, potentially lending more support to the LMXB hypothesis as other LMXB ULXs are also transients (e.g. Middleton et al. 2012).

Previous studies have shown that many ULXs appear to have blue counterparts (e.g. Roberts et al. 2008). Many of our potential counterparts appear to agree with this, when using simple stellar templates. If we now consider this on a ULX by ULX basis (instead of each candidate counterpart in turn), we find that OB companions are possible in all but one instance - NGC 253 ULX2. This make this a key source for further study. While B-type companions are viable for all but one other ULX (NGC 5204 X-1), while we can rule out O-type companions in 20 cases. However, the apparent blue colour of these ULX optical counterparts may be due to the presence of a strong, blue, accretion disc component (e.g. Copperwheat et al. 2005; 2007). Since we did not include an accretion disc in these stellar template fits,  finding possible red counterparts is of interest. If these classifications are correct, it would suggest that the star is dominant in these filters, as any disc emission would be intrinsically blue. This implies that either the ULX was dim in X-rays during these optical observations, or that these red objects are not the true counterparts to these ULXs. 
 
\subsection{Possible optical variability; is this impacting our study?}
\label{subsection:op var}

ULXs are observed to vary in X-rays on time-scales of days or more (e.g. Roberts et al. 2006), with many showing suppressed variability on time-scales of hours (e.g. Heil et al. 2009). In Galactic X-ray binary systems, the optical emission has also been seen to vary in a way that is related to its X-ray emission (e.g. Charles \& Coe 2006). Investigations of the optical variability of these sources can be very beneficial, made evident by the recent work of Tao et al. (2011), and we consider this type of analysis a valid next step for this sample, but it is beyond the scope of the current work. This sample contains only one exposure of each band for each ULX.  Our concern here is to see if any variability in the emission from these sources is negatively impacting our analysis. We assume that observations taken within 24 hours should be minimally impacted, so we investigate sources that were observed in more than one epoch with \emph{HST}. By fitting each epoch separately, we can investigate the potential counterpart variability on these time-scales, by checking for changes in their SEDs. 

Of the 31 ULX fields considered (NGC 598 ULX1 and IC 342 ULX2 having been removed), multiple epochs were observed in 17 cases.  Ignoring cases where no counterpart was visible, or where we only detect the candidate counterpart in one band, there are only eight potential counterparts to consider; NGC 4190 X-1 (1), NGC 253 ULX3 (1), M81 X-6 (1) \& NGC 1313 X-1 (1), along with IC 342 X-1 (1) and (2) and NGC 5204 X-1 (1) and (2). To look for variability, 
we fit the SEDs from each epoch of observations separately with the same $\chi^2$ test outlined above, noting any changes in the preferred stellar type. 

NGC 1313 X-1 (1) is observed in five different energy bands, four during the same epoch. This only allows for additional fitting of one epoch. With the removal of the F606W band we find no change, suggesting  little to no variation in the SED.
 A similar result is found when fitting NGC 4190 X-1 (1). The full SED of this source contains four filters, three of which were scheduled together. By fitting only these three bands, the best fitting stellar type appears  slightly redder (B9 V). Such changes are not significant (within errors), but variations like this have been noted before in NGC 1313 X-2 (1) by Mucciarelli et al. (2007) and NGC 5055 X-2 (Roberts et al. 2008), with variations attributed to non-stellar processes (possibly from the accretion disc). 

Although IC 342 X-1 and NGC 253 ULX3 have multiple bands for each observational epoch, each of their potential counterparts are visible in only one band from a different epoch. If we remove that one band and refit, we find a statistically similar fit, with the same range of stellar templates allowable (within errors).

NGC 5204 X-1 (1) and (2) have two bluer bands from one observation (09370\_01), whilst two redder bands are from another (08601\_39). For each candidate counterpart we fit each observational epoch separately. (1) does not show any significant changes in fitting, while (2) shows an extreme change. This candidate counterpart has previously been identified and discussed by Goad et al. (2002) as a star cluster (09370\_01). When Liu et al. (2004; 08601\_39) revisited this source, they incorporated higher resolution data from the ACS HRC, which was able to resolve the source into two components, revealing the presence of an O5 V star and a redder star cluster. We find that the complete data set is best fit by a template for an F8 V star, but that the complete data is not well described by any stellar type, with fits of the blue and red bands showing a two-component fit. In the blue bands we are seeing emission from primarily the young O-type star, whilst the redder bands contain emission from both the star and the nearby cluster. This appears to confirm the suggestions of Liu et al. (2004).

Finally we return to M81 X-6.  The data for its candidate counterpart is unlike any of our available templates. This spectrum appears to be bright at both the red and the blue ends. The initial fit for all data is a B2 III, whilst two of the separate observations (09073\_01 \& 10584\_18) are fit by  an A3 III (O to F) and M0 V (O to early M) respectively. One way to explore this further is to split the spectrum by wave band, fitting the F336W, F435W and F555W photometric magnitudes in the first instance and F555W, F606W and F814W in the second. By doing this we note a large difference in the spectral types observed, and obtain the best fits for an O8 f and M0 V template star respectively. This is too large a discrepancy to be explained by variability, and would seem to indicate some form of source confusion/contamination  (c.f. NGC 5204 X-1 (2)) by the combination of emission from two stars, or a two component spectrum that could be explained by an irradiated disc and a red supergiant, the second of which is an intriguing option. Further analysis is required to confirm either scenario. 

Of the 8 potential counterparts discussed above, we find minimal impact from variability in 6 cases. In the remaining 2 cases, we find that the most extreme variations can be more easily explained by the presence of a two component spectrum. This could encompass a star and an accretion disc, or it could be a the presence of multiple stars (or a star + stellar cluster). This shows the importance of both SED construction from a single observation, and for variability studies. Each can give us valuable information on the optical counterpart of the ULX, but combining multiple epochs within a single SED can lead to misinterpretation.  

\subsection{Comparisons to previous studies}
\label{subsection:previous}

Table \ref{tab:mags} notes any previous identifications and source classifications. Where more than one candidate counterpart is present in our sample, we list the previous identification alongside the counterpart matching that referred to in the literature. Thirteen potential counterparts have been previously identified from our sample (discounting NGC 598 ULX1), although IC 342 X-2 (1) was unable to be classified in previous works. We find that we are still unable to classify it with current archival data. Although NGC 5204 X-1 (1) \& (2)  and M81 X-6 (1) were previously classified, and are discussed in detail in Section \ref{subsection:op var}. As a result we will not discuss them further in this section.

Of the nine cases remaining, we find that four of our stellar type ranges are in agreement with previous results (Ho IX X-1 (1), NGC 1313 X-2 (1), NGC 2403 X-1 (1) \& NGC 5408 X-1 (1)\footnote{A recent study by Tao et al. (2011) proposes that the emission from some of these sources may be non-stellar in origin. Here we are only comparing stellar type classifications, we will discuss other forms of optical emission later in the text, hence we do not discuss their work further at this point.}). In two further cases, the authors attempt to take intrinsic reddening into account, altering their result. This occurs for IC 342 X-1 (1) and Holmberg II X-1 (1). We find these sources to be of types KO IV (IC 342 X-1 (1), possible types cover G -- K range) and A3 III (Holmberg II (1), with errors covering B -- A types), whilst previous work found these to be an F type supergiant (Feng \& Kaaret 2008; Roberts et al. 2008) and an OB-type possible supergiant (Kaaret et al. 2004) respectively. This demonstrates that intrinsic reddening can have a large impact in the classification of such objects.

We are unable to classify two of the remaining cases. The first of these is NGC 5128 X-1, for which we detect no counterpart. The candidate counterpart to this source was initially identified by Ghosh et al. (2006), as an OB star. However, Ghosh et al. (2006) performed their own data reduction to get the deepest image possible. This would suggest that the data retrieved from the HLA is not maximised for depth of image, so if we wish to consider the fainter sources the data should be reduced accordingly. The same problem arises with the faint candidate counterpart to Circinus ULX1, where we are unable to obtain good constraints on the magnitudes of this candidate counterpart. 

A recent study by Yang et al. (2011) identified a new potential counterpart to NGC 1313 X-1, labeled as NGC 1313 X-1 (1) in this study. The authors studied all of the available \emph{HST} data on this source, finding some variability in the F555W band on inter-observational time-scales. They attributed this to variations in the accretion disc, and so used only the redder bands to type the companion star to this object. Their analysis indicated the presence of a late-type giant, possibly a K5-M0 II star. However, the absolute F555W magnitude obtained from this work suggests that this source is too bright to be explained by a star of this class. It requires a younger bright giant or supergiant to explain the observed luminosity (assuming only a stellar origin). As our initial fitting used the entire \emph{HST} SED, our fitting is dominated by the blue component, which affects our classification. Their work highlights the need to consider the variability of these systems (which we tested in Section \ref{subsection:op var}), and the need to consider the presence of the accretion disc in these extreme systems. Each of these are discussed further in the next section of this paper.

\subsection{Accretion disc emission}
\label{subsection:disc}

The final point that should be considered when looking at the magnitudes and typing of these systems is the presence of an accretion disc. In accreting X-ray binary systems, optical radiation is released from both the companion star and the accretion disc. The presence of such a disc would increase the emission and change the shape of the source spectrum. The colour of the star and disc will also be changed by the X-ray irradiation of the disc and companion star. In order to explore this further we apply current theoretical models designed to describe such systems. Attempts have been made to create and apply such models (e.g. Copperwheat et al. 2005, 2007; Patruno \& Zampieri 2008; Madhusudhan et al. 2008),  which indicated that the most likely counterpart to a ULX would be a high-mass donor performing mass transfer via Roche lobe overflow, although their findings differ dramatically for the resulting black hole masses with some suggesting MsBHs while others prefer IMBHs. 

Copperwheat et al. (2005) used irradiation models, in combination with models of OB main sequence stars and four supergiants ranging from F to M, to explore the resulting emission from the system. Their work indicated that the emission from ULXs would be impacted greatly, observing a large brightening in the observed magnitude due to the irradiation of the disc and companion star (a change of $\sim$ 0.5 -- 5 magnitudes, depending on the disc size, companion star type, X-ray hardness and the filter band). Such a change in the absolute magnitude of these systems could help to explain some of the $\Delta m_{555}$ values observed in our sample, possibly even including those in the Circinus galaxy. Copperwheat et al. (2007) applied this model to the candidate ULX optical counterparts known at the time, to constrain the parameters of those systems. This assumes that we are observing a binary system that contains a compact object and a companion star, with the accretion disc being fed by Roche lobe overflow (irrespective of the companion star's mass). No assumption is made on the mass of the compact object, with masses spanning the MsBH and IMBH range (10 -- 1000 $M_\odot$), with a wide range of stellar masses and radii also available. Mass accretion rates are inferred from the X-ray luminosity of the system, with the optical emission incorporating light from both the irradiated star and the accretion disc. We refer the reader to Copperwheat et al. (2005; 2007) for a more detailed discussion of the model and its application.

Here we apply the same model to our current sample. To do this, we require X-ray flux ratios for each ULX for which we have possible optical counterparts. Ideally this should be derived from data taken concurrently with that of the optical data. This ideal case would allow us to understand the X-ray emission of the system at the time our optical data were observed, which would have implications on the amount of X-ray re-ionisation. However, since we are using archival data, this is generally not possible, so we work on a best efforts basis, combining the X-ray and optical data in order to obtain some constraints on the nature of the system. Phenomenological models can provide general constraints on the shape of low quality spectra, but the absorption columns can vary widely depending on model choice and data quality. We use published results, searching for statistically sound fits to either \emph{Chandra} or \emph{XMM-Newton} data with more physically motivated models. Whenever these are unavailable, we consider phenomenological fits to the data. These models are then read into \textsc{xspec} to derive flux ratios for each source. The model flux is obtained for the 0.3 -- 1.0 \& 1.0 -- 10.0 keV ranges, first with the Galactic absorption column removed and then with the intrinsic model fit. The derived flux ratios are listed in Table \ref{tab:flux rat}, along with the relevant models and references. 

The resulting Galactic absorption corrected flux ratios are combined with the Galactic extinction corrected optical magnitudes to provide a multi-wavelength view of the emission from these systems. Likewise, the intrinsic X-ray flux ratios were also combined with the intrinsic optical magnitudes for fitting. Both Galactic corrected and intrinsic values are fit with models from Copperwheat et al. (2005; 2007) to consider each system in two states: one in which the donor is in superior conjunction and the inclination is cos({\it i}) $=$ 0.0 (the optical light is dominated by the irradiated star, there is no optical disc emission); and the same conjunction with an inclination of cos({\it i}) $=$ 0.5 (both a star and disc contribution, although the ratio of these components will vary in general the disc contribution tends to become dominant). This is carried out for each potential counterpart for which we have available X-ray spectra and multiple optical bands.

We have obtained magnitudes in multiple optical bands for 25 candidate counterparts of 18 ULXs. We list constraints on the binary parameters for the Galactic-corrected optical magnitudes (Table \ref{tab:copperwheat1}) and intrinsic magnitudes (Table \ref{tab:copperwheat2}).  In each case we consider the candidate to be the true counterpart to the ULX, we treated cos({\it i}) $=$ 0.0 and cos({\it i}) $=$ 0.5 cases separately. We select some of the more well known ULXs and some interesting cases from both Galactic corrected and intrinsic case and provide their full fits in Figure \ref{fig:copperwheat}. In each case, the figure captions contain the main findings, constructed using both the Figures and Tables \ref{tab:copperwheat1} \& \ref{tab:copperwheat2}. 

Initially we consider only the fits from the Galactic corrected optical magnitudes. As previously stated, this can be considered a lower limit for the extinction of these sources. 

\begin{figure*}
\leavevmode
\begin{center}

\includegraphics[height=135mm, angle=0]{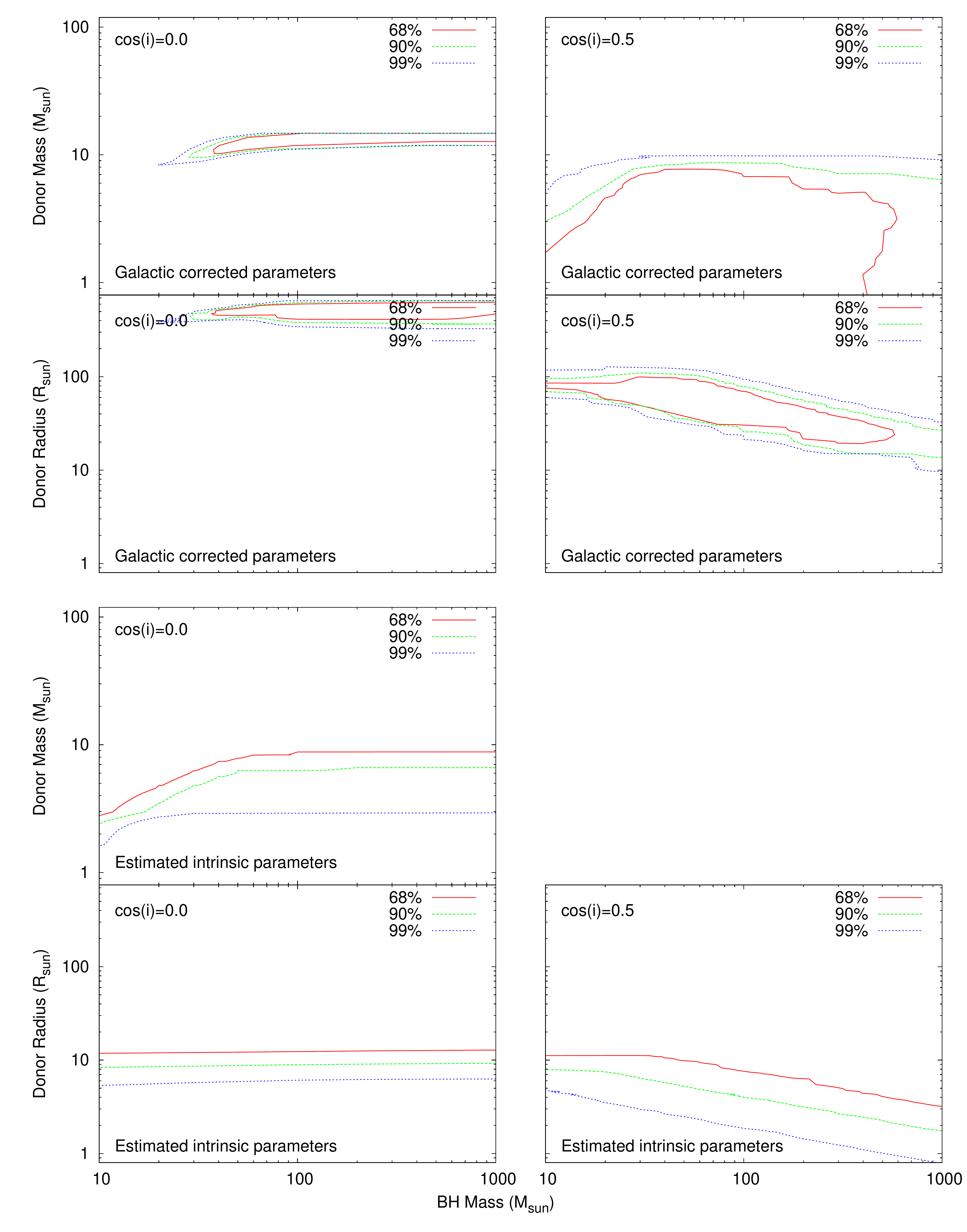}

\end{center}
\caption{
\small{
{\it pg. 1} -  Confidence contours for select candidate ULX counterparts. Parameters are plotted in two groups of four for each ULX; top four containing fits achieved using Galactic corrected data while the bottom four relate to intrinsic magnitudes. Left-hand panels represent a superior conjunction with an inclination of cos({\it i}) $=$ 0.0 (observed emission dominated by irradiated star). Right hand panels are for an inclination of cos({\it i}) $=$ 0.5, again at superior conjunction (emission from both irradiated star \& disc, although generally dominated by the disc). Panels are missing when no constraints are obtained. Here are fits for NGC 253 ULX2 (1). Galactic extinction/absorption corrected data suggests that $M_{BH} \ga 38 M_\odot$ for an edge on system, where the companion mass and radius constraints indicate a late-type giant to be a likely counterpart (comparing to Zombeck 1990). When the inclination is set to cos({\it i}) $=$ 0.5, then $M_{BH} \la 590 M_\odot$ at the 1 $\sigma$ level. Here, stellar constraints suggest either an OB-type companion (in which case the source would need to be heavily reddened) or a later-type giant. Switching now to the intrinsic fits,  we lose many constraints. We only obtain lower limits on the star's radius when the system is inclined, while we obtain lower limits on both the stellar mass and radius at cos({\it i}) $= 0.0$. The mass constraints show that the system cannot be explained by a LMXB, but that either an intermediate or high-mass companion is possible, where stellar radius constraints tell us that we are observing either an OB star or a giant or supergiant (Zombeck 1990). }}
\label{fig:copperwheat}
\end{figure*}

For 10 of the potential counterparts, we are able to not only constrain the mass and radius of the companion at the $1 \sigma$ level, but also the mass of the black hole in the system, for certain assumed inclinations. They are NGC 55 ULX1 (1) \& (2), NGC 253 ULX1 (3), NGC 253 ULX2 (1), NGC 253 XMM6 (1),  NGC 253 XMM6 (2), Ho IX X-1 (1) \& (3), IC 342 X-1 (1) and NGC 5204 X-1 (2). These findings are summarised in Table \ref{tab:BH mass}. When comparing these to the rest of the population of candidate counterparts, we find that they generally have smaller errors than the other potential counterparts. Eight of these also seem to be redder in colour, with the exceptions being Ho IX X-1 (1)  \& (3).

Two of these sources, NGC 253 ULX2 (1) and NGC 5204 X-1 (2), show both upper and lower black hole mass constraints, although each are provided for a different inclination. We find that we are only able to constrain the lower black hole mass limit from NGC 253 ULX2 (1) when we assume that all of the observed optical emission is from the irradiated companion star (cos({\it i}) $=$ 0.0). Here, fitting implies that the black hole must be greater than $\sim$ 37.5 $M_\odot$, a mass that is larger than any observed MsBH seen to date, but still covering both the massive stellar and intermediate mass regimes. When we consider the alternative scenario, in which we observe emission from both the companion star and the accretion disc (cos({\it i}) $=$ 0.5), we obtain an upper limit of 590 $M_\odot$. Again, this covers all classes  of black hole.  

In the case of NGC 5204 X-1 (2), we see the reverse. We obtain an upper limit on the mass when only considering emission from the companion, while we obtain a lower limit when cos({\it i}) $=$ 0.0. These are $M_{BH} \la 20 M_\odot$ \& $M_{BH} \ga 15.9 M_{\odot}$, respectively. In the first case, this would imply that we are observing a regular sMBH, similar to those seen in our own galaxy, but accreting at a much higher accretion rate (if this is the correct companion). However, if the inclination is increased, so that we also see some of the accretion disc in the optical bands, the lower limit allows for any category of black hole. However, its two component nature is thought to be a result of a star cluster and an O5 star. As a result it is not considered the likely companion to this ULX.

In two cases, we obtain lower-mass limits at an angle of cos({\it i}) $=$ 0.0. These are Ho IX X-1 (1) and IC 342 X-1 (1). In each case we obtain a lower limit of $\sim$ 19 $M_{\odot}$, placing their lowest mass at the upper end of those observed in our own galaxy. In each of the remaining six cases, we obtain upper black hole mass limits, under the assumption that the observed optical data incorporates both emission from the irradiated disc and companion star (cos({\it i}) $=$ 0.5). NGC 55 ULX (1) \& (2), NGC 253 ULX1 (3) and NGC 253 XMM6 (2) all have upper limits that are approximately $<$ 400~$M_{\odot}$, such that no classification on black hole type can be made. However, NGC 253 XMM6 (2) and Ho IX X-1 (3) have upper mass limits that lie solely within the range of massive MsBHs, with masses $< 85 M_\odot$, although we should note that Ho IX X-1 (1) is thought to be the more likely counterpart, due to He II emission in its optical spectra (e.g. Roberts et al. 2011).

In cases where we have only constraints on the companion, we find some general trends emerging. When the system is inclined such that the disc is edge-on (so that emission is purely from the companion star), we find that the mass and radius ranges of the star tend to increase with increasing black hole mass. The opposite trend is present when emission is also thought to come from the disc. There are a few instances when this is not the case. In four cases (NGC 253 ULX3 (1), M81 X-6 (1), NGC 1313 X-1 (1) \& NGC 1313 X-2 (1)) we see the opposite trends occurring, while Ho IX X-1 (1) \& IC 342 X-1 (1) show approximately the same values across the range of black hole masses.

What can these stellar mass ranges tell us about the system? Using the following approximate mass ranges, we are able to classify the potential companion stars of these ULXs. Low mass stars are considered to be those of $\la 1 M_\odot$, those in the range of $1 \la M_* \la 10 M_\odot$ are intermediate, while those with $M_* \ga 10 M_\odot$ are considered high-mass stars. We find that, although it was previously thought that these systems are HMXB, the presence of low-mass stars cannot be ruled out in 27 cases of the 52 tested (26 potential ULX counterparts in 2 scenarios).  Intermediate mass stars are possible in 41 cases, while high-mass stars cannot be ruled out in 34 cases. If we fold in the observed absolute magnitudes of these candidates (from Table \ref{tab:mags}), we can again compare to Zombeck (1990) to see how many of these ULXs could be playing host to OB companions. Of the 18 ULXs that had potential counterparts available for fitting, all can hold B stars (depending on choice of inclination and black hole mass), while only 8 can contain O stars. 

Due to the increased errors on our intrinsic magnitudes, Table \ref{tab:copperwheat2} shows that we are only able to obtain one black hole mass constraint from the 52 cases considered. The fits to Ho IX X-1 (3) provide an upper bound of 350 $M_\odot$ in the case of cos({\it i}) $=$ 0.5. This is also listed in Table \ref{tab:BH mass}. We are also unable to obtain mass constraints on the companion in 30 of the 52 cases, 14 of which also give no constraints on the radius of the companion. Where constraints are achieved, they follow the same trends as those outlined from Galactic extinction corrections, however we have many more lower limits on the star's mass and radius, as constraints on their upper bounds are lost. Where we have constraints, we attempt to classify the candidate counterparts as low, high or intermediate mass, we find that 7 cases can be described by a system containing a low-mass companion, that an intermediate mass star cannot be ruled out in 20 cases, while 22 cases may contain a high-mass companion.

\section{Summary}
\label{section:summary}

Here we present the findings of our survey of the potential optical counterparts to ULXs, that combines data from both the \emph{Hubble} Legacy Archive and the \emph{Chandra Space Telescope}. We collate information pertaining to those ULXs residing within $\sim$ 5 Mpc, and search for any potential counterparts. We find that from our initial sample of 45 ULXs, 12 have no archival data. In the remaining 33 cases we collated data from each telescope and correct the astrometry of the downloaded data. By cross-correlating \emph{Chandra} and \emph{HST} field, we found two of the sample to reside within the nucleus of their host galaxies, they were therefore removed from our analysis. We find 22 of the 31 remaining ULXs show the presence of candidate optical counterparts, with 13 ULXs having a single optical candidate in the X-ray error region. Nine of our sample have no observed counterpart within the error region, although as Ptak et al. (2006) highlighted in some cases this will be due to insufficient depth in the exposures of these fields. The remaining 22 ULXs have a total of 40 potential counterparts, 26 of which are observed in multiple bands affording us the opportunity to attempt classification. It is obvious that not all can be the true counterparts to these ULXs, so we derive the number of chance coincidences to remove these from our sample. This suggests that 13 $\pm$ 5 of the detected counterparts are correct for the 22 ULXs considered. When we remove Circinus 1 and NGC 3034 sources from the catalogue, this changes to 15 $\pm$ 4 for 19 ULXs.

We find that initial identifications of potential counterparts show no prevalence of a single stellar type. Classifications cover the wide range of types from blue OB stars to red M types, and range in size from main sequence to supergiants (that are possibly reddened). When considering the derived absolute magnitudes of these sources in the F555W filter band ($\simeq M_V$), the results are more suggestive of giants/bright-giants/supergiants in the majority of cases, although some appear too bright to be explained by even the most luminous stars. The presence of such luminous objects indicates that, in some cases we are either observing foreground sources that are not related to the ULX, or that the stellar emission is enhanced by emission from an irradiated star and/or accretion disc. Such emission could easily brighten the system by up to $\sim$ 5 magnitudes (Copperwheat et al. 2007), in agreement with the observed disparity in optical flux. If instead we combine the range of possible stellar types with the derived absolute magnitudes, this indicates that we are mainly observing OB-type stars, with OB stars ruled out for only 1 ULX - NGC 253 ULX2. This source has only one detected counterpart, a red SED that can only be explained by late K or M stars, while its absolute V magnitude is $\sim$ - 6.2. However, this magnitude and SED fitting was obtained from the Galactic-corrected magnitudes, so would it be reasonable in the intrinsic case?  If we compare galactic and total columns we find that $N_H$ goes from 3 to 20 ($\times 10^{20}$ cm$^{-2}$). This means that $E(B - V)$ changes from $\sim$ 0.06 to $\sim$ 0.8, a relatively small change. However, the errors on the derived $E(B - V)$ values would be considerably larger due to the large uncertainties in $N_H$. This means that we are unable to explore this option at present. To test this further we would need great constraints on the extinction/absorption of this system, constraints that could be achieved using deeper X-ray observations.

The application of X-ray irradiation models provides constraints on the black hole mass in only 10 cases, when fitting each of the potential counterparts using only the Galactic corrected X-ray and optical values (assuming in each case that this is the correct counterpart to the ULX). However, constraints are limited. In one case, the limit suggests a sMBH (NGC 5204 X-1 (2), cos ($i$) = 0.0), while in another the companion can be either a sMBH or a MsBH(Ho IX X-1 (3), cos ($i$) = 0.5). These are interesting results as they agree with the current theory regarding these more standard ULXs. However, in another instance, only a sMBH is ruled out (MsBH \& IMBHs allowable; NGC 253 ULX2 (1), cos ($i$) = 0.0). We find that the fits from five of these cases provide an upper limit on the black hole mass of the order of hundreds of $M_{\odot}$ (NGC 55 ULX1 (1) \& (2), cos ($i$) = 0.5; NGC 253 ULX1 (3), cos ($i$) = 0.5; NGC 253 ULX2 (1), cos ($i$) = 0.5) , while the three remaining cases cannot rule out any classification of black hole (Ho IX X-1 (1), cos ($i$) = 0.0; IC 342 X-1 (1), cos ($i$); NGC 5204 X-1 (2), = 0.0, cos ($i$) = 0.0). We lose almost all constraint in the intrinsic case, obtaining only 1 upper limit of 350 $M_\odot$ for Ho IX X-1 counterpart 3 when the system is inclined. 

We also obtain companion stellar constraints in some cases for both galactic extinction/absorption corrected values and intrinsic data. We find that, although it was previously thought that these systems are HMXBs, the presence of low-mass stars cannot be ruled out in 27 cases of the 52 tested (26 potential ULX counterparts in 2 scenarios) for Galactic corrected values, while 7 show that low-mass companions lie within acceptable mass and radius ranges for the intrinsic case. Intermediate mass stars are possible in 41 cases, while high-mass stars cannot be ruled out in 34 cases for Galactic corrected magnitude/flux ratio fitting, while 20 \& 33 intrinsic cases can be explained by intermediate or high mass stars respectively. 

This work has also highlighted several sources for which additional photometric or spectroscopic analysis could provide interesting science. NGC 253 is a galaxy containing two transients that may have been turned off at the time of the archival \emph{HST} observations. Another interesting thing to note for the companions in this galaxy, is that they appear to be very red, and well-fit by later-type companions. Follow-up photometric analysis of the stars in this galaxy could give greater constraints on possible companion types, while new deeper observations, taken with simultaneous X-ray data would confirm the level of X-ray emission from the transient ULXs, and show any change in optical emission from these sources. This analysis has also revealed several good candidates for optical spectroscopic follow-up, 5 of which have been successfully awarded time with the Gemini Observatory as part of our ongoing programme (NGC 1313 X-2 (1),  NGC 5204 X-1 (1) \& Ho IX X-1 (1), NGC 4395 X-1 (1), \& NGC 253 ULX2 (1), a number of which will be discussed in Gladstone et al. (\emph{in prep}), while two others have been studied by alternate groups (NGC 5408 X-1 (1), e.g. Cseh et al. 2011 \& Gris{\'e} et al. 2012; Ho II X-1 (1) PI: Liu). 

\section*{Acknowledgements}

We thank the anonymous referee for their helpful comments in improving the content of this paper. JCG gratefully acknowledges funding from the Avadh Bhatia Fellowship and from an Alberta Ingenuity New Faculty Award (PI CH). CMC was funded by grant ST/F002599/1 from the Science and Technology Facilities Council (STFC). COH acknowledges funding from Alberta Ingenuity and from NSERC Discovery Grants, and conversations with G.~R. Sivakoff.

This work is based on data from the \emph{Chandra} satellite, which is operated by the National Aeronautics and Space Administration (NASA). It is also based on observations made with the NASA/ESA Hubble Space Telescope, and obtained from the Hubble Legacy Archive, which is a collaboration between the Space Telescope Science Institute (STScI/NASA), the Space Telescope European Coordinating Facility (ST-ECF/ESA) and the Canadian Astronomy Data Centre (CADC/NRC/CSA). Funding for the SDSS and SDSS-II has been provided by the Alfred P. Sloan Foundation, the Participating Institutions, the National Science Foundation, the U.S. Department of Energy, the National Aeronautics and Space Administration, the Japanese Monbukagakusho, the Max Planck Society, and the Higher Education Funding Council for England. The SDSS Web Site is http://www.sdss.org/. The SDSS is managed by the Astrophysical Research Consortium for the Participating Institutions. The Participating Institutions are the American Museum of Natural History, Astrophysical Institute Potsdam, University of Basel, University of Cambridge, Case Western Reserve University, University of Chicago, Drexel University, Fermilab, the Institute for Advanced Study, the Japan Participation Group, Johns Hopkins University, the Joint Institute for Nuclear Astrophysics, the Kavli Institute for Particle Astrophysics and Cosmology, the Korean Scientist Group, the Chinese Academy of Sciences (LAMOST), Los Alamos National Laboratory, the Max-Planck-Institute for Astronomy (MPIA), the Max-Planck-Institute for Astrophysics (MPA), New Mexico State University, Ohio State University, University of Pittsburgh, University of Portsmouth, Princeton University, the United States Naval Observatory, and the University of Washington.

\onecolumngrid


\begin{figure*}
\leavevmode
\begin{center}

\includegraphics[height=64mm, angle=0]{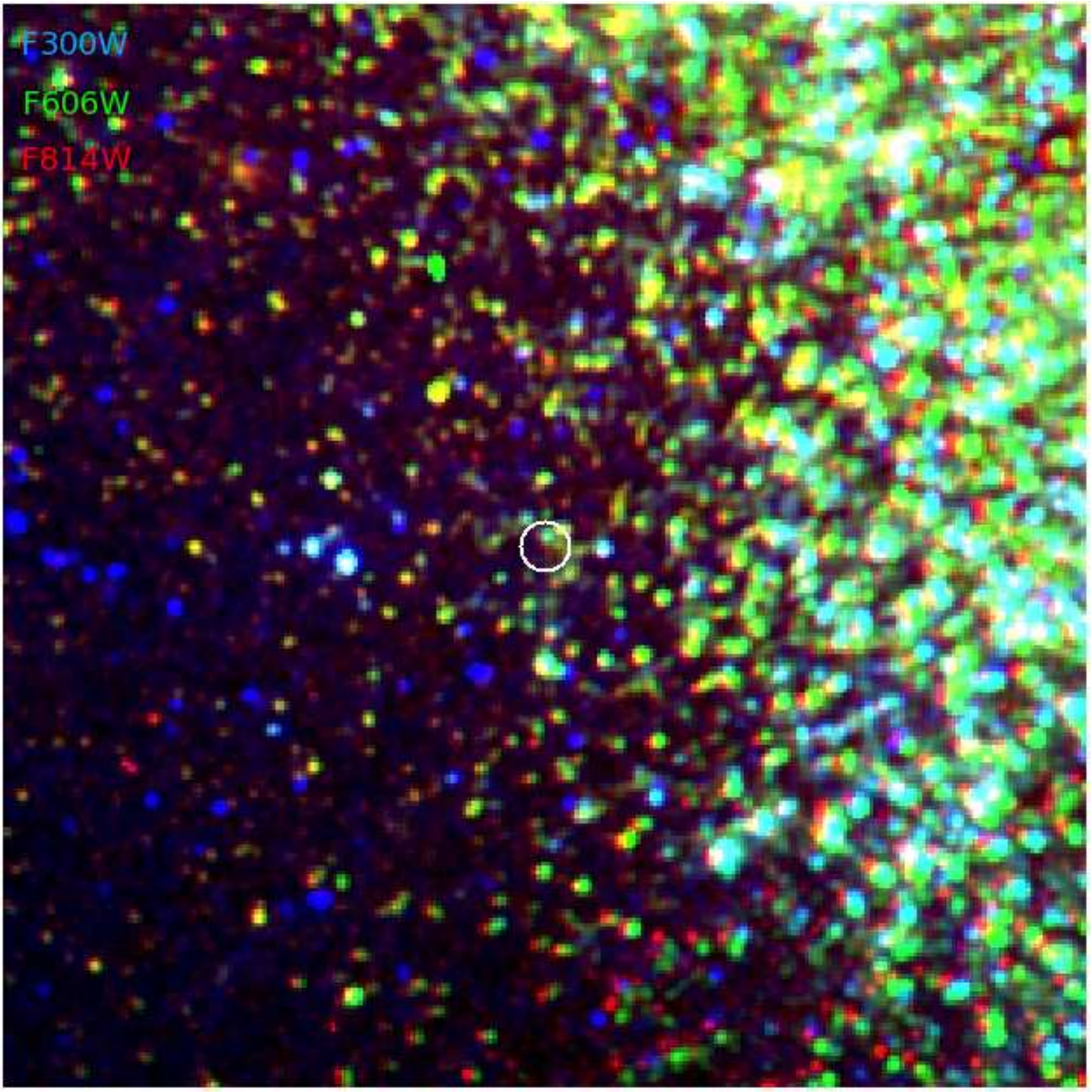} \hspace*{0.4cm}
\includegraphics[height=64mm, angle=0]{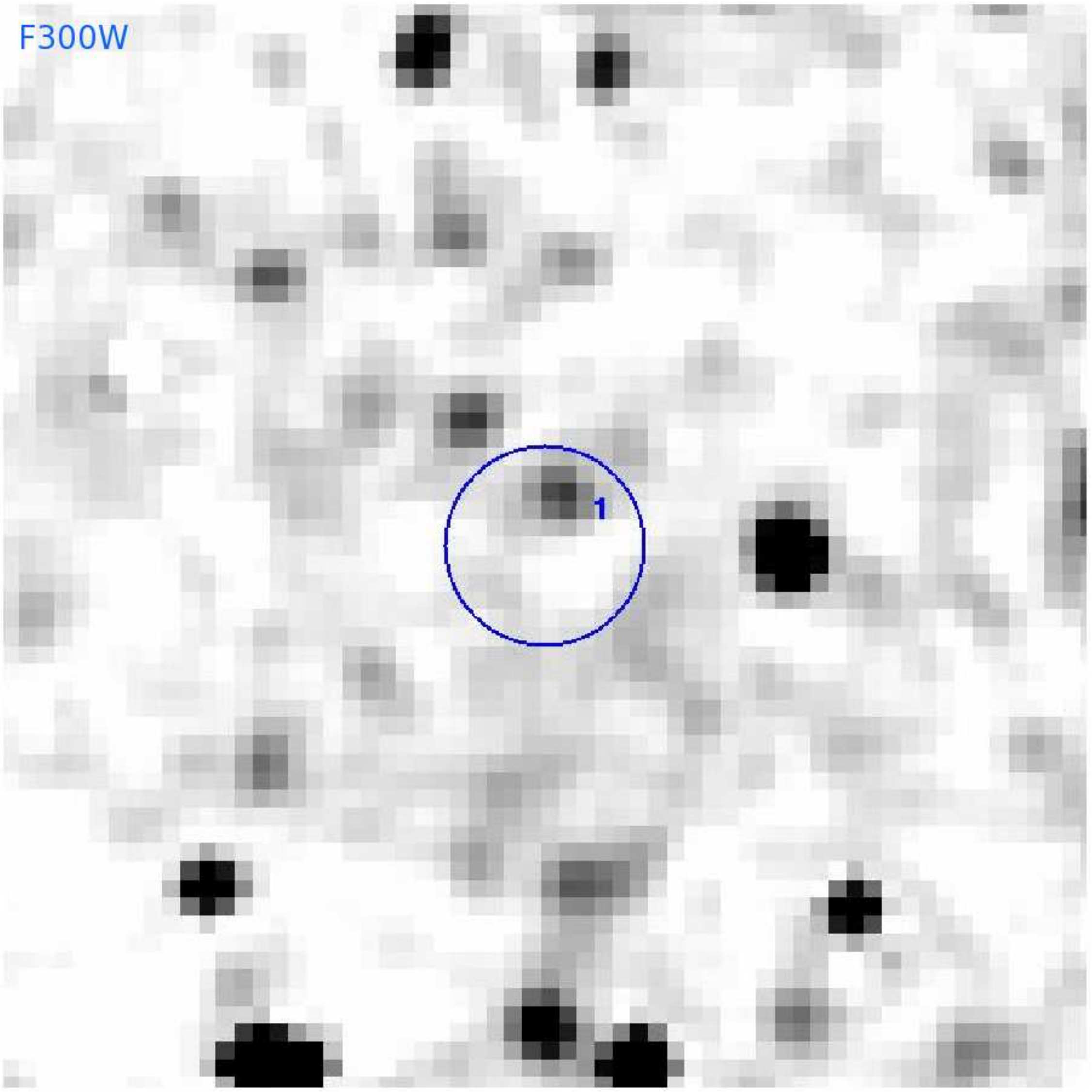} 
\vspace*{0.3cm} 

\includegraphics[height=64mm, angle=0]{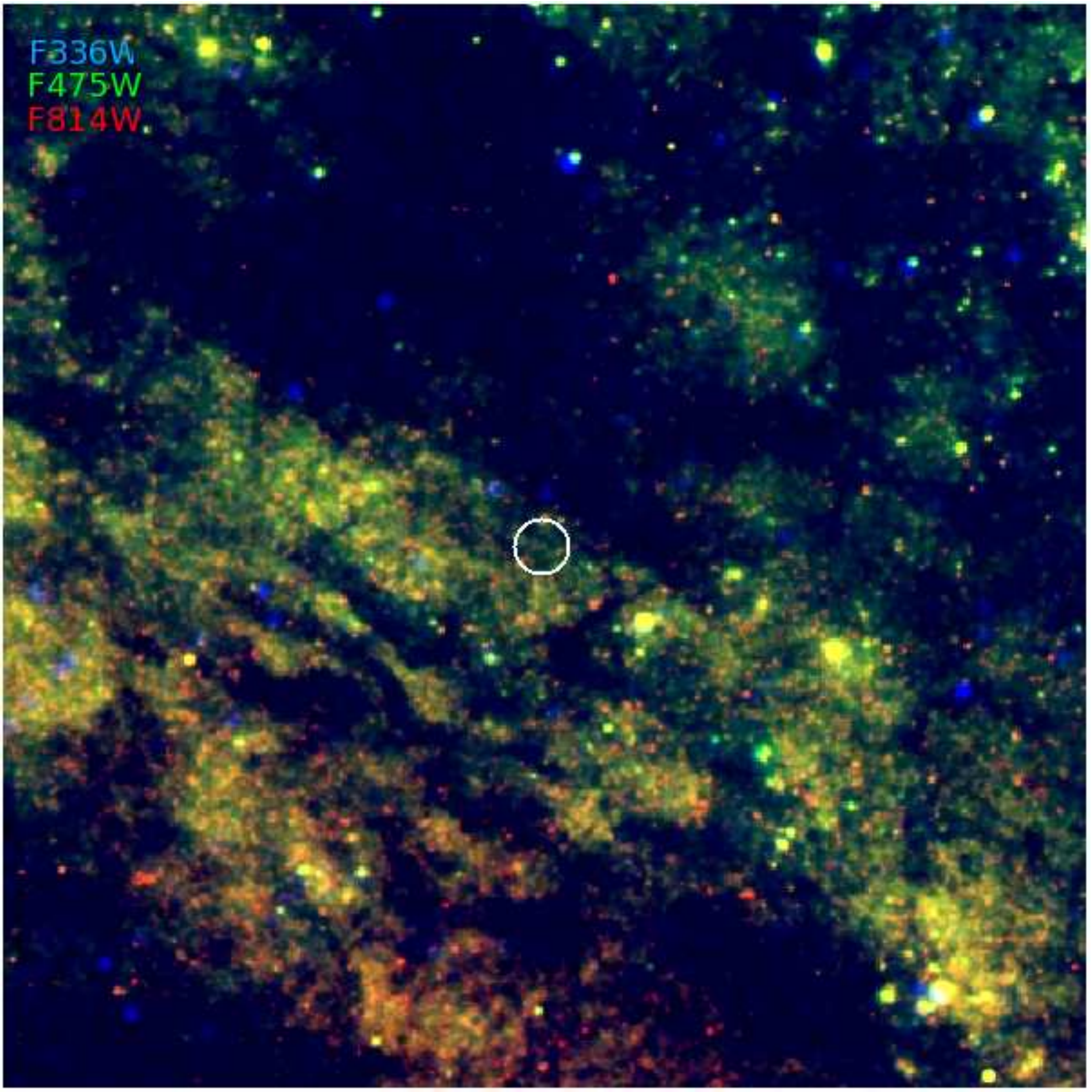} \hspace*{0.4cm}
\includegraphics[height=64mm, angle=0]{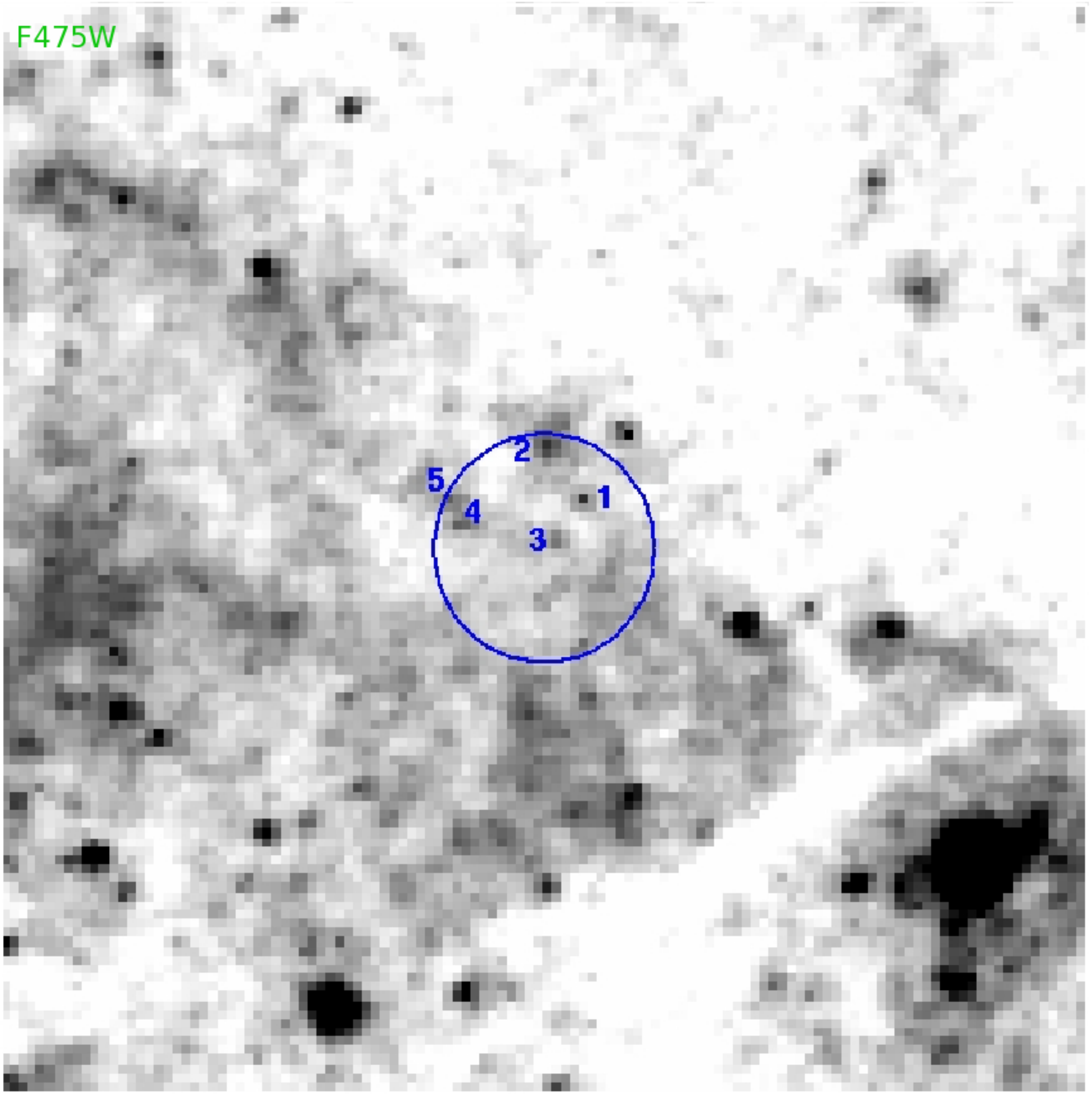} 
\vspace*{0.3cm} 

\includegraphics[height=64mm, angle=0]{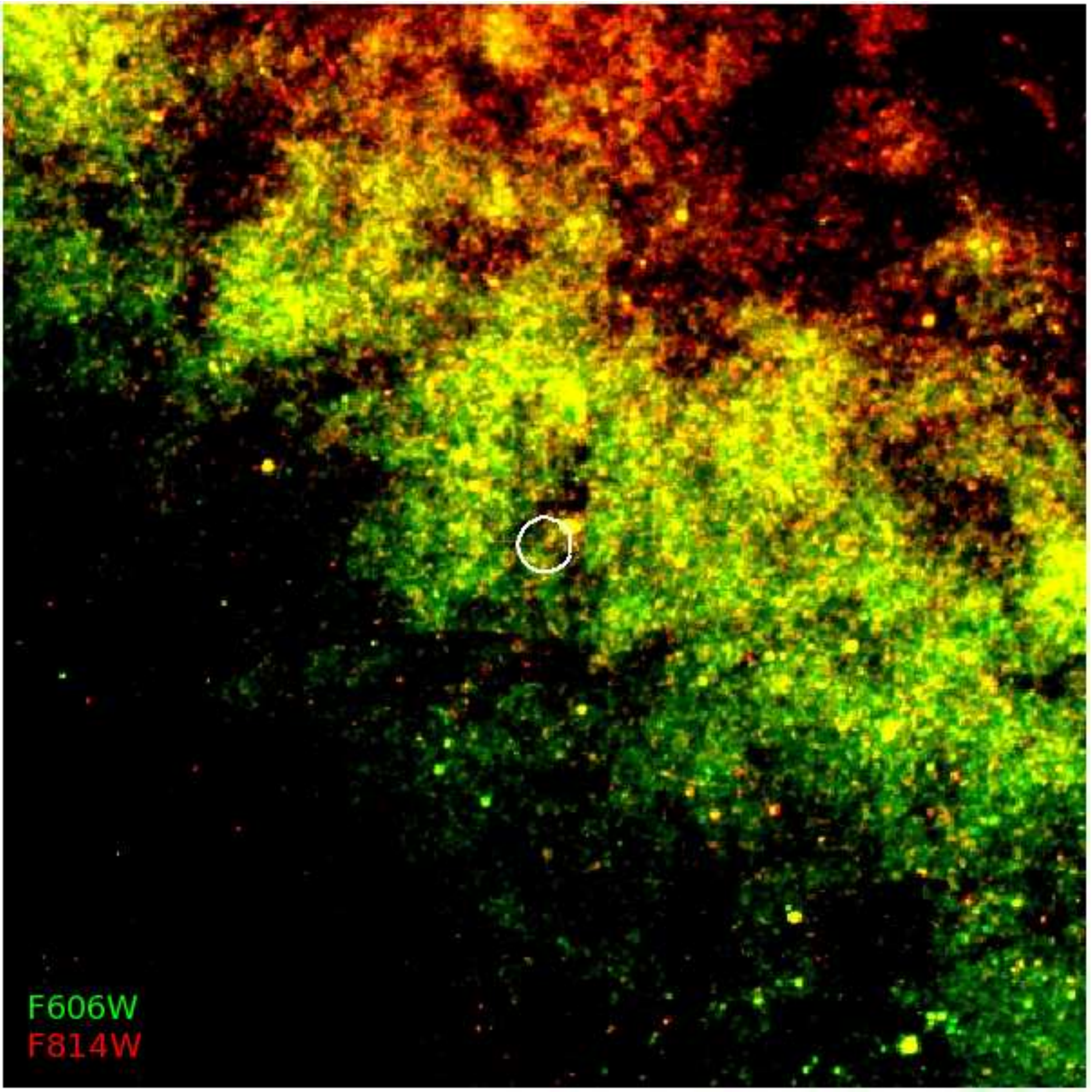} \hspace*{0.4cm}
\includegraphics[height=64mm, angle=0]{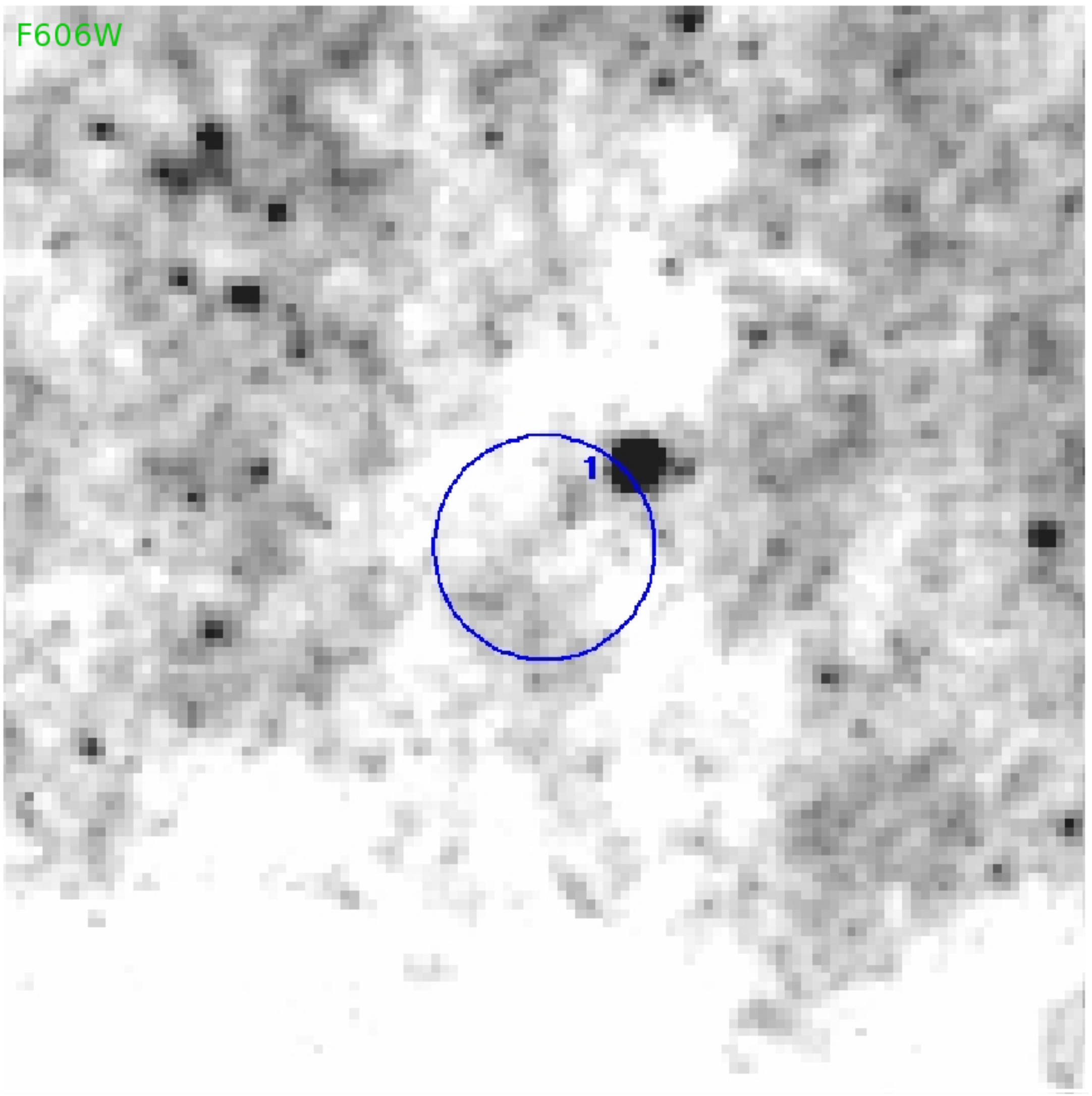} 

\end{center}
\addtocounter{figure}{-4}
\caption{\small{\emph{continued, pg. 2} -  Specific notes: displayed ULX regions are, from top to bottom, NGC 4190 X-1 (F300W), NGC 253 ULX 1 (F475W) \& NGC 253 ULX2 (F606W). }}
\label{fig:pictures}
\end{figure*}

\begin{figure*}
\leavevmode
\begin{center}

\includegraphics[height=64mm, angle=0]{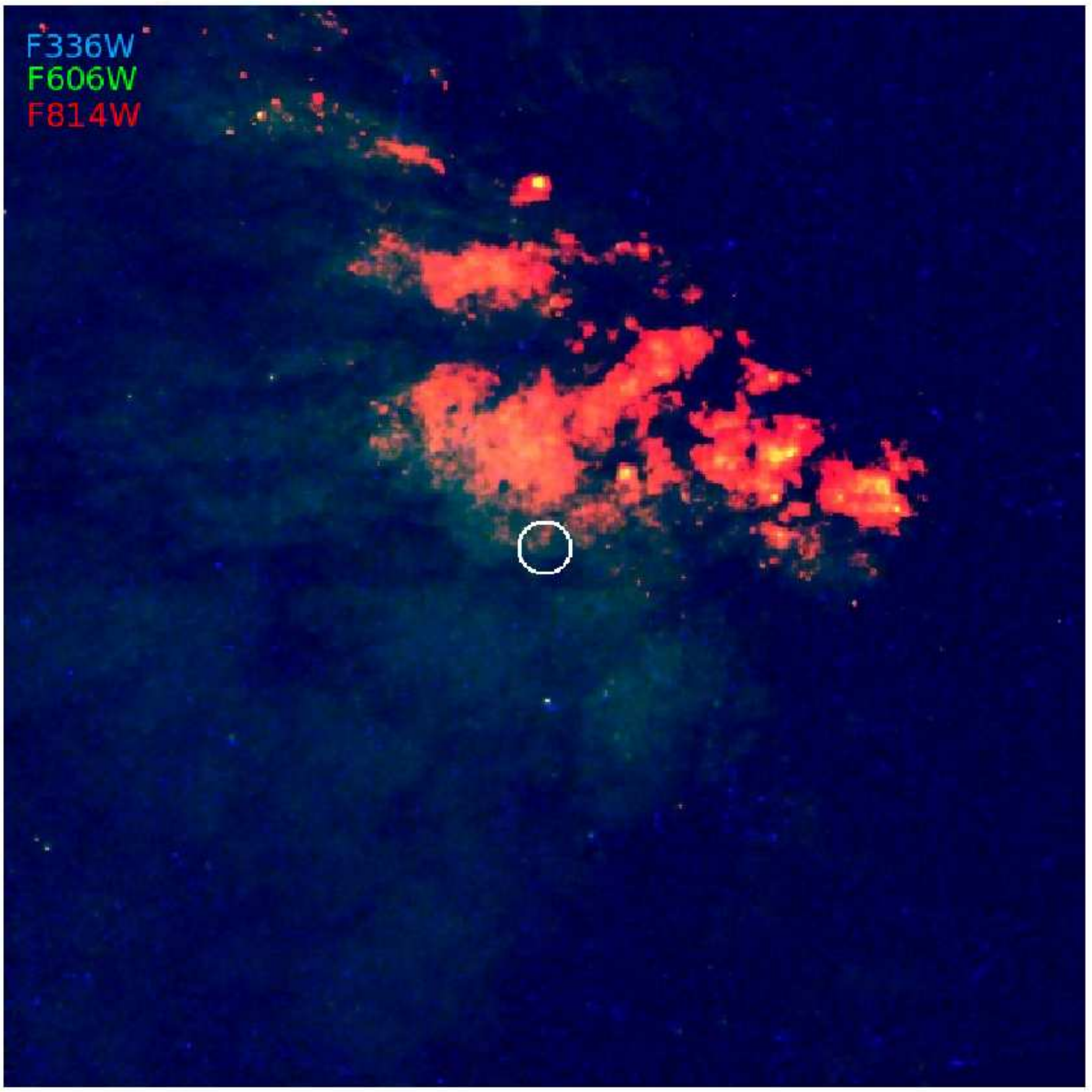} \hspace*{0.4cm}
\includegraphics[height=64mm, angle=0]{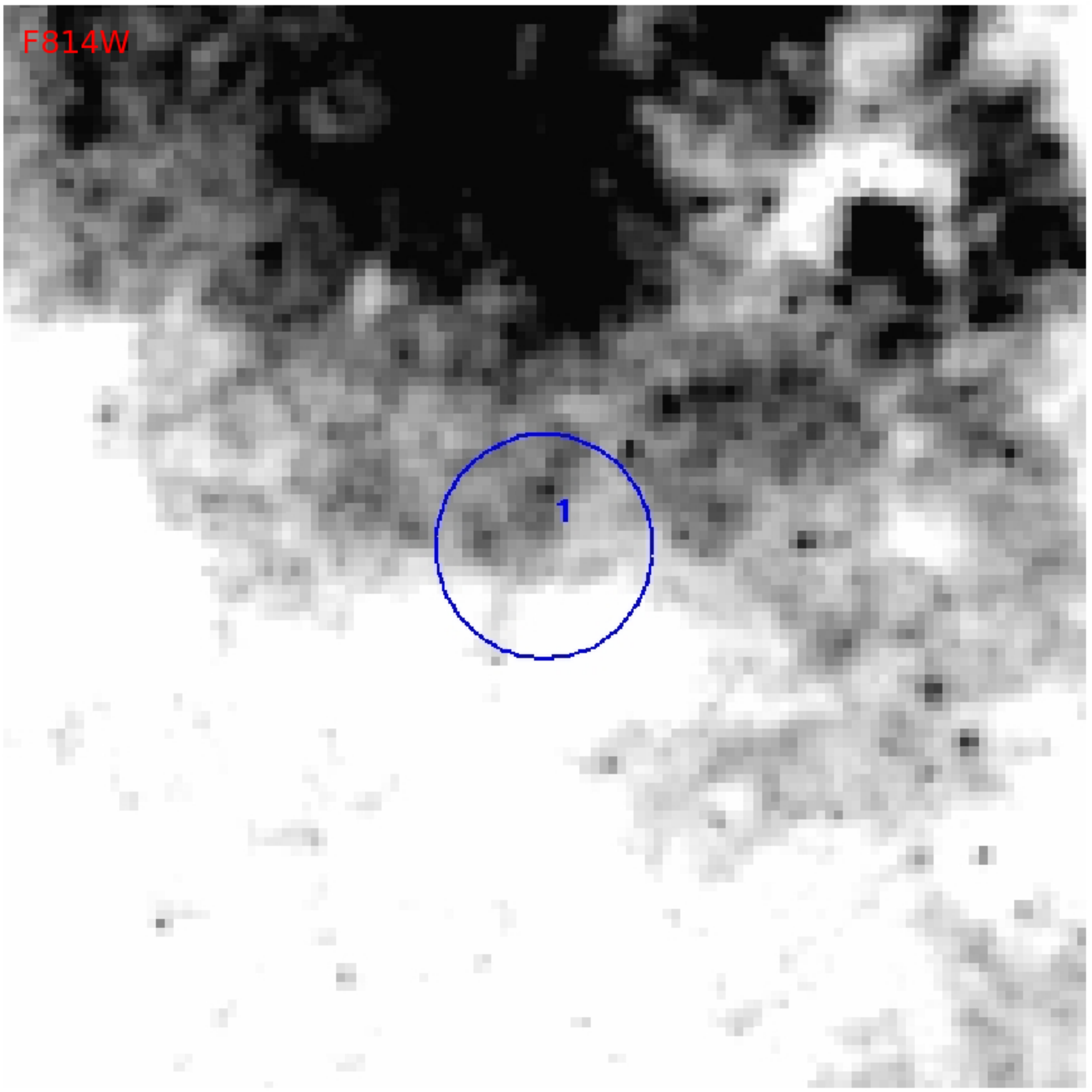} 
\vspace*{0.3cm} 

\includegraphics[height=64mm, angle=0]{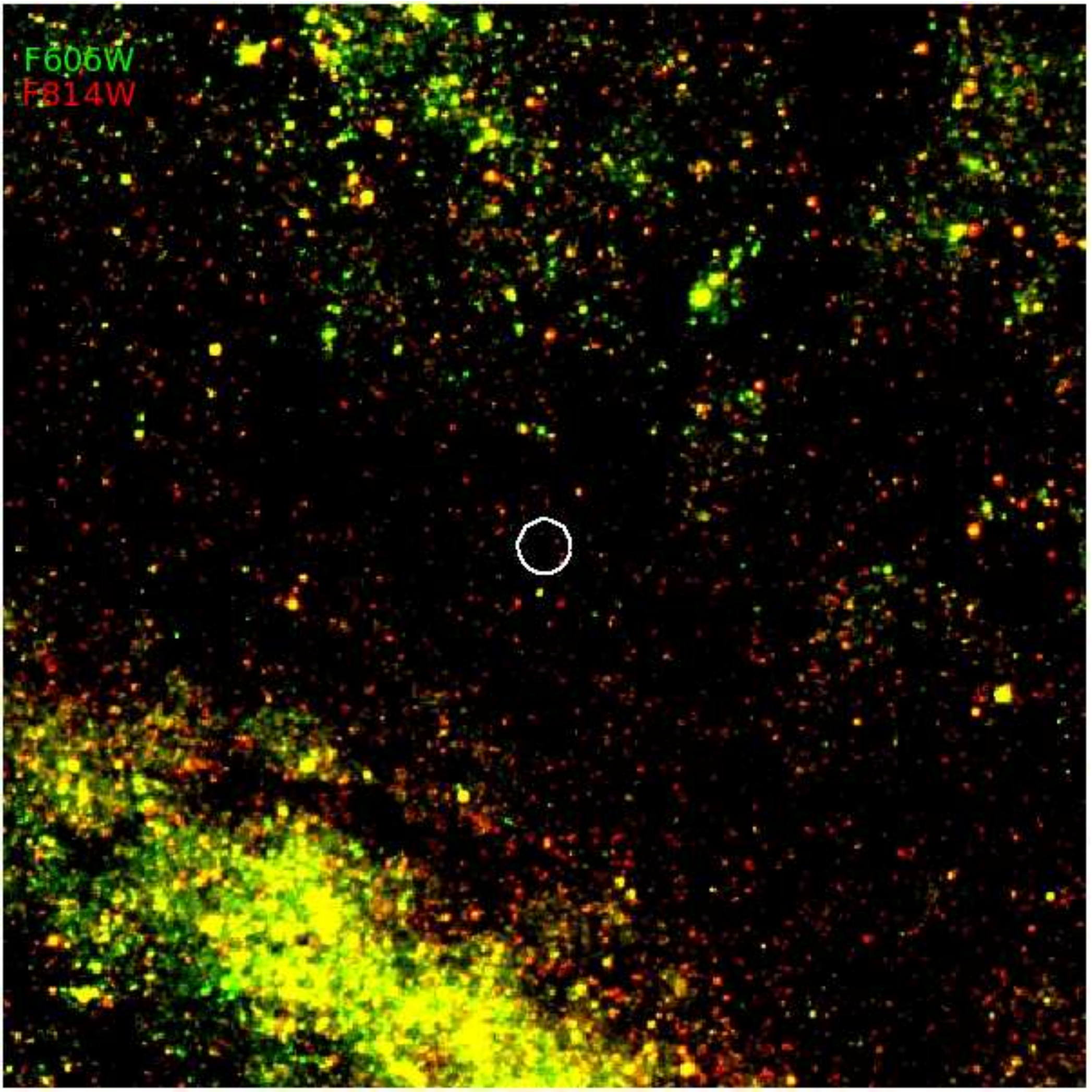} \hspace*{0.4cm}
\includegraphics[height=64mm, angle=0]{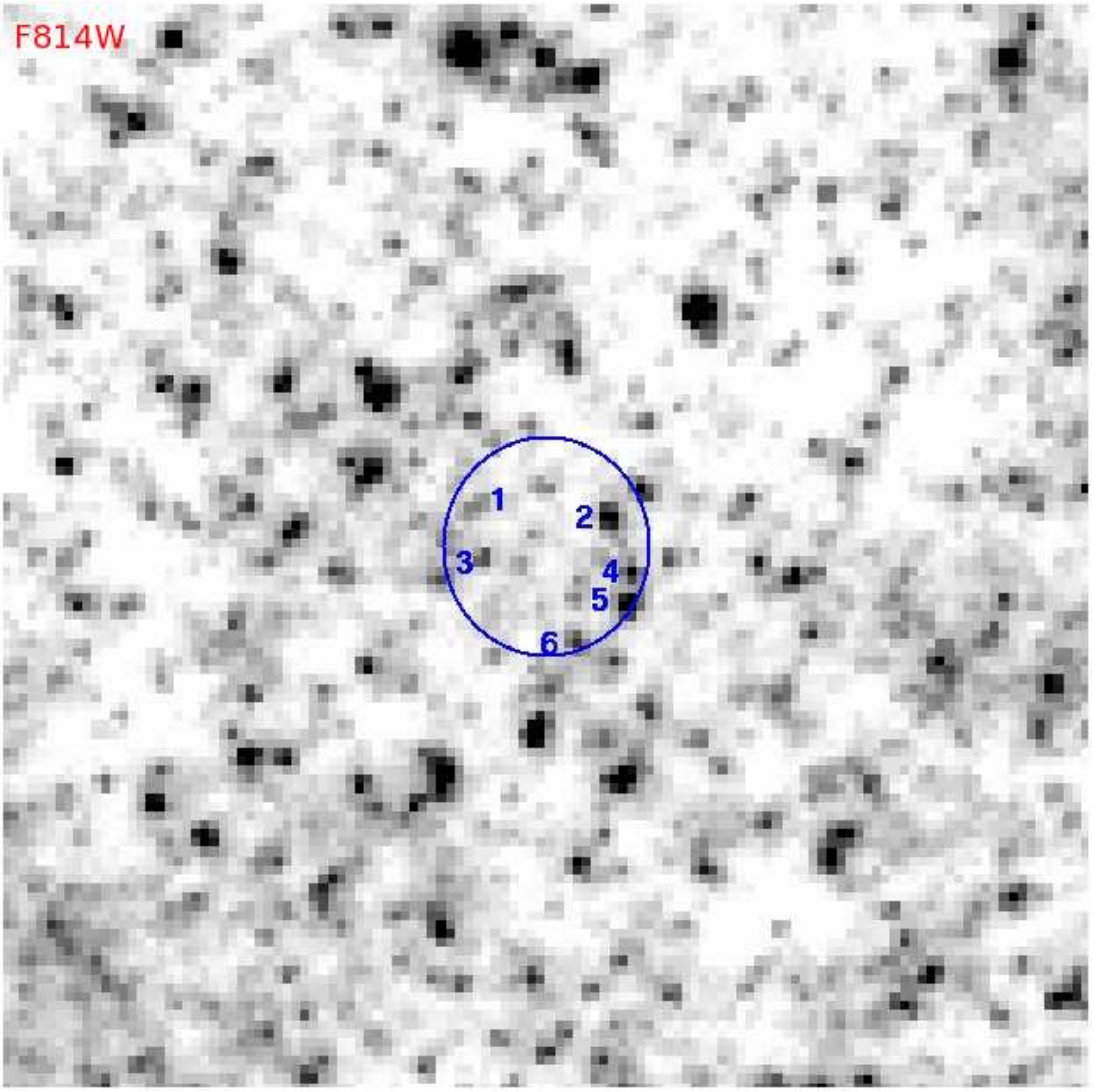} 
\vspace*{0.3cm} 

\includegraphics[height=64mm, angle=0]{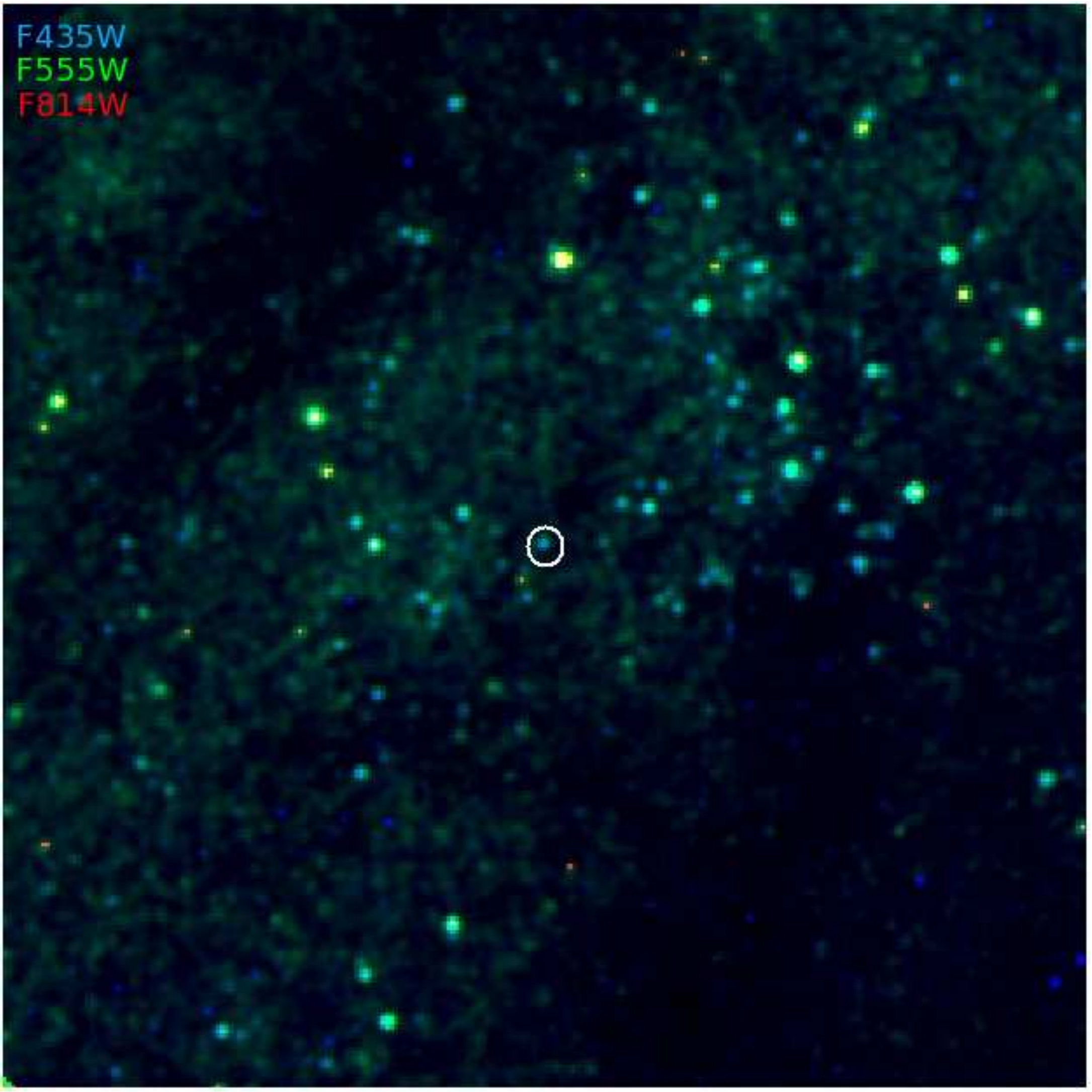} \hspace*{0.4cm}
\includegraphics[height=64mm, angle=0]{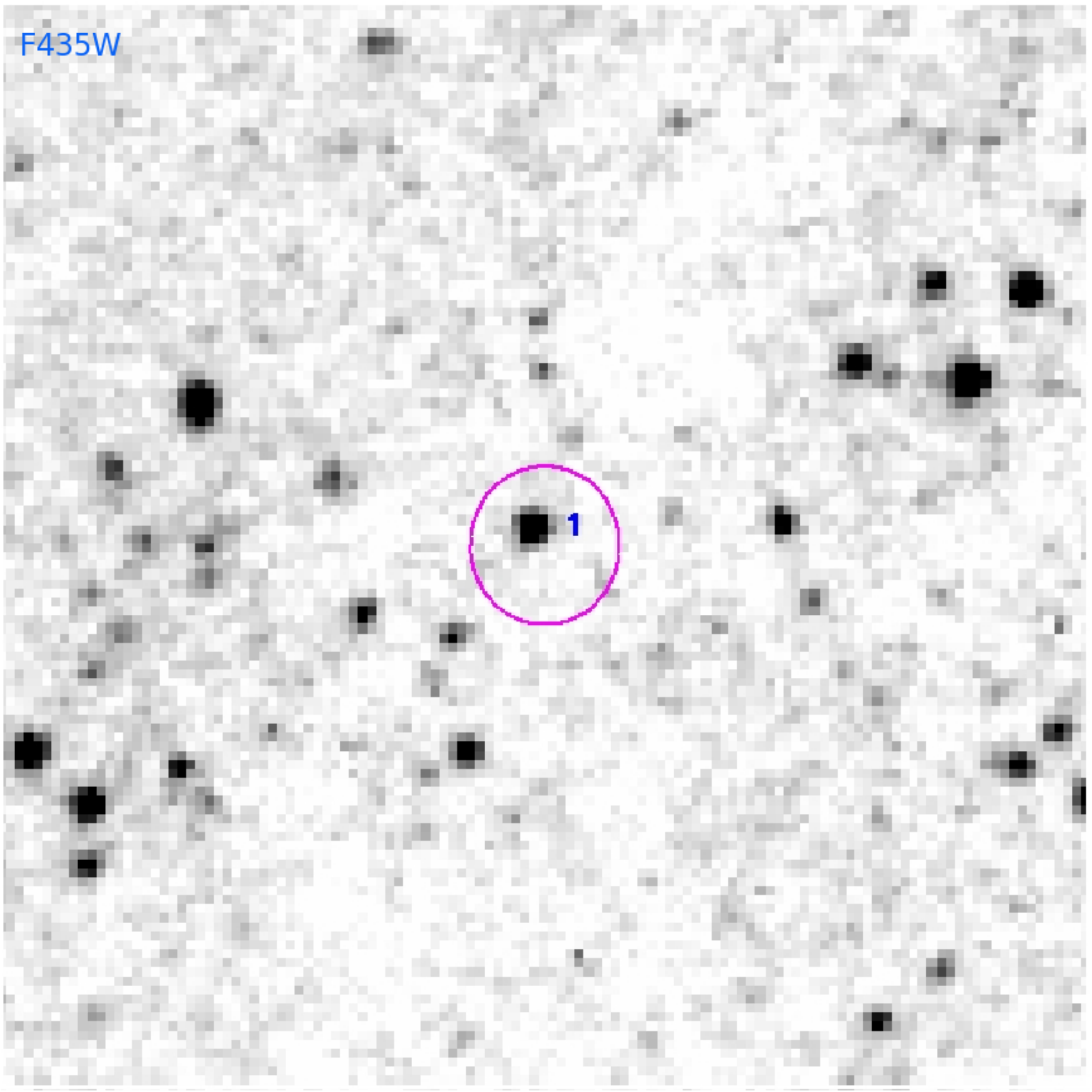} 

\end{center}
\addtocounter{figure}{-1}
\caption{\small{\emph{continued, pg. 3} -  Specific notes: displayed ULX regions are, from top to bottom, NGC 253 ULX3 (F814W), NGC 253 XMM6 (F814W) \& M81 X-6 (F435W). }}
\label{fig:pictures}
\end{figure*}

\begin{figure*}
\leavevmode
\begin{center}

\includegraphics[height=64mm, angle=0]{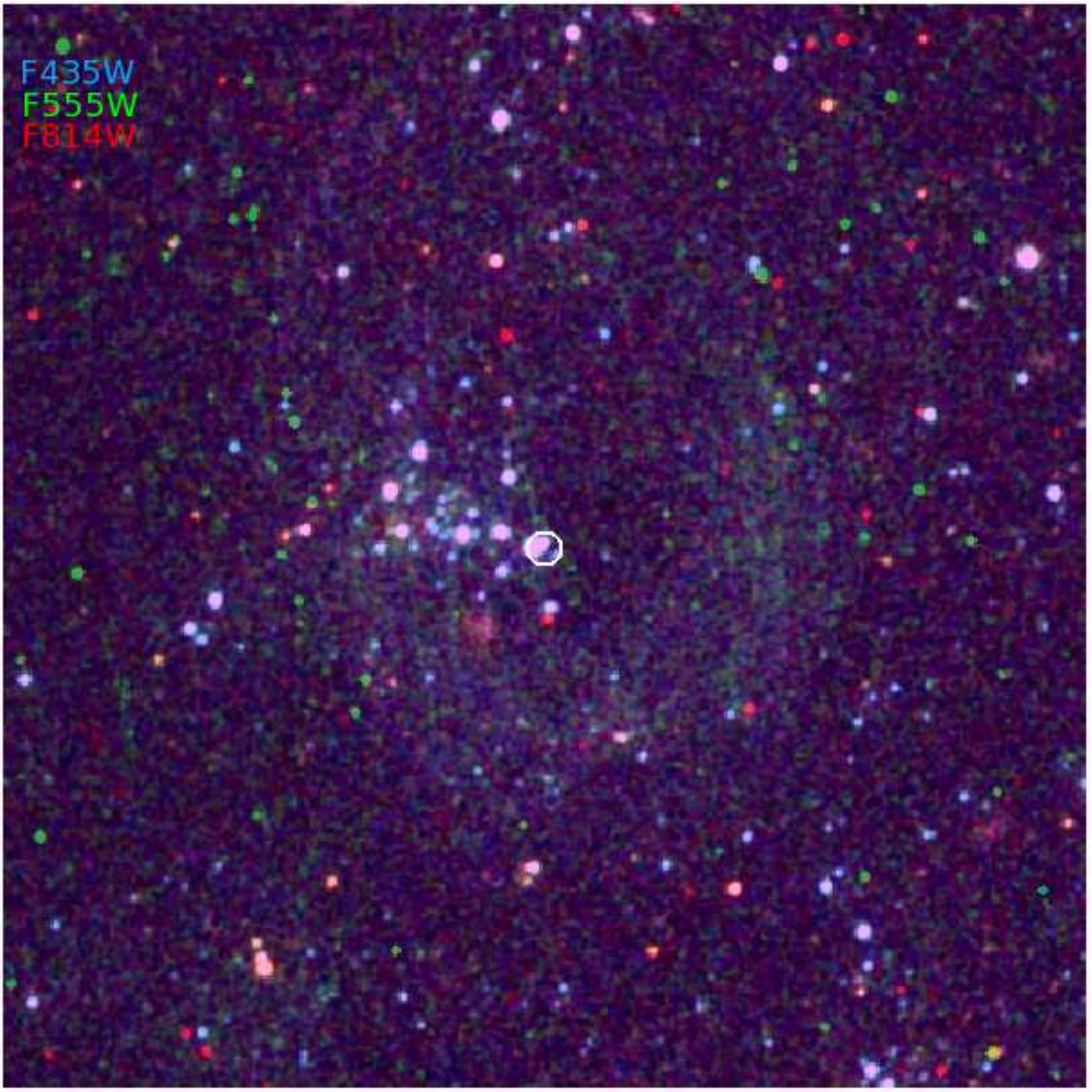} \hspace*{0.4cm}
\includegraphics[height=64mm, angle=0]{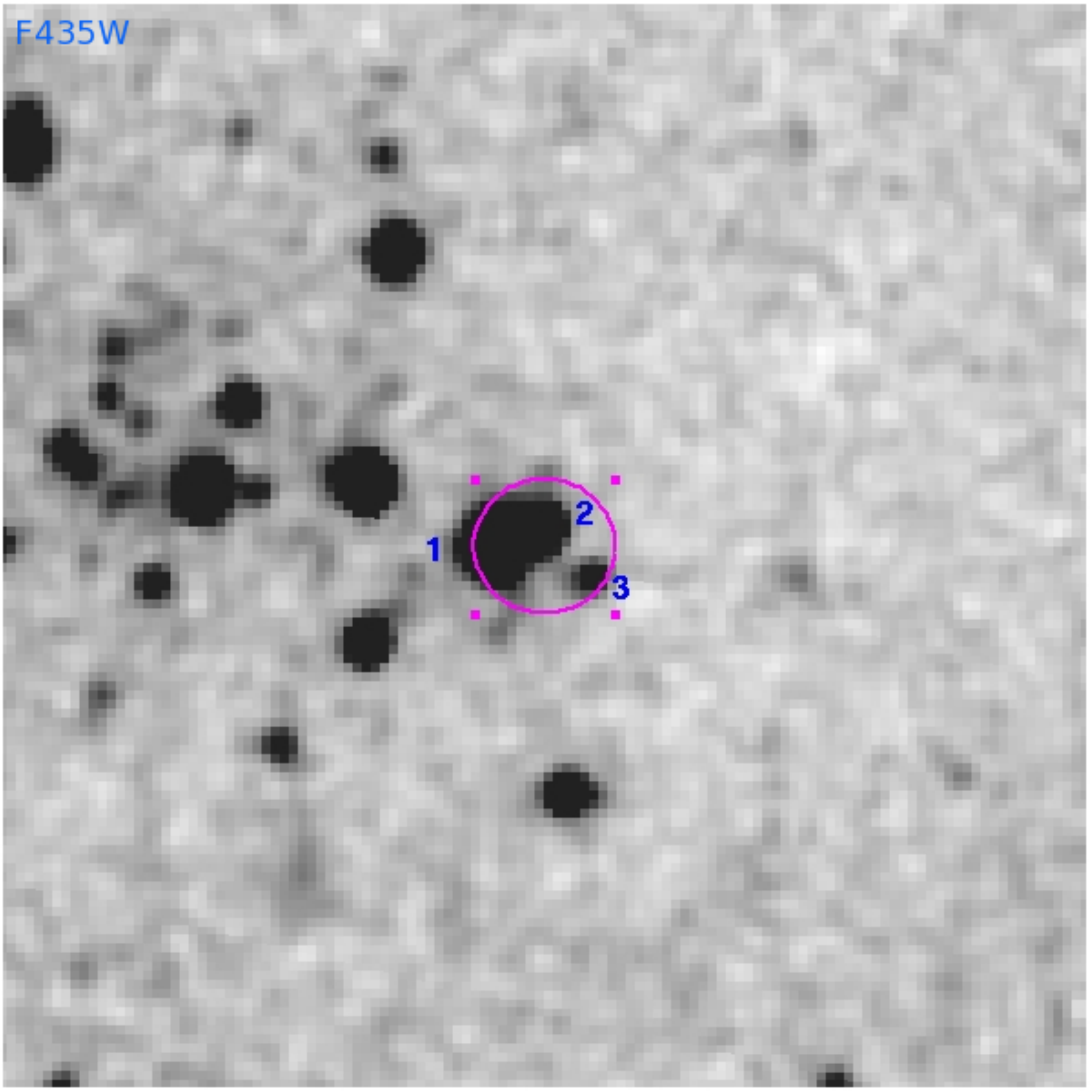} 
\vspace*{0.3cm} 

\includegraphics[height=64mm, angle=0]{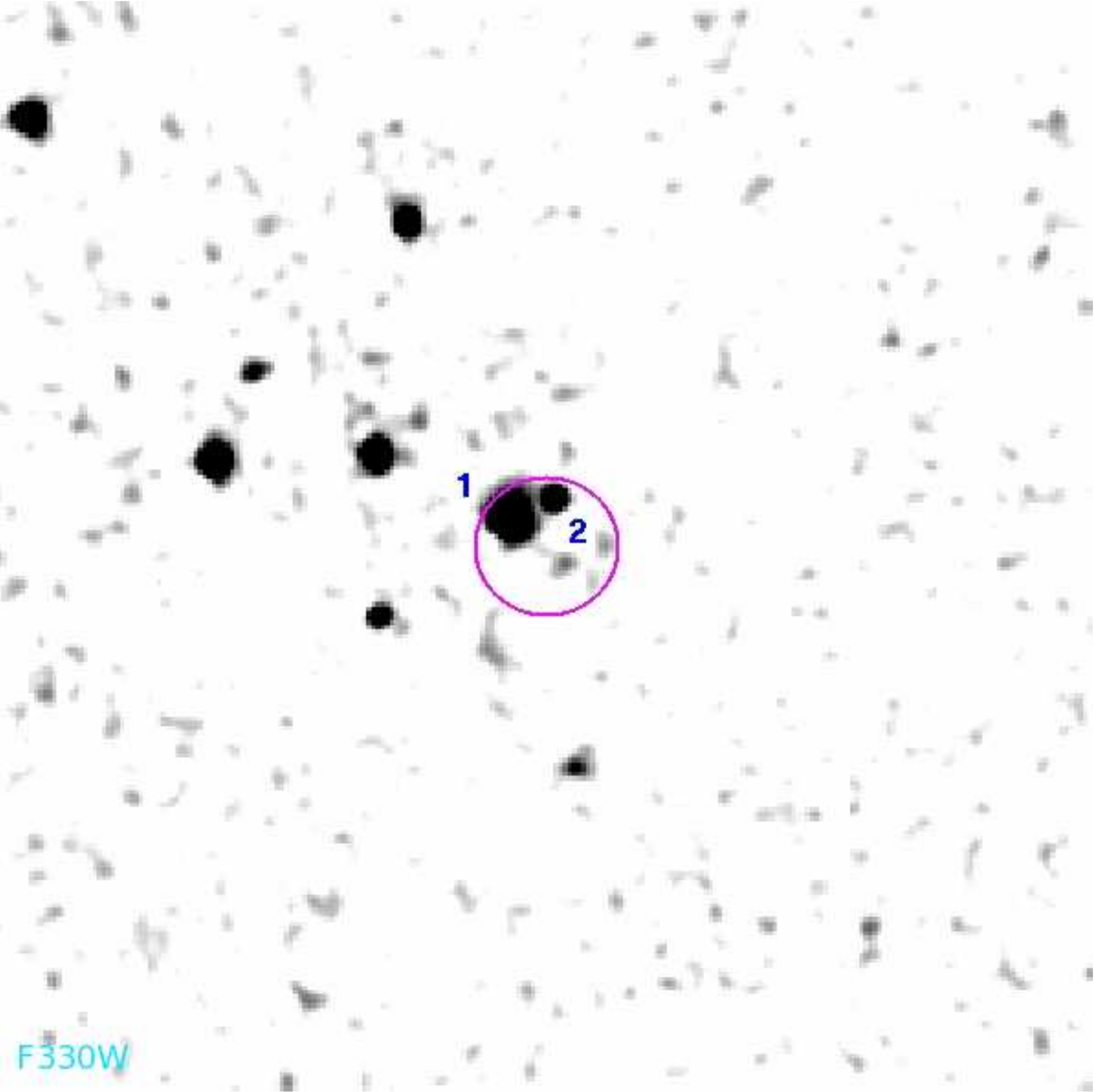} 
\vspace*{0.3cm} 

\includegraphics[height=64mm, angle=0]{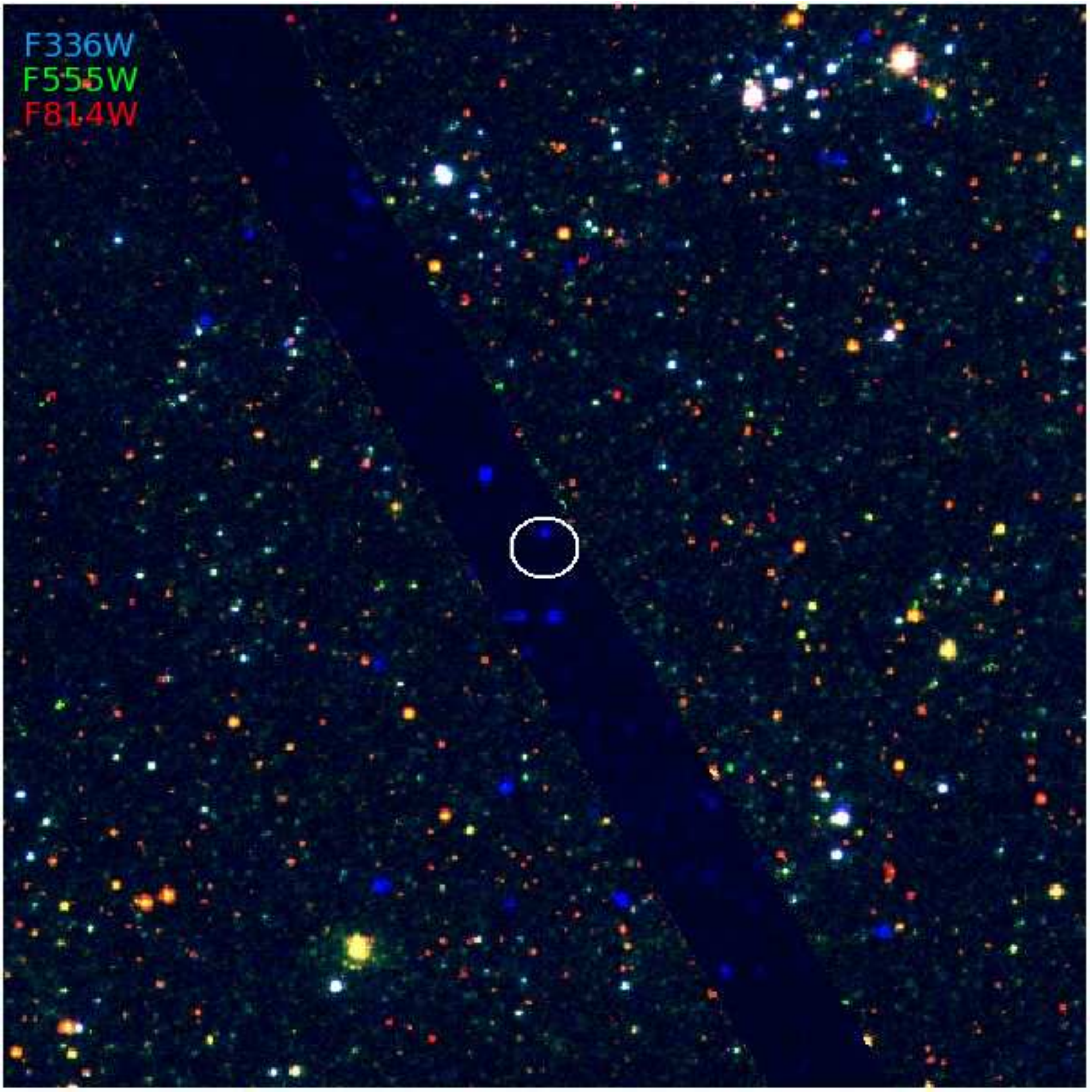} \hspace*{0.4cm}
\includegraphics[height=64mm, angle=0]{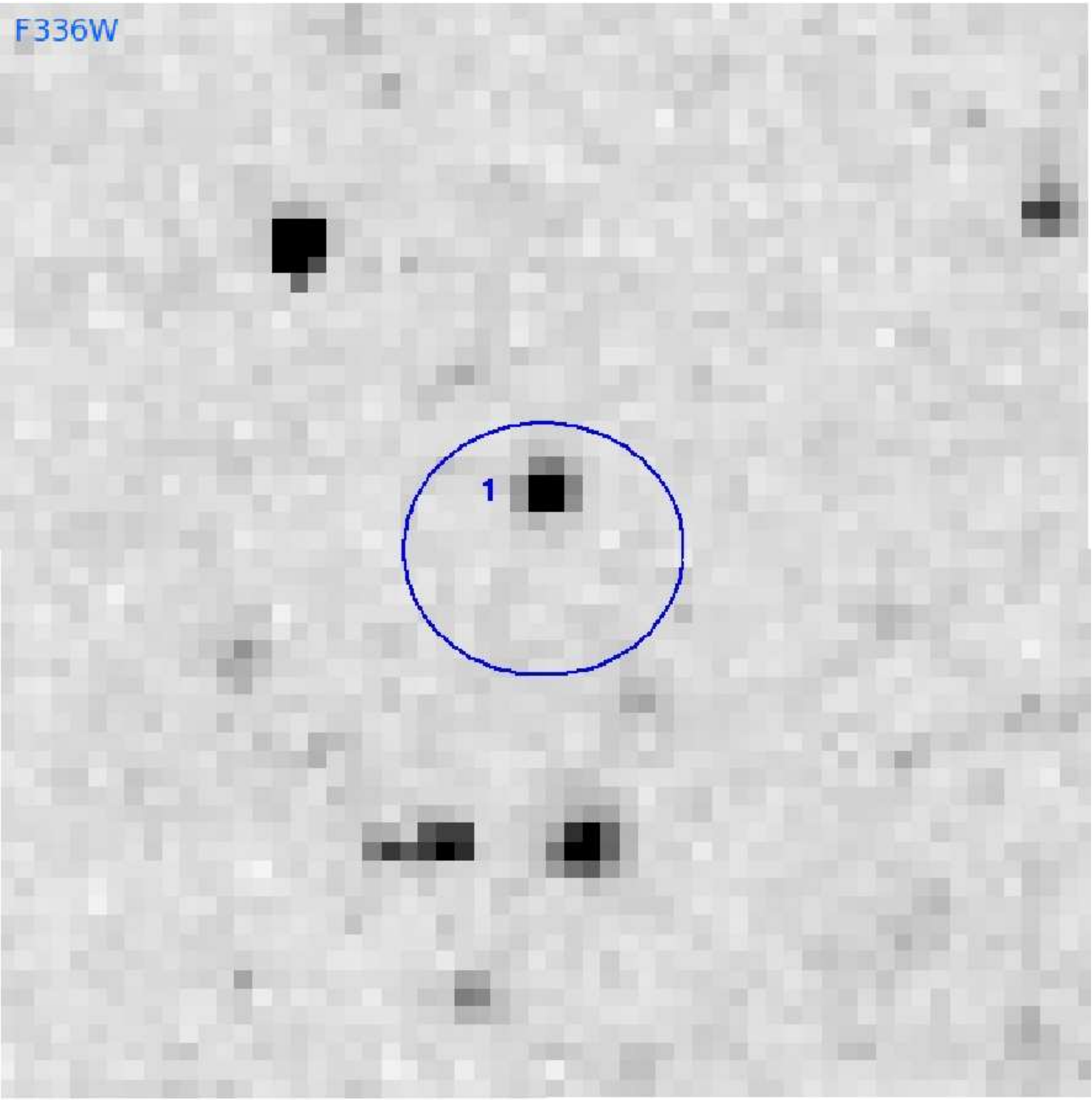} 

\end{center}
\addtocounter{figure}{-1}
\caption{\small{\emph{continued, pg. 4} -  Specific notes: displayed ULX regions are, from top to bottom, Holmberg IX X-1 (F435W \& F330W) \& NGC 4395 ULX1 (F336W). Two finders are provided for Ho IX X-1 as source 1 and 2 are only clearly separated in the F330W band due to the greater spacial resolution of HRC. However, source 3 is not detected in this band. }}
\label{fig:pictures}
\end{figure*}

\begin{figure*}
\leavevmode
\begin{center}

\includegraphics[height=64mm, angle=0]{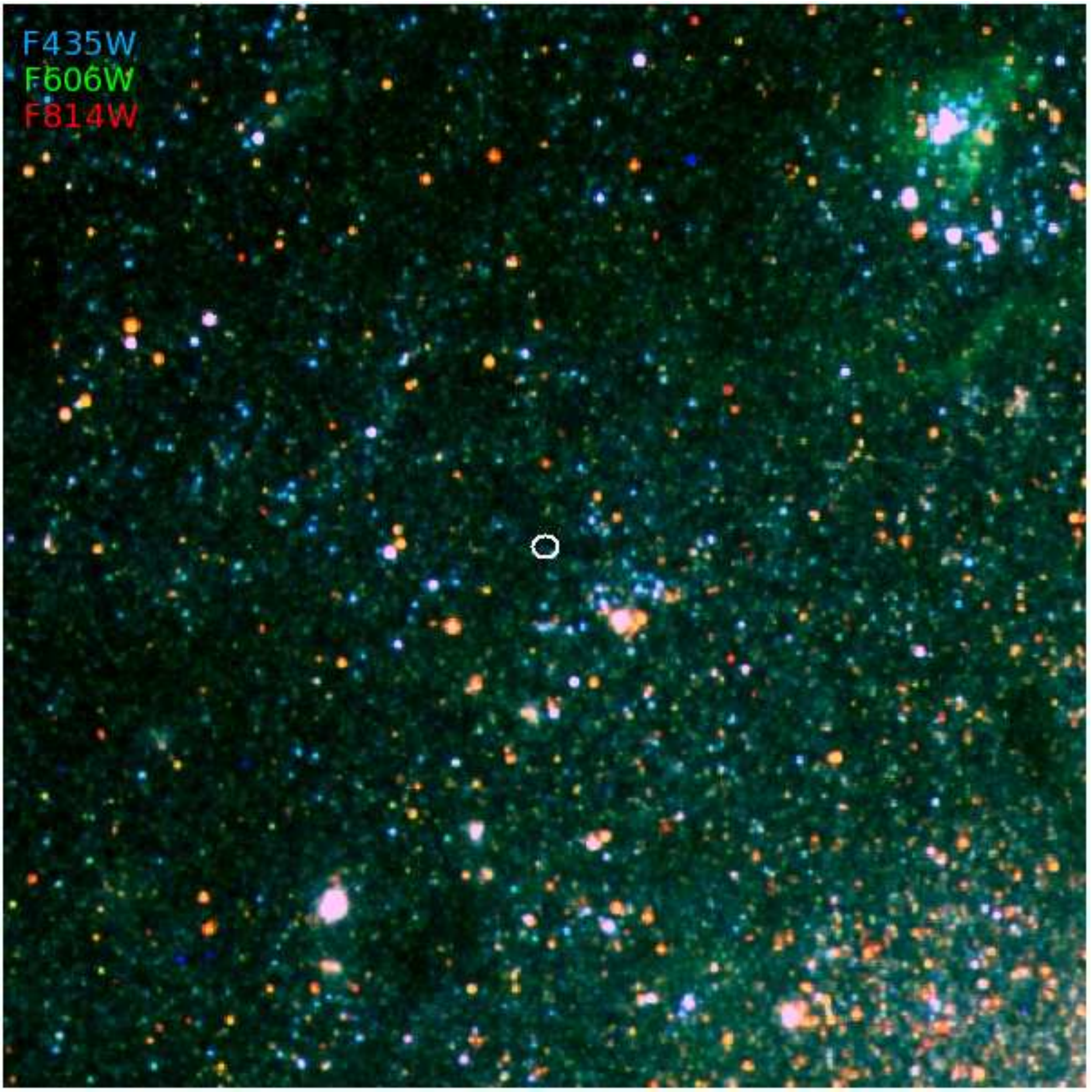} \hspace*{0.4cm}
\includegraphics[height=64mm, angle=0]{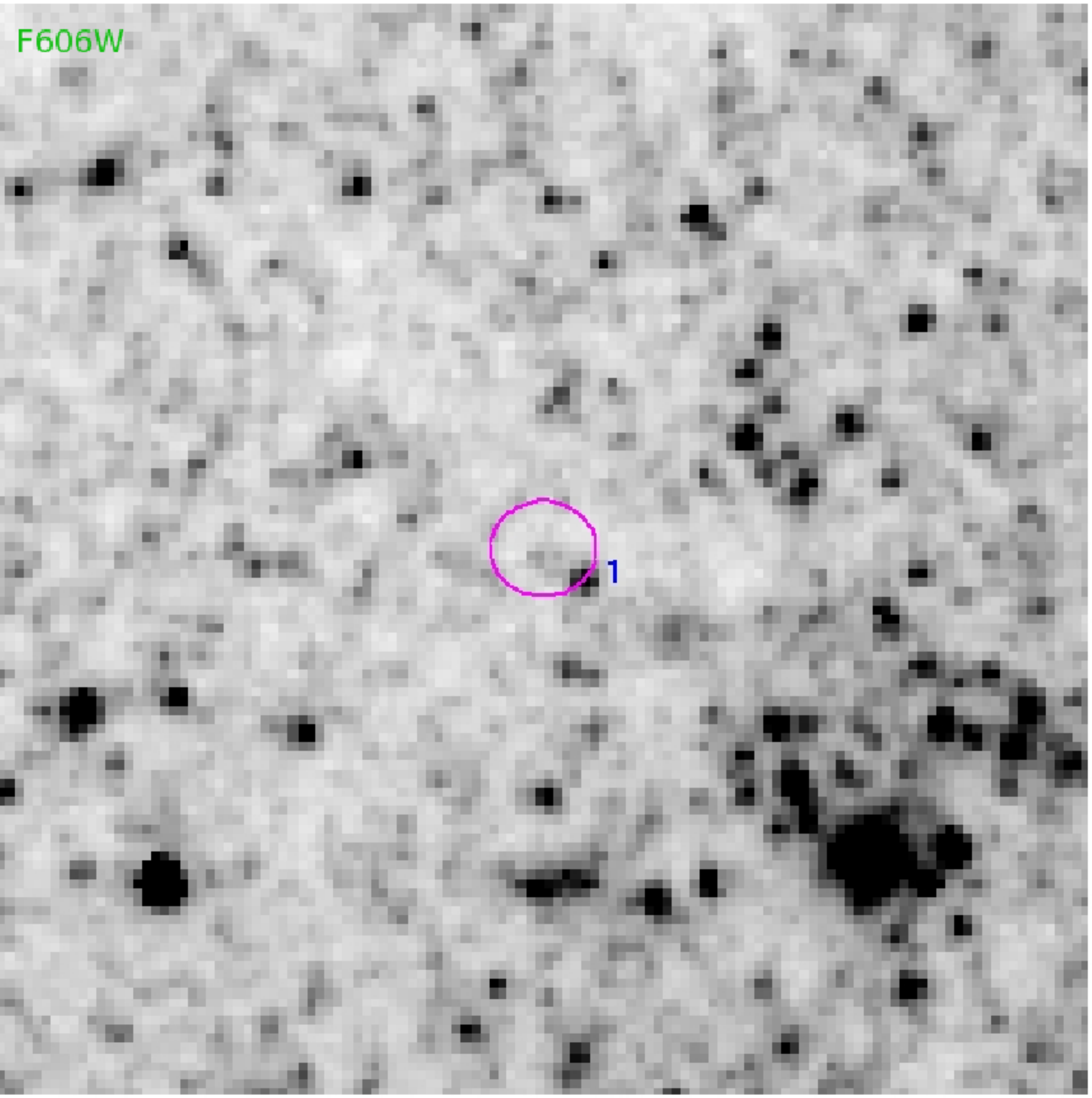} 
\vspace*{0.3cm} 

\includegraphics[height=64mm, angle=0]{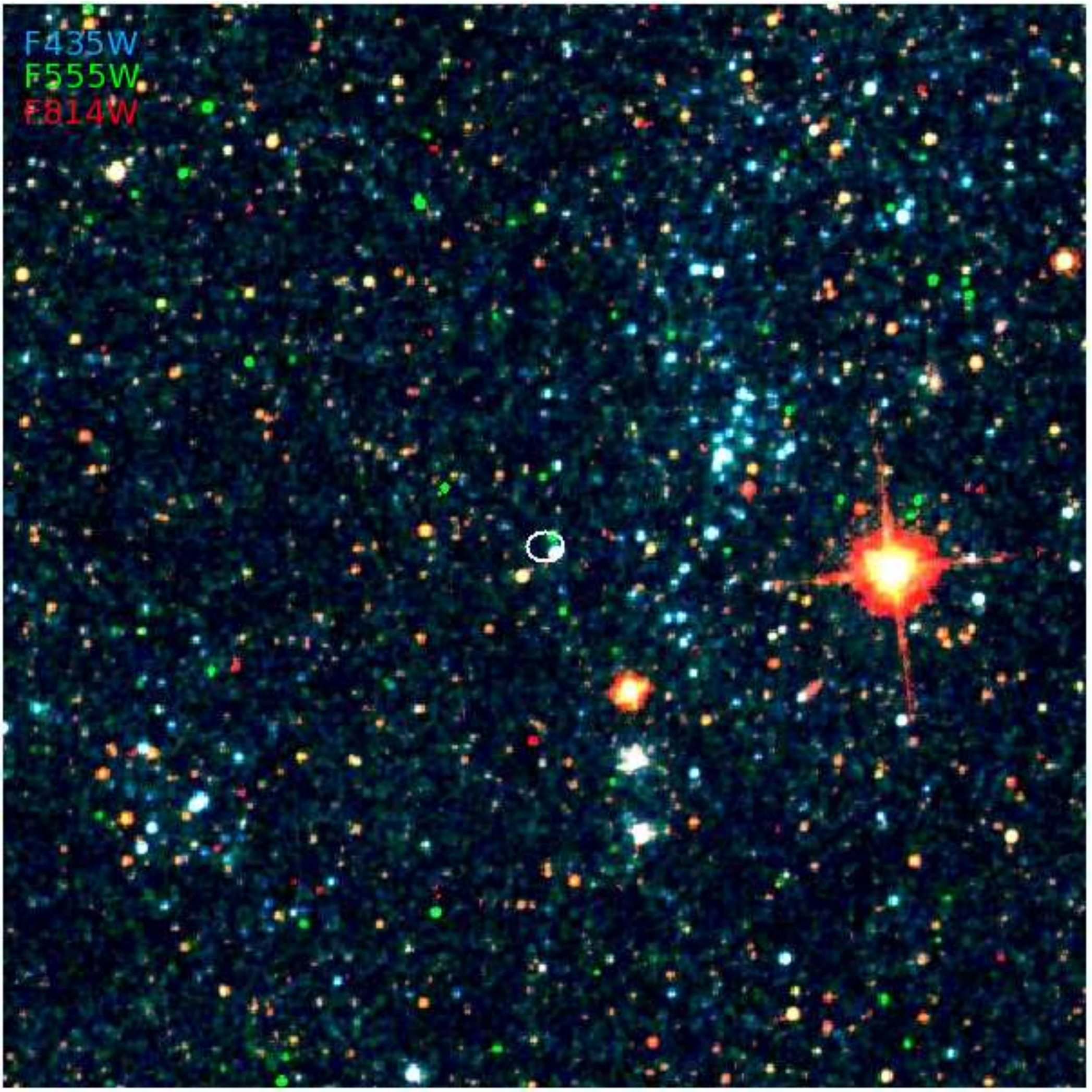} \hspace*{0.4cm}
\includegraphics[height=64mm, angle=0]{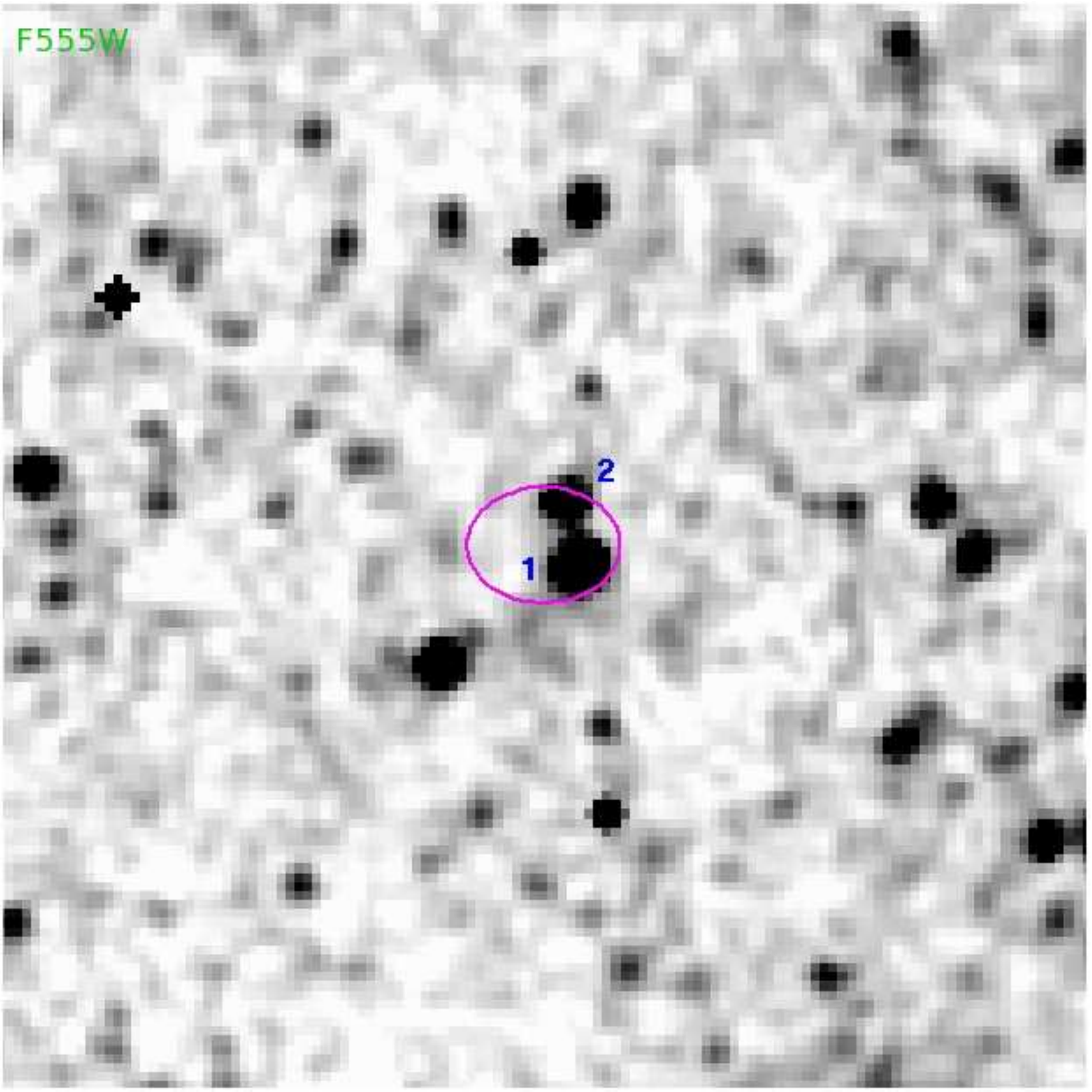} 
\vspace*{0.3cm} 

\includegraphics[height=64mm, angle=0]{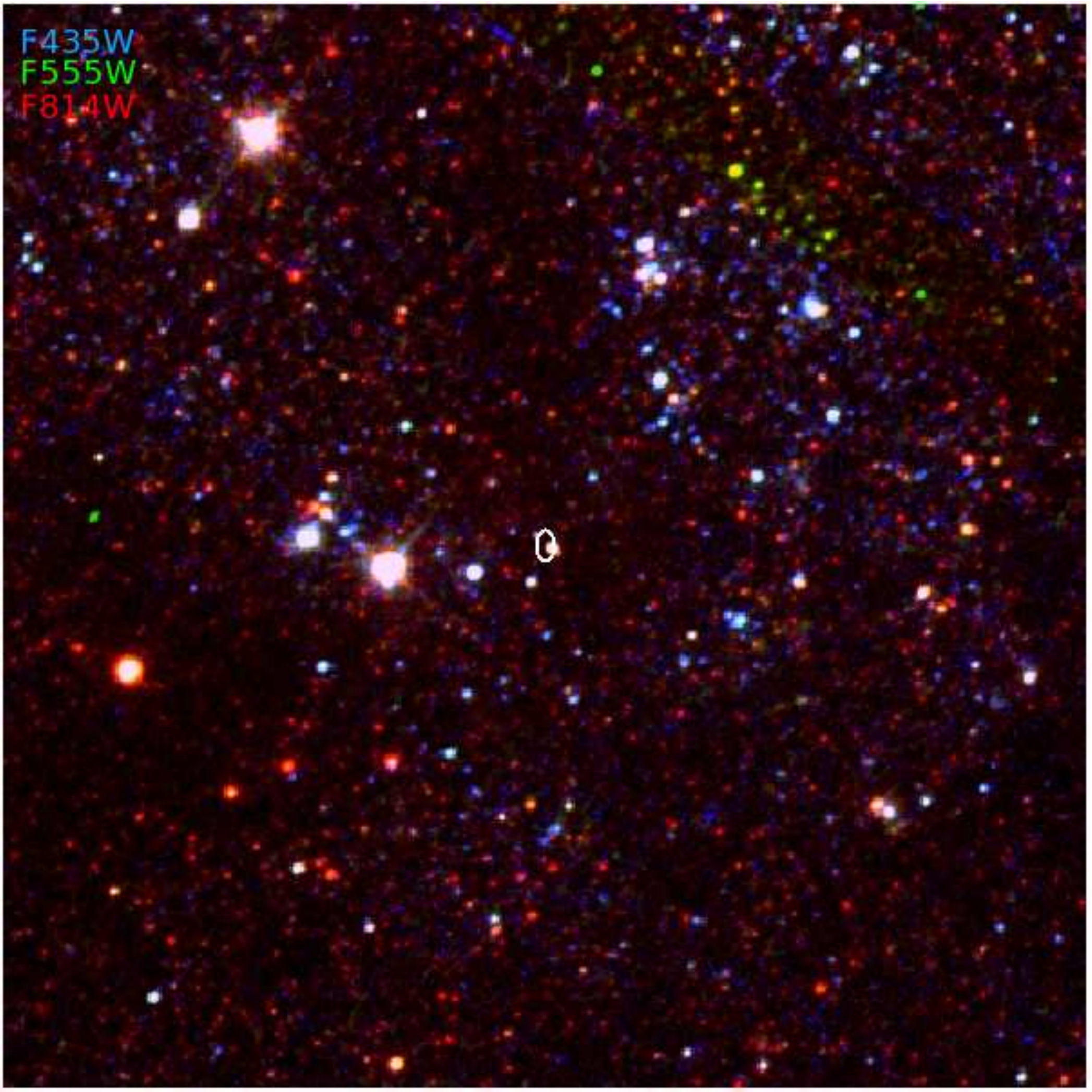} \hspace*{0.4cm}
\includegraphics[height=64mm, angle=0]{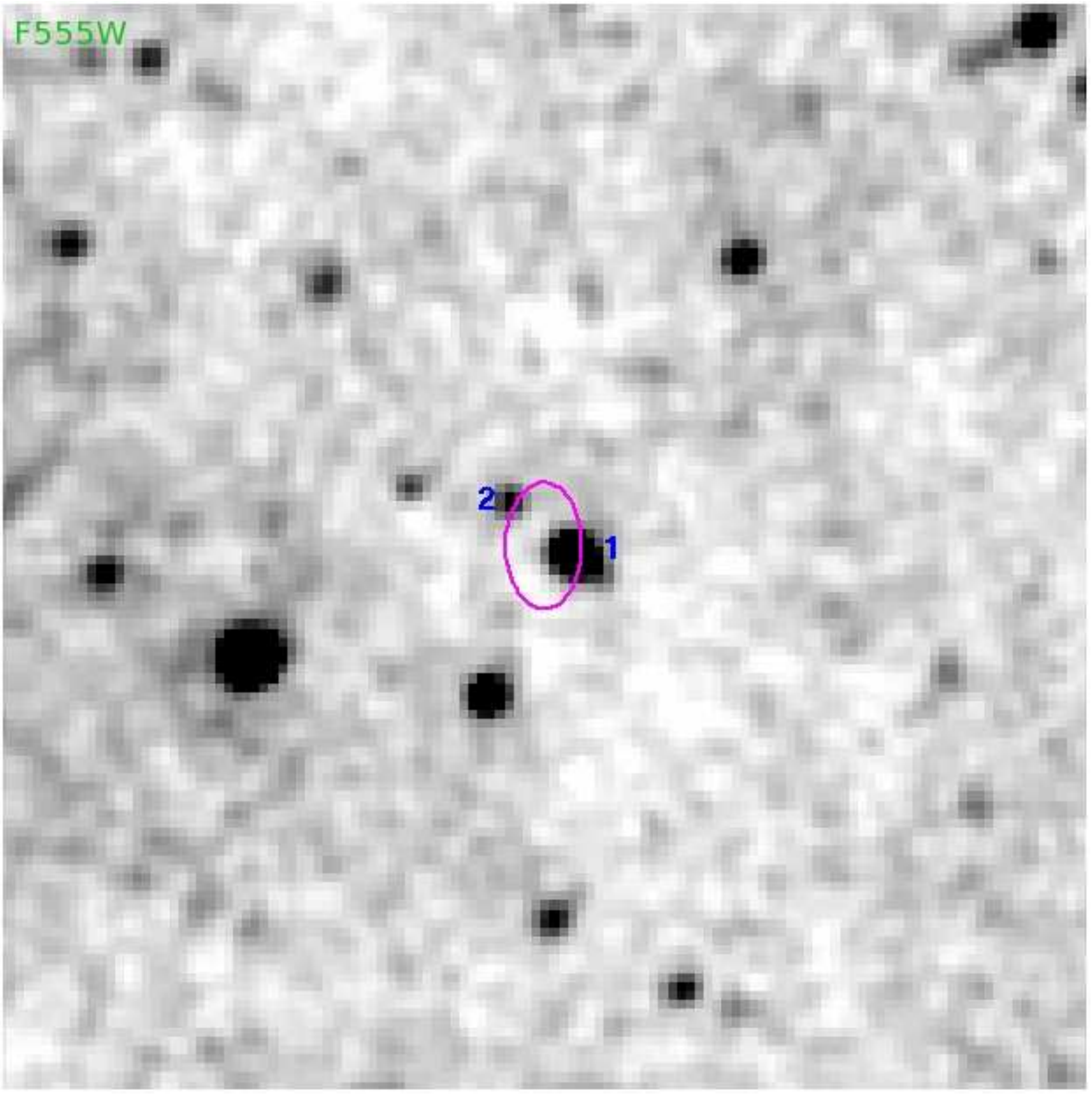} 

\end{center}
\addtocounter{figure}{-1}
\caption{\small{\emph{continued, pg. 5} -  Specific notes: displayed ULX regions are, from top to bottom, NGC 1313 X-1 (F606W), NGC 1313 X-2 (F555W) \& IC 342 X-1 (F555W). }}
\label{fig:pictures}
\end{figure*}

\begin{figure*}
\leavevmode
\begin{center}

\includegraphics[height=64mm, angle=0]{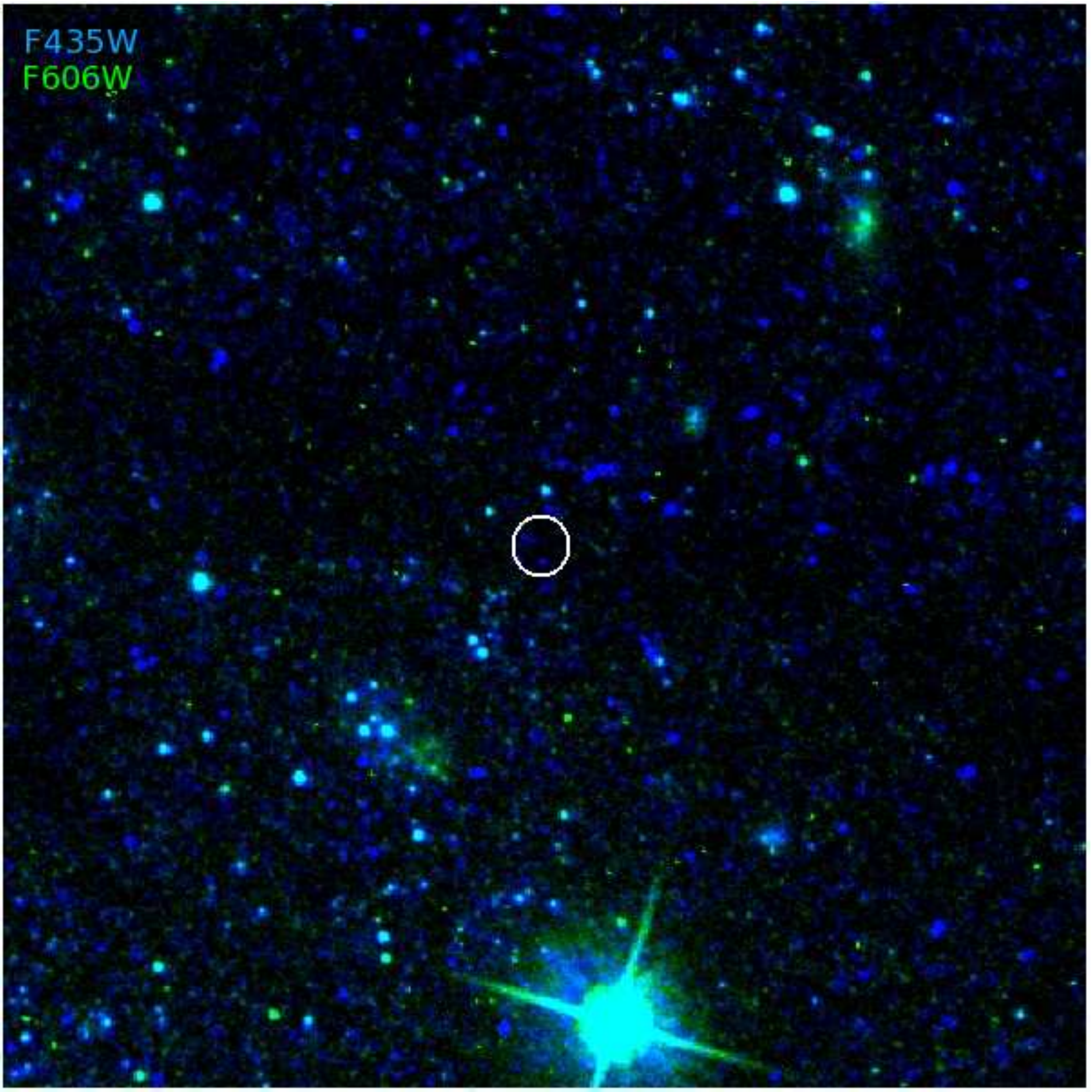} \hspace*{0.4cm}
\includegraphics[height=64mm, angle=0]{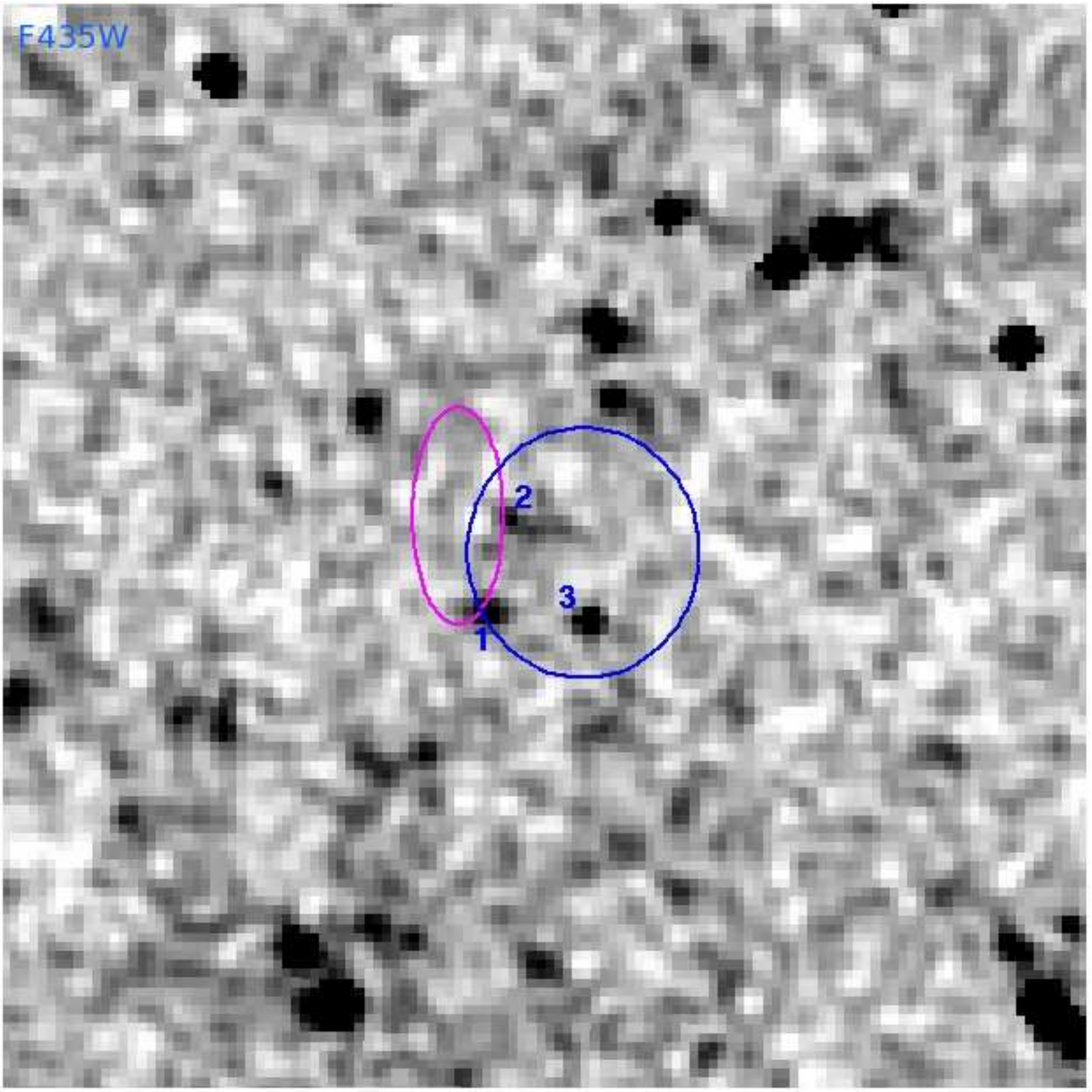} 
\vspace*{0.3cm} 

\includegraphics[height=64mm, angle=0]{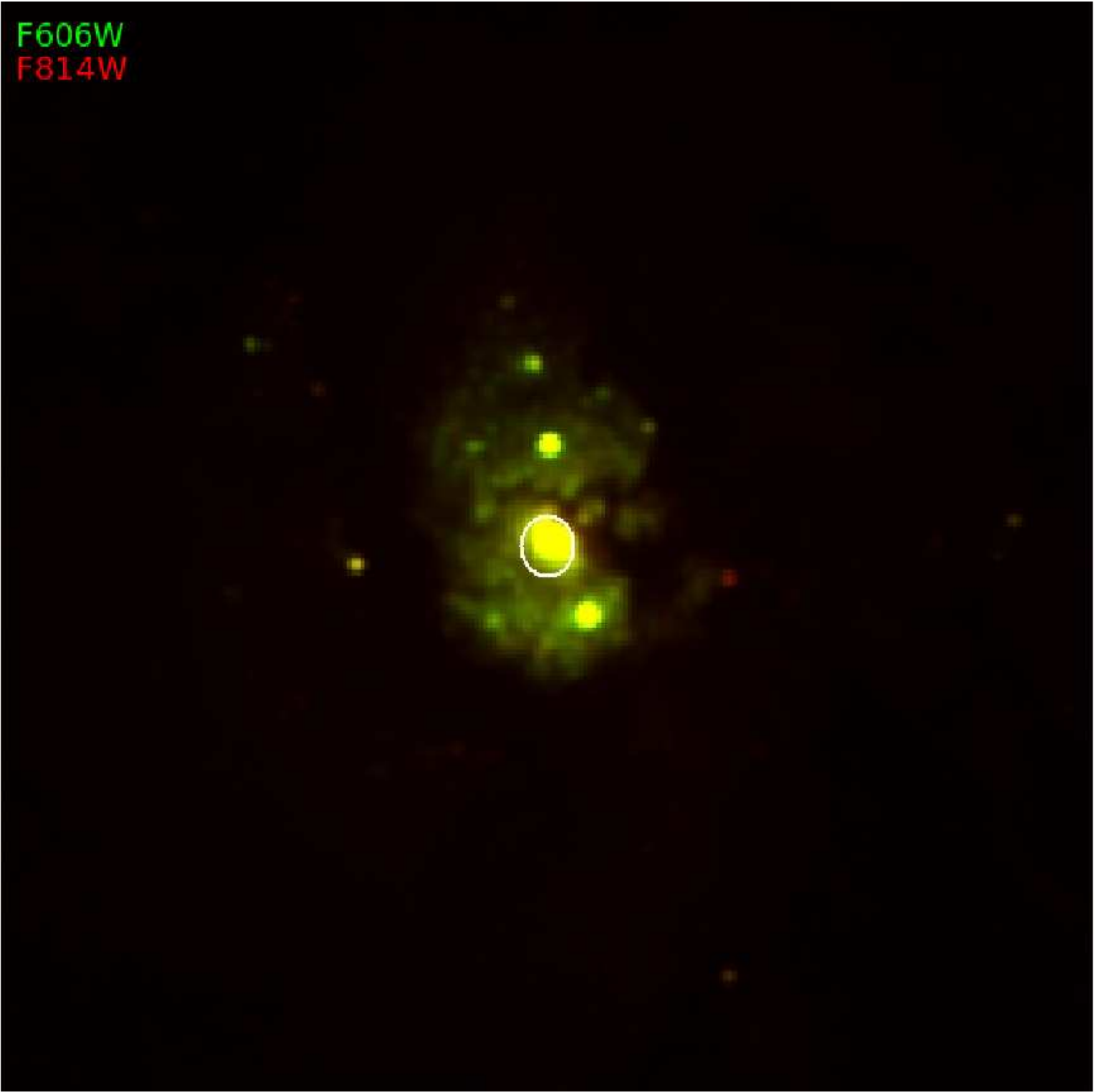} \hspace*{0.4cm}
\includegraphics[height=64mm, angle=0]{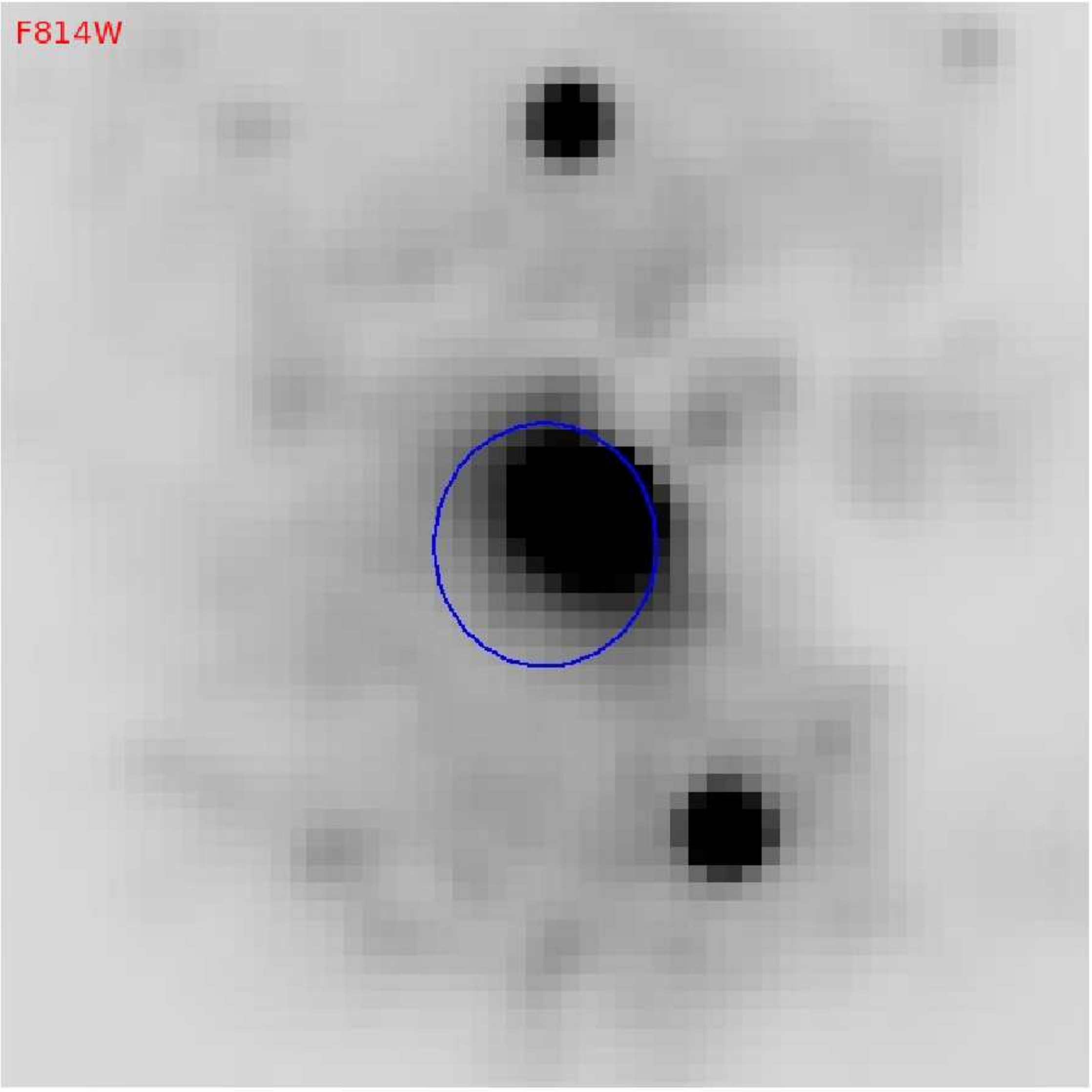} 
\vspace*{0.3cm} 

\includegraphics[height=64mm, angle=0]{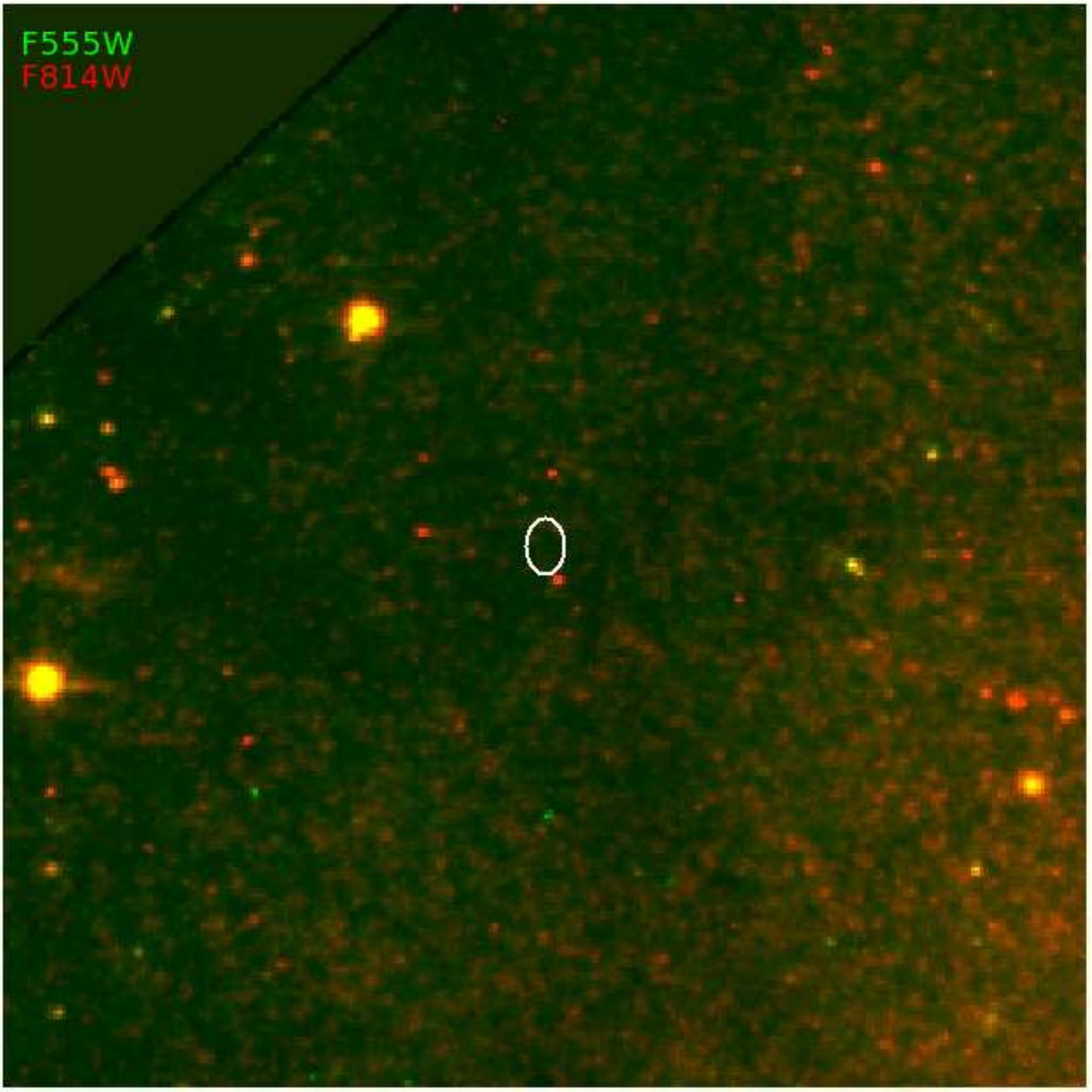} \hspace*{0.4cm}
\includegraphics[height=64mm, angle=0]{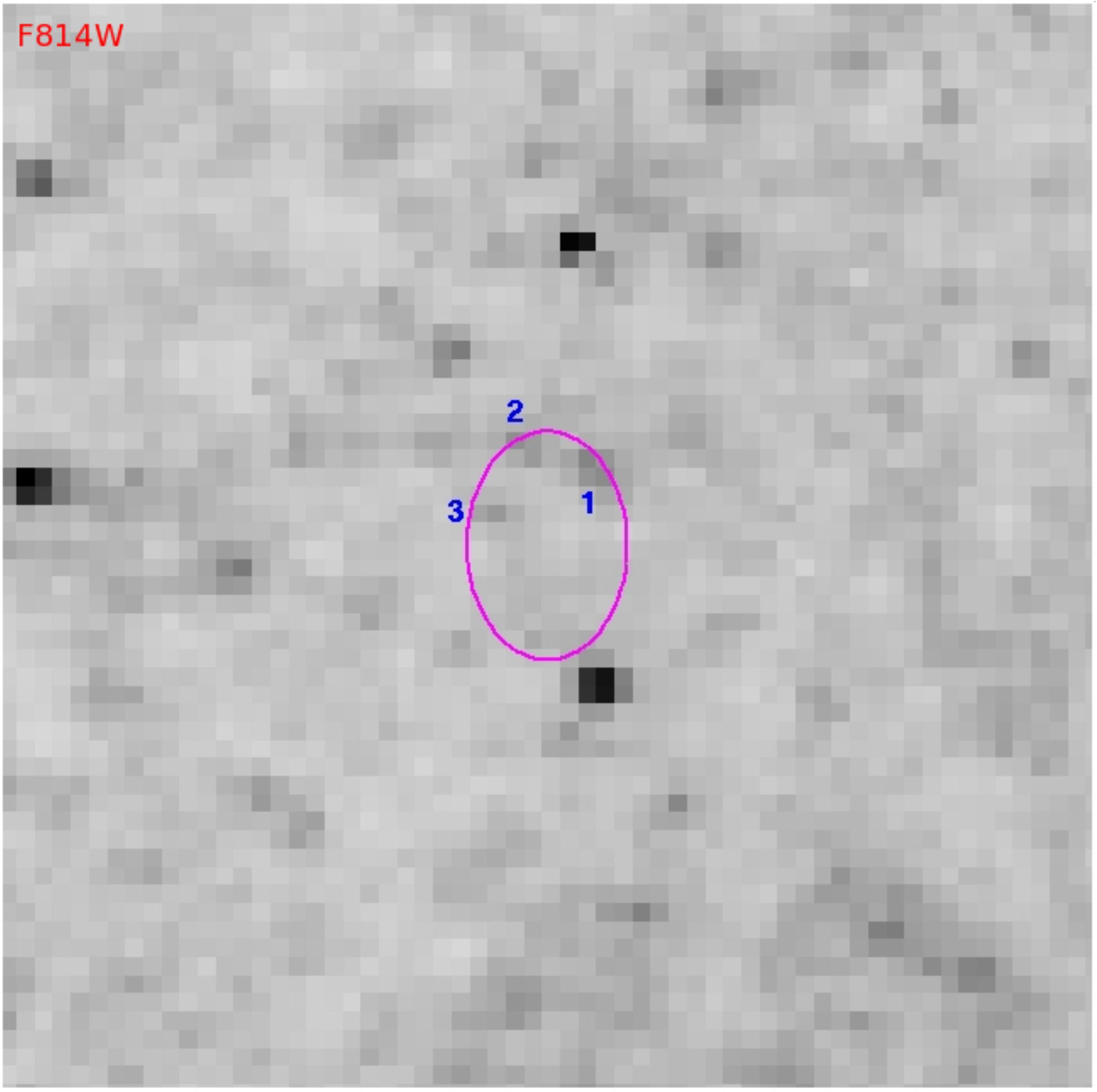} 

\end{center}
\addtocounter{figure}{-1}
\caption{\small{\emph{continued, pg. 6} -  Specific notes: displayed ULX regions are, from top to bottom, IC 342 X-2 (F435W), IC 342 ULX2 (F814W), \&  IC 342 X6 (F814W),. IC 342 X-2 has three potential counterparts within the blue error ellipse, while only two are contained within the magenta region. As a result we consider sources 1 \& 2 to be more likely counterparts, but retain 3 in our analysis due to uncertainties on cross-matching fields. Cross-matching Chandra positions with \emph{HST} data shows that IC 342 ULX2 is associated with the nucleus of the galaxy, and therefore should not be considered a ULX. It is most likely that this is a low luminosity AGN.}}
\label{fig:pictures}
\end{figure*}

\begin{figure*}
\leavevmode
\begin{center}

\includegraphics[height=64mm, angle=0]{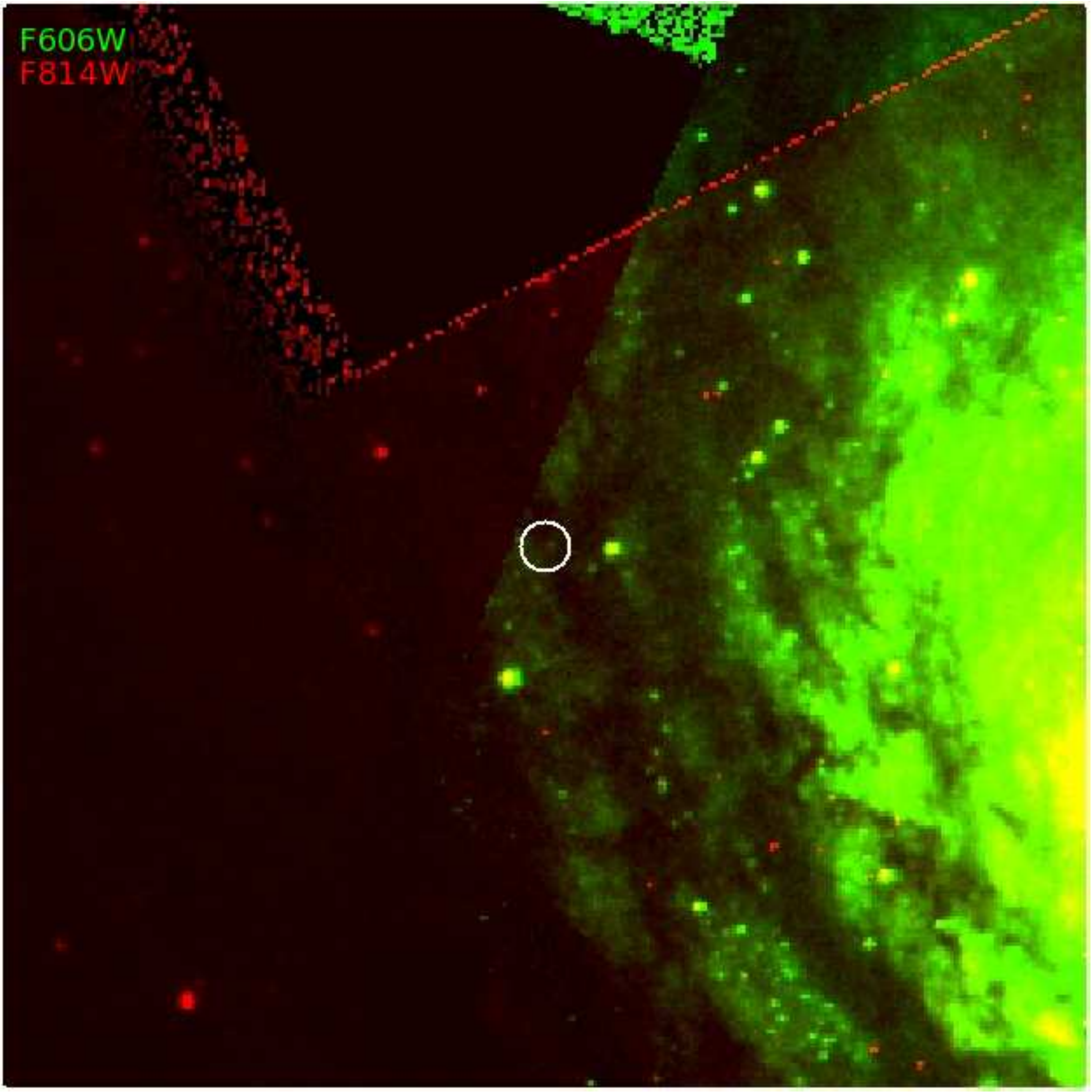} \hspace*{0.4cm}
\includegraphics[height=64mm, angle=0]{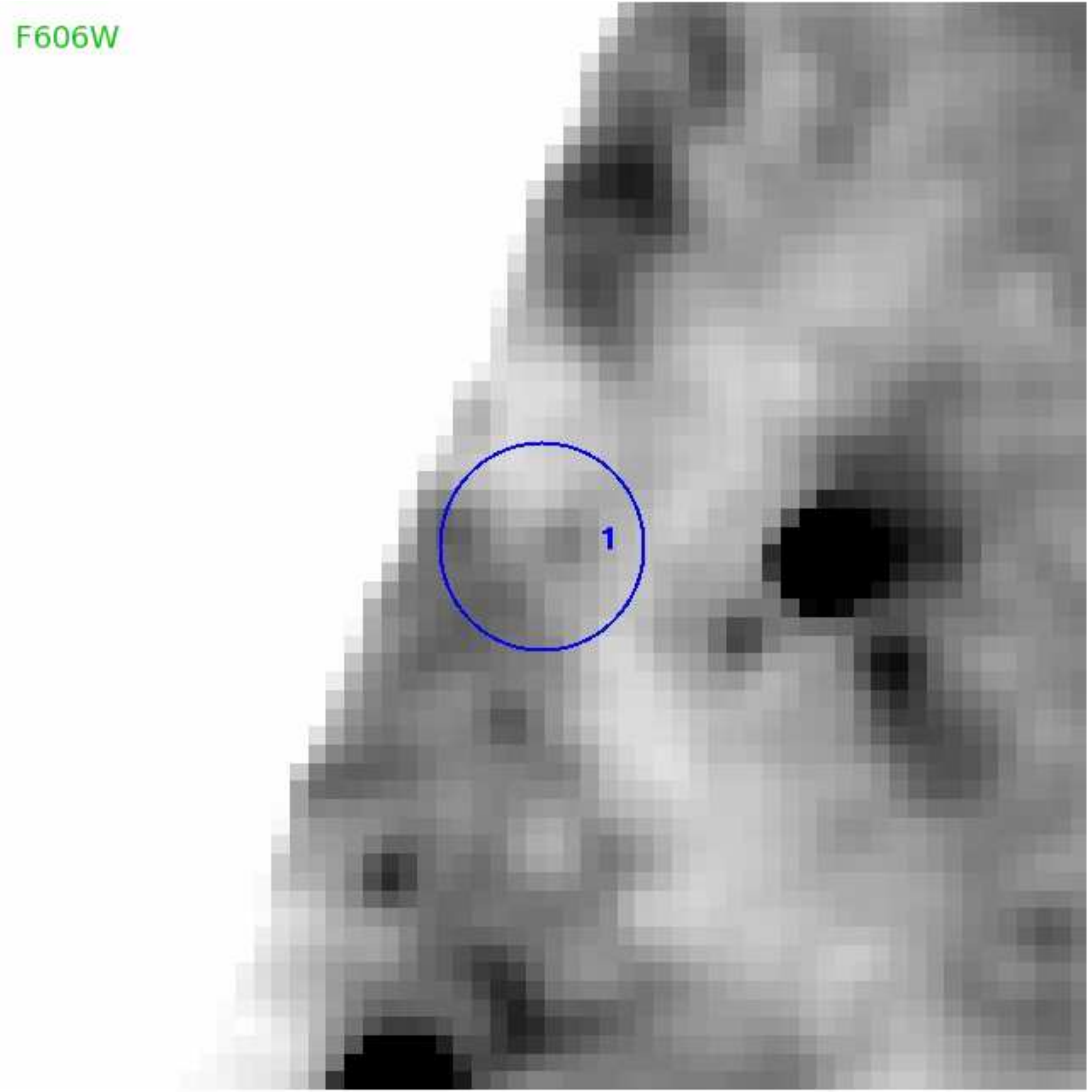} 
\vspace*{0.3cm} 

\includegraphics[height=64mm, angle=0]{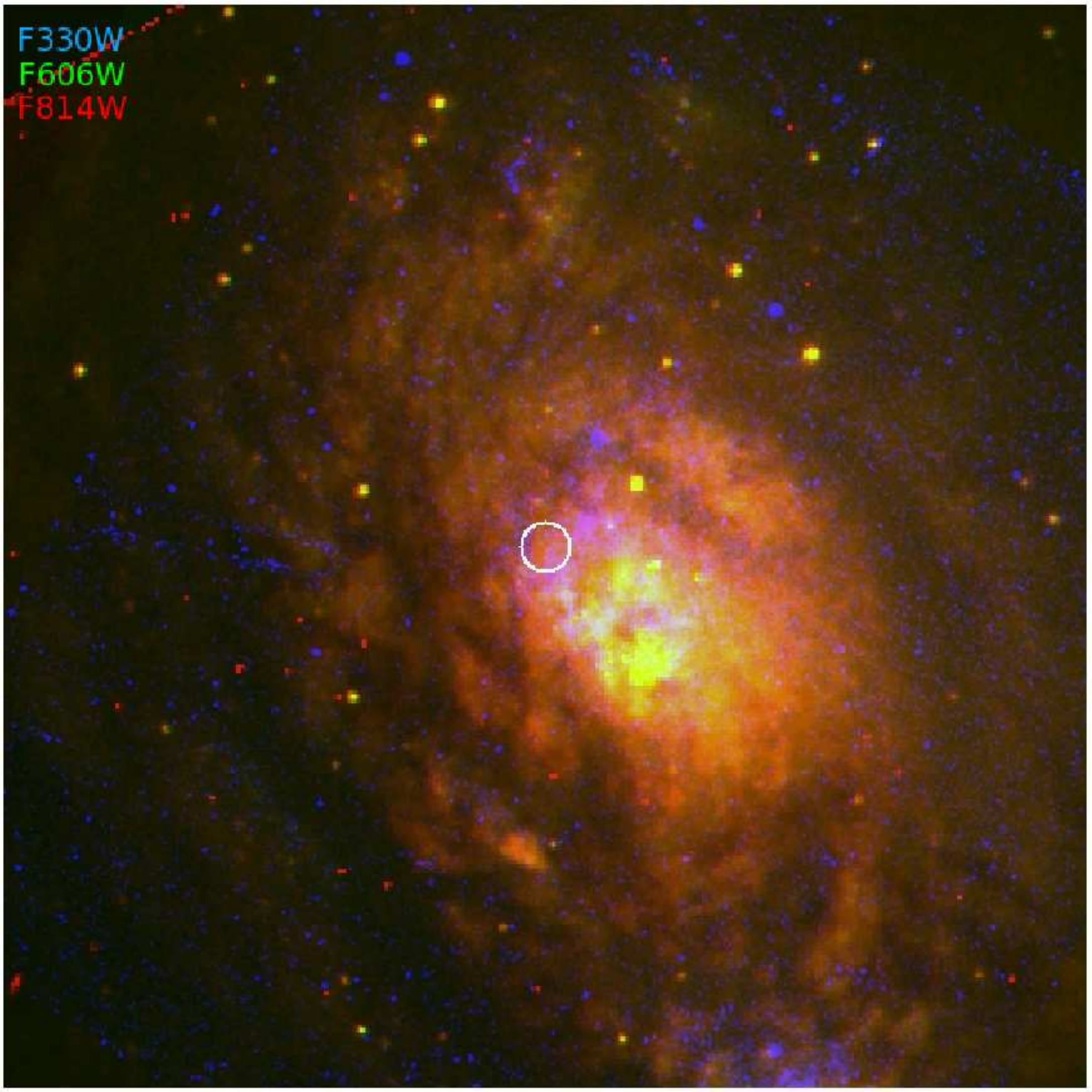} \hspace*{0.4cm}
\includegraphics[height=64mm, angle=0]{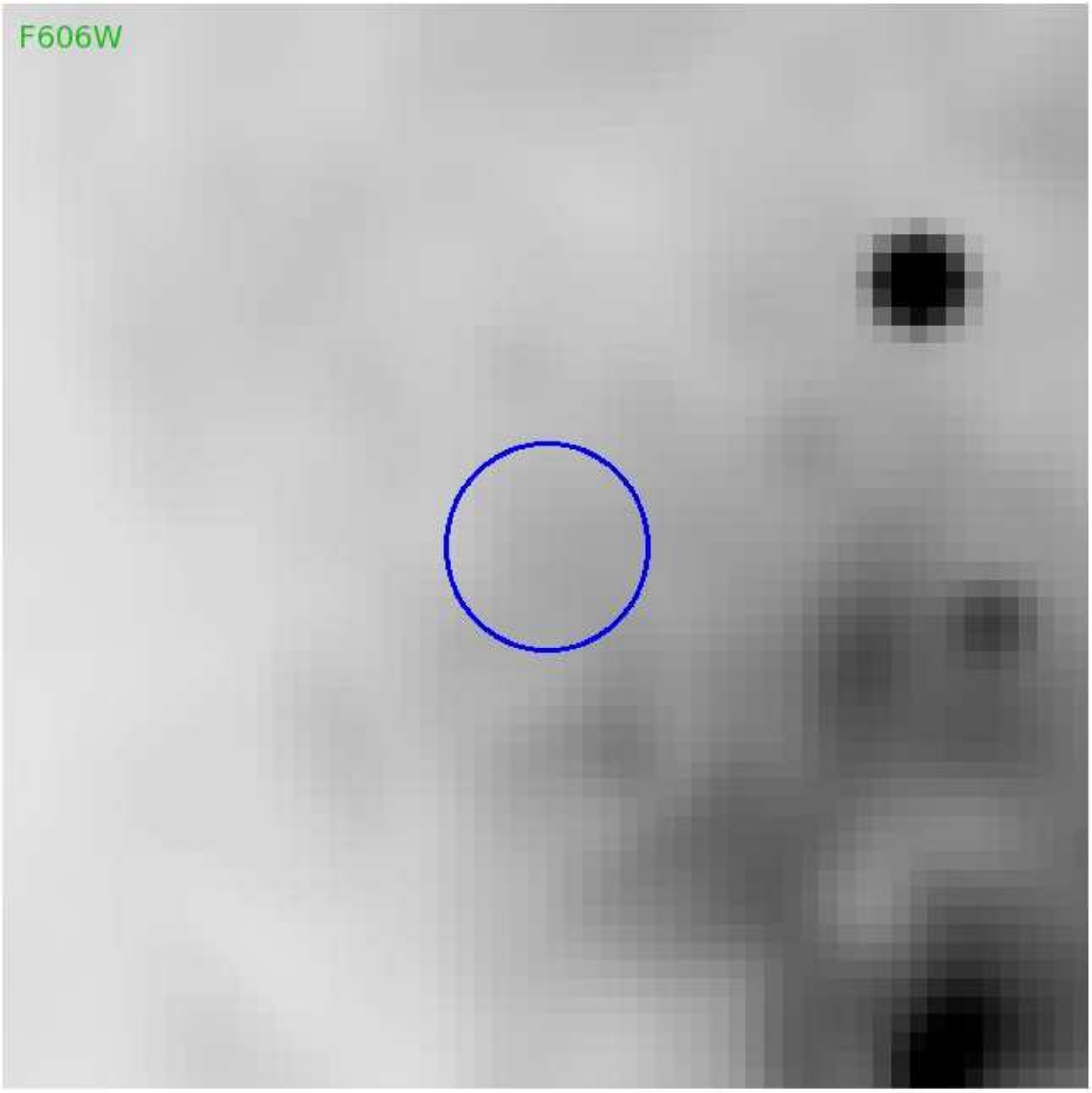} 
\vspace*{0.3cm} 

\includegraphics[height=64mm, angle=0]{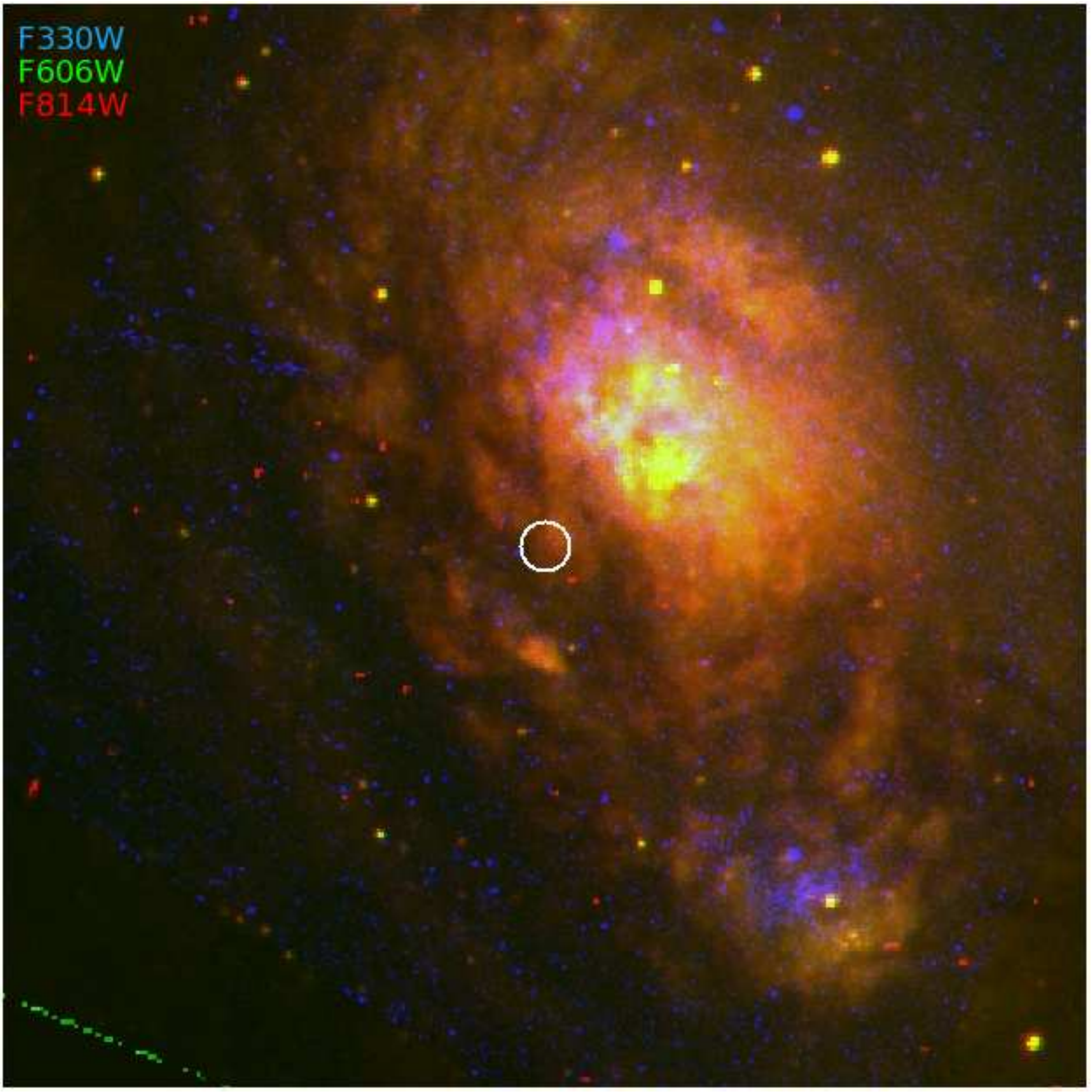} \hspace*{0.4cm}
\includegraphics[height=64mm, angle=0]{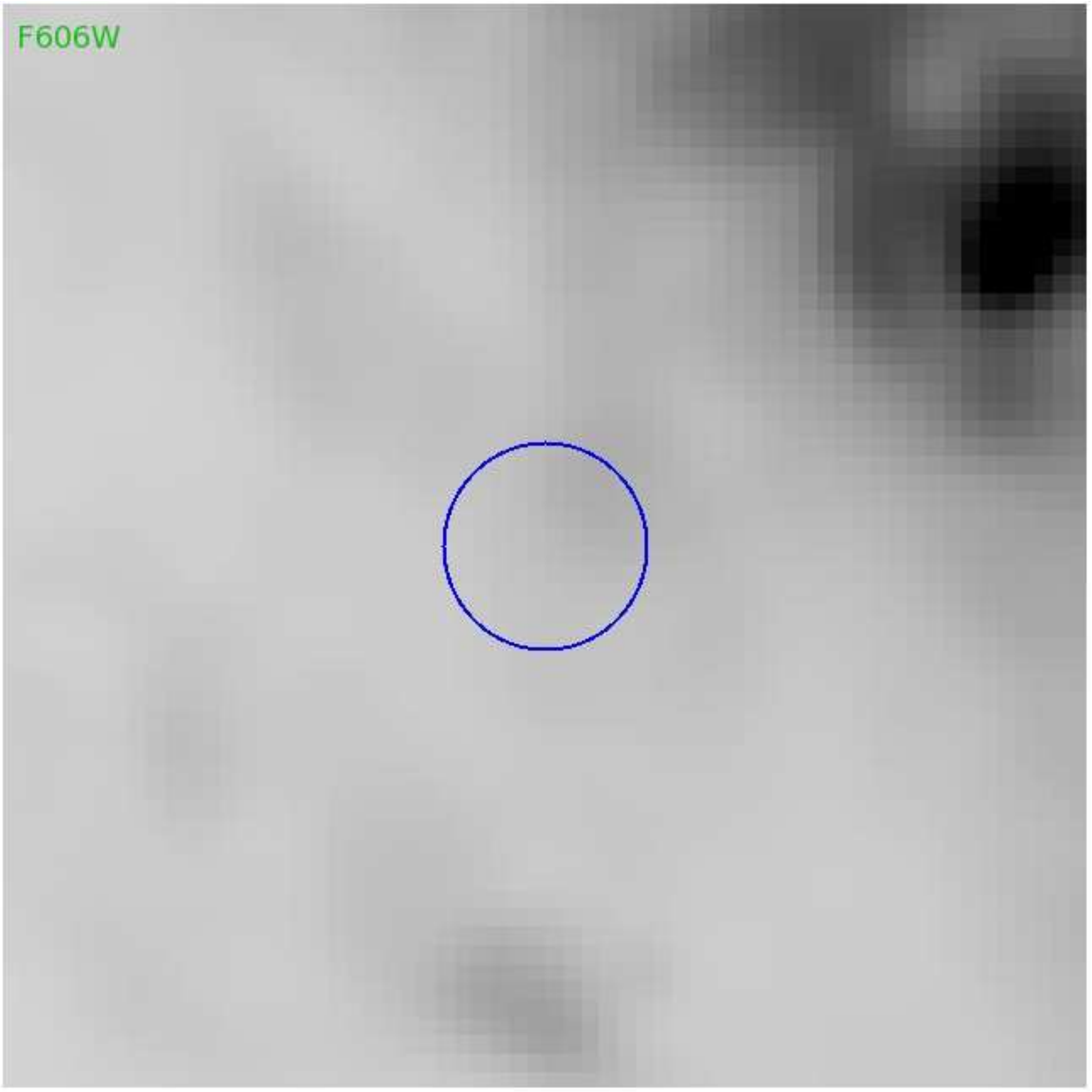} 

\end{center}
\addtocounter{figure}{-1}
\caption{\small{\emph{continued, pg. 7} -  Specific notes: displayed ULX regions are, from top to bottom, Circinus ULX1 (F606W), Circinus ULX3 (F606W) \& Circinus ULX4 (F606W).}}
\label{fig:pictures}
\end{figure*}

\begin{figure*}
\leavevmode
\begin{center}

\includegraphics[height=64mm, angle=0]{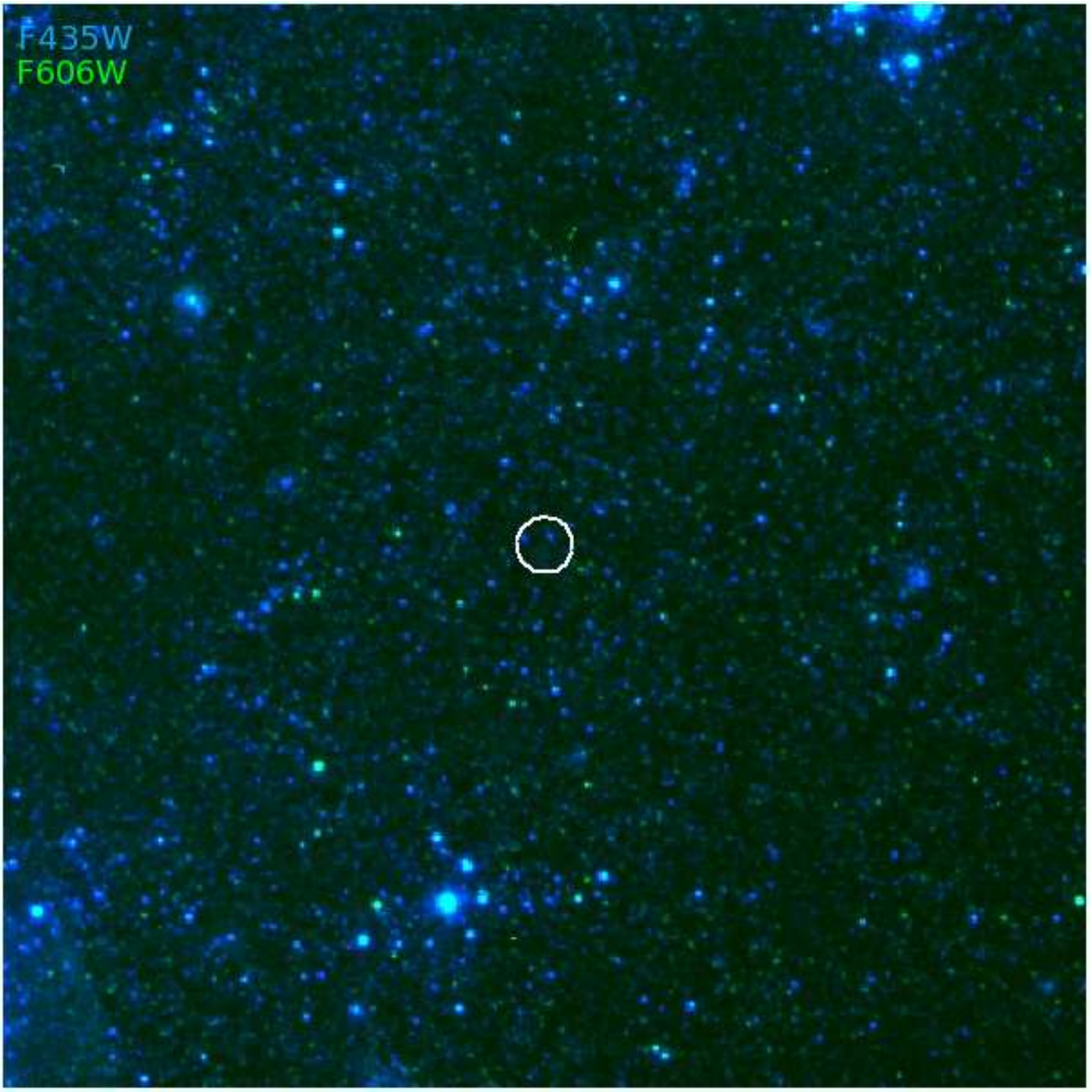} \hspace*{0.4cm}
\includegraphics[height=64mm, angle=0]{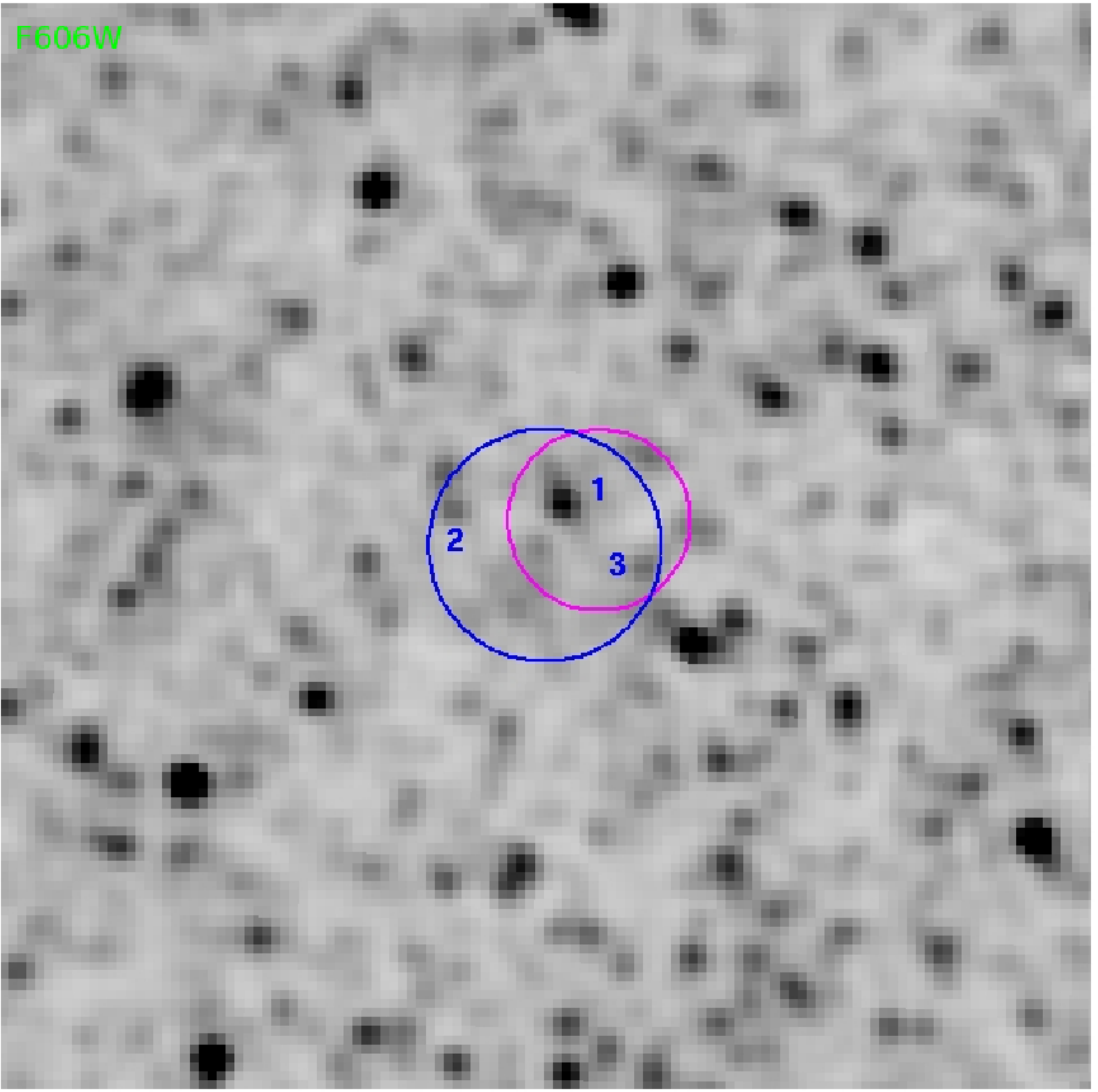} 
\vspace*{0.3cm} 

\includegraphics[height=64mm, angle=0]{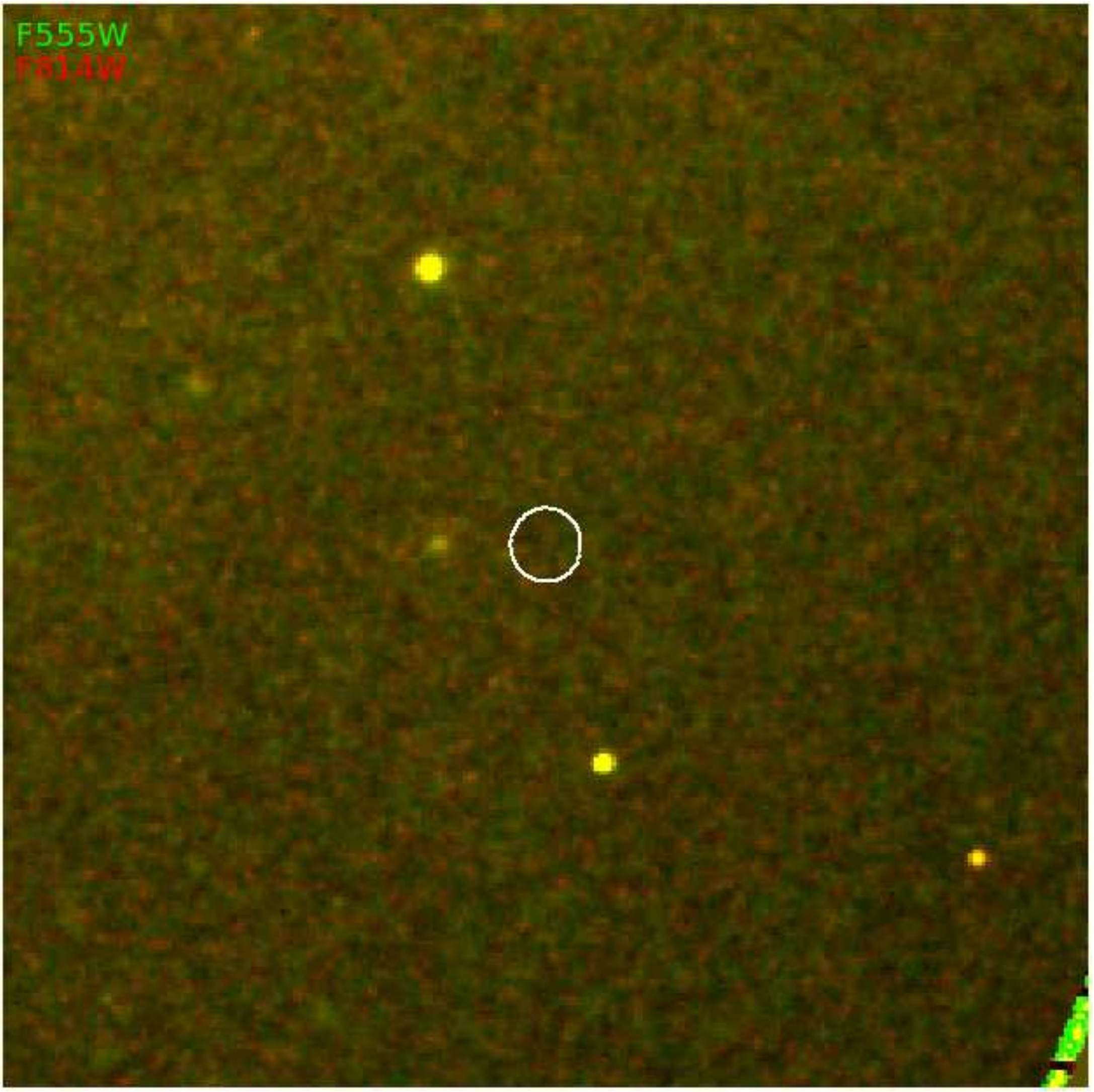} \hspace*{0.4cm}
\includegraphics[height=64mm, angle=0]{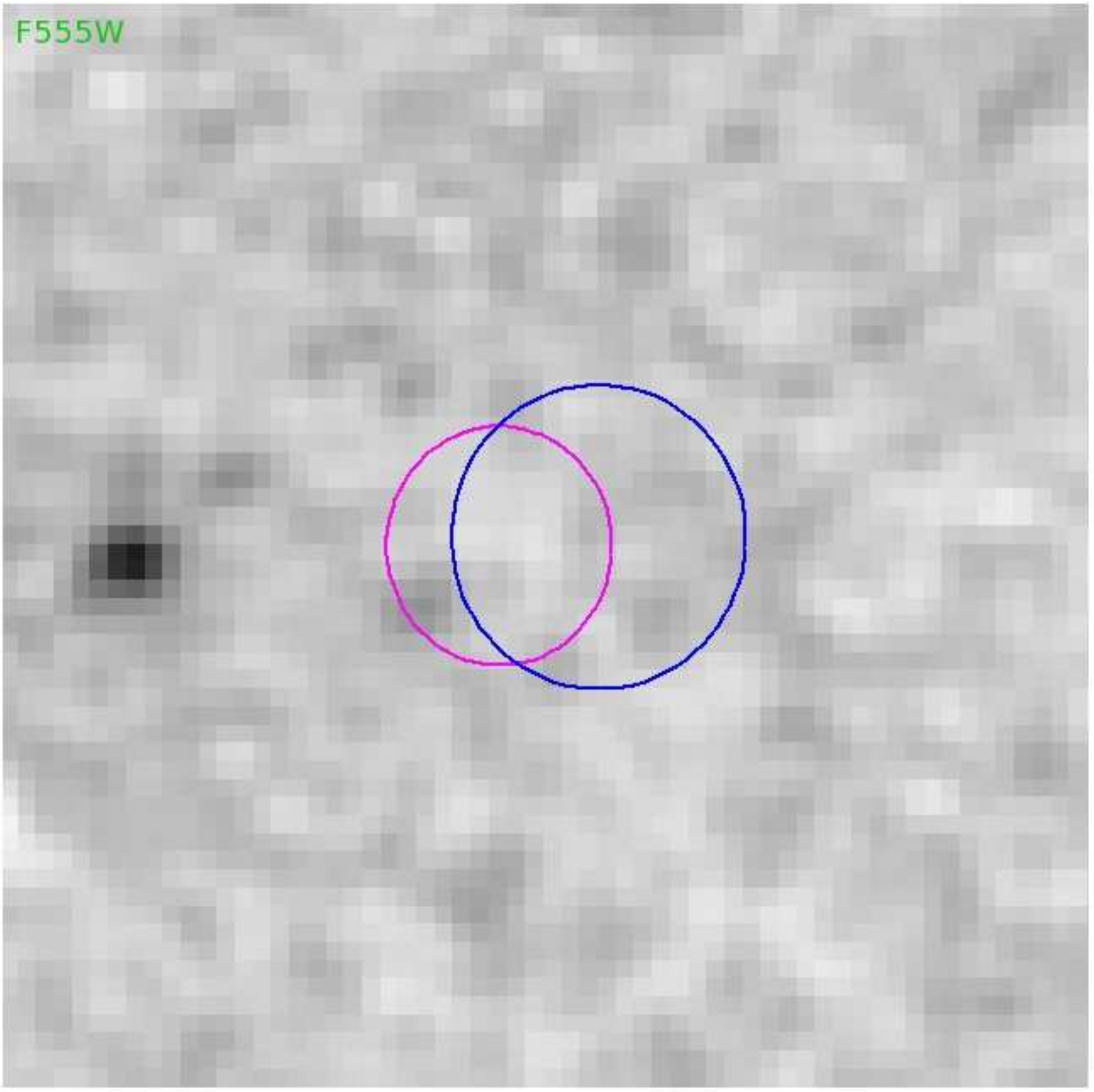} 
\vspace*{0.3cm} 

\includegraphics[height=64mm, angle=0]{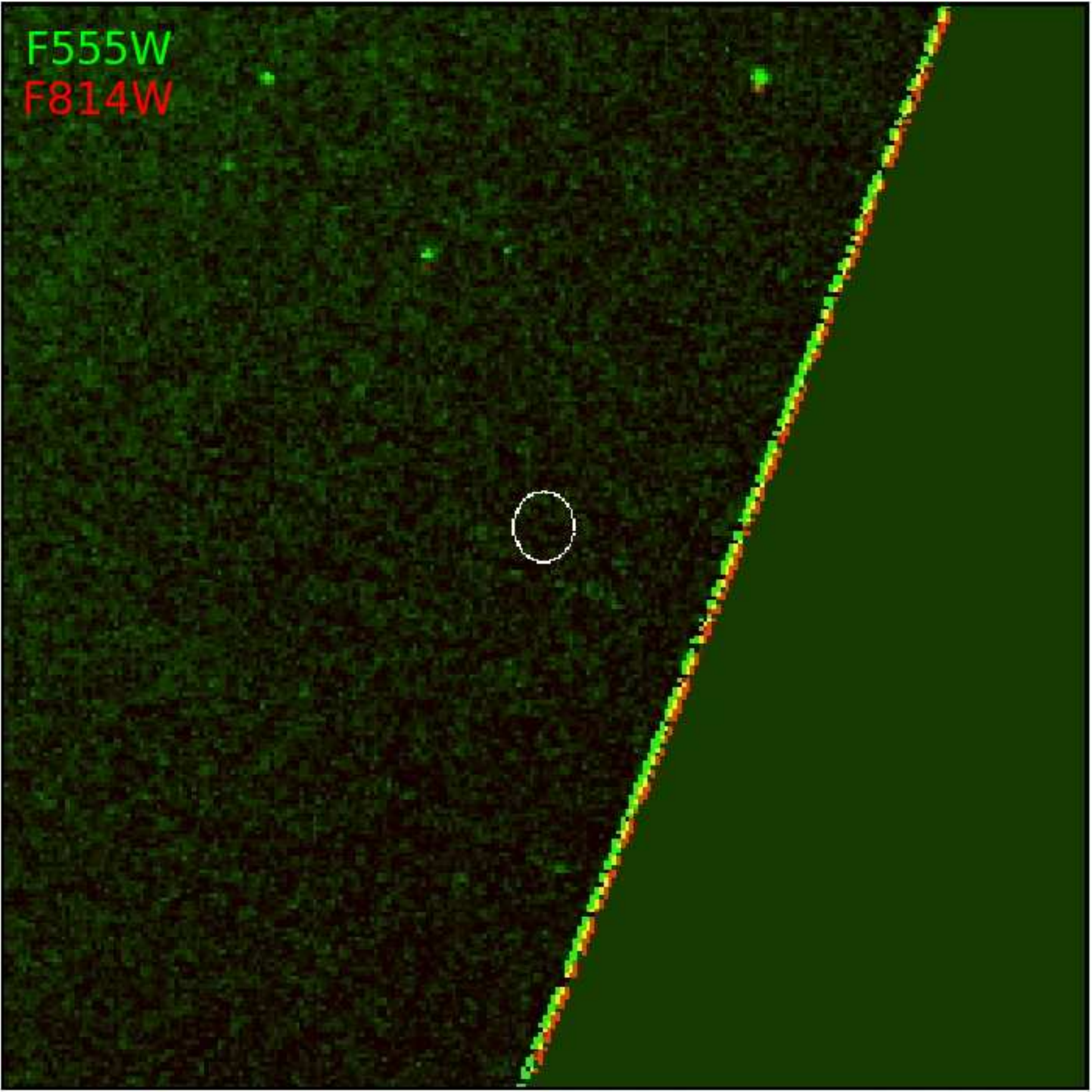} \hspace*{0.4cm}
\includegraphics[height=64mm, angle=0]{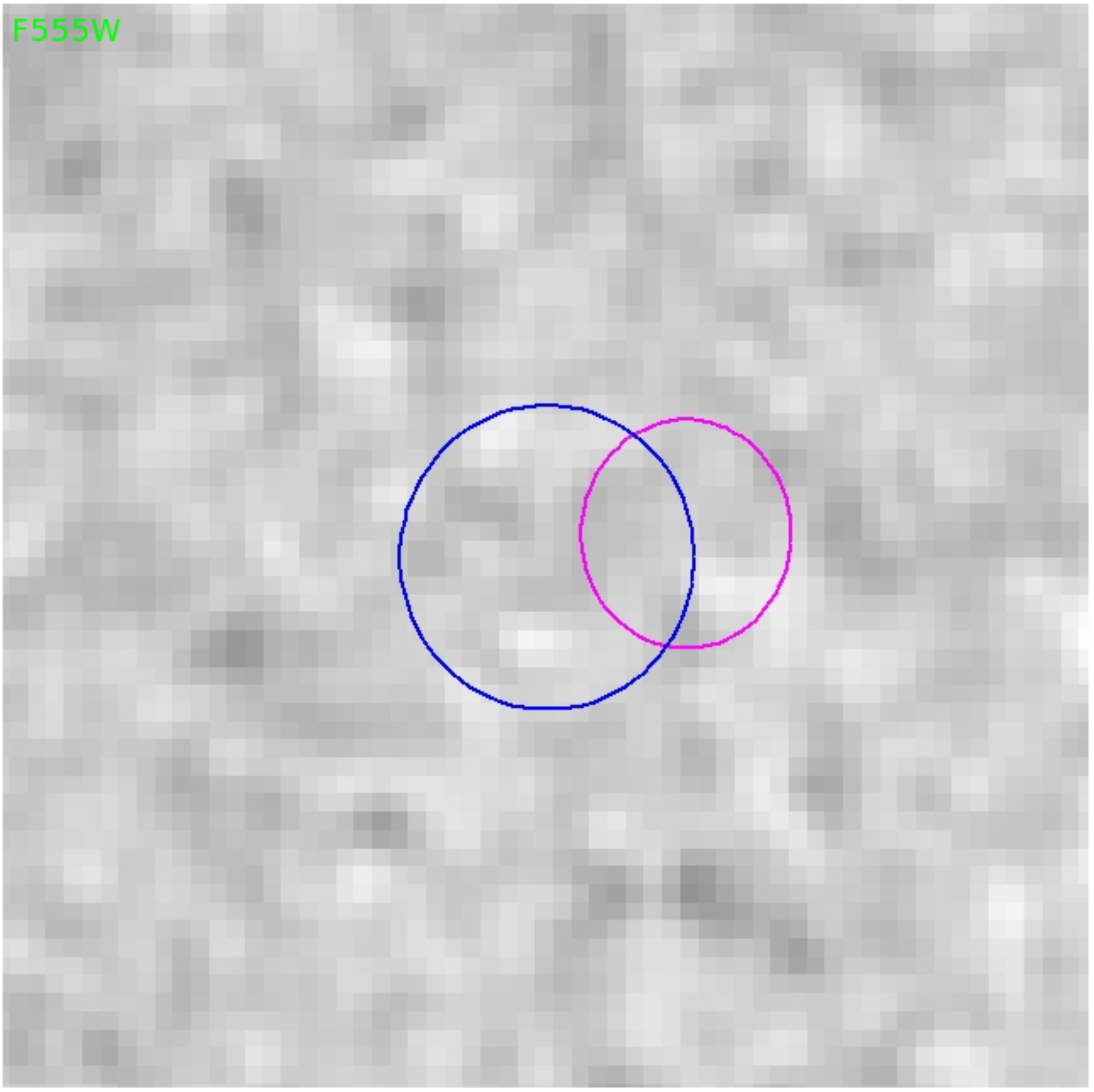} 

\end{center}
\addtocounter{figure}{-1}
\caption{\small{\emph{continued, pg. 8} -  Specific notes: displayed ULX regions are, from top to bottom, NGC 2403 X-1 (F606W), NGC 5128 ULX1 (F555W) \& CXOU J132518.3-430304 (F555W). Potential counterparts 2 \& 3 to NGC 2408 X-1 are ruled out as they reside outside the error region in other bands. }}
\label{fig:pictures}
\end{figure*}

\begin{figure*}
\leavevmode
\begin{center}

\includegraphics[height=64mm, angle=0]{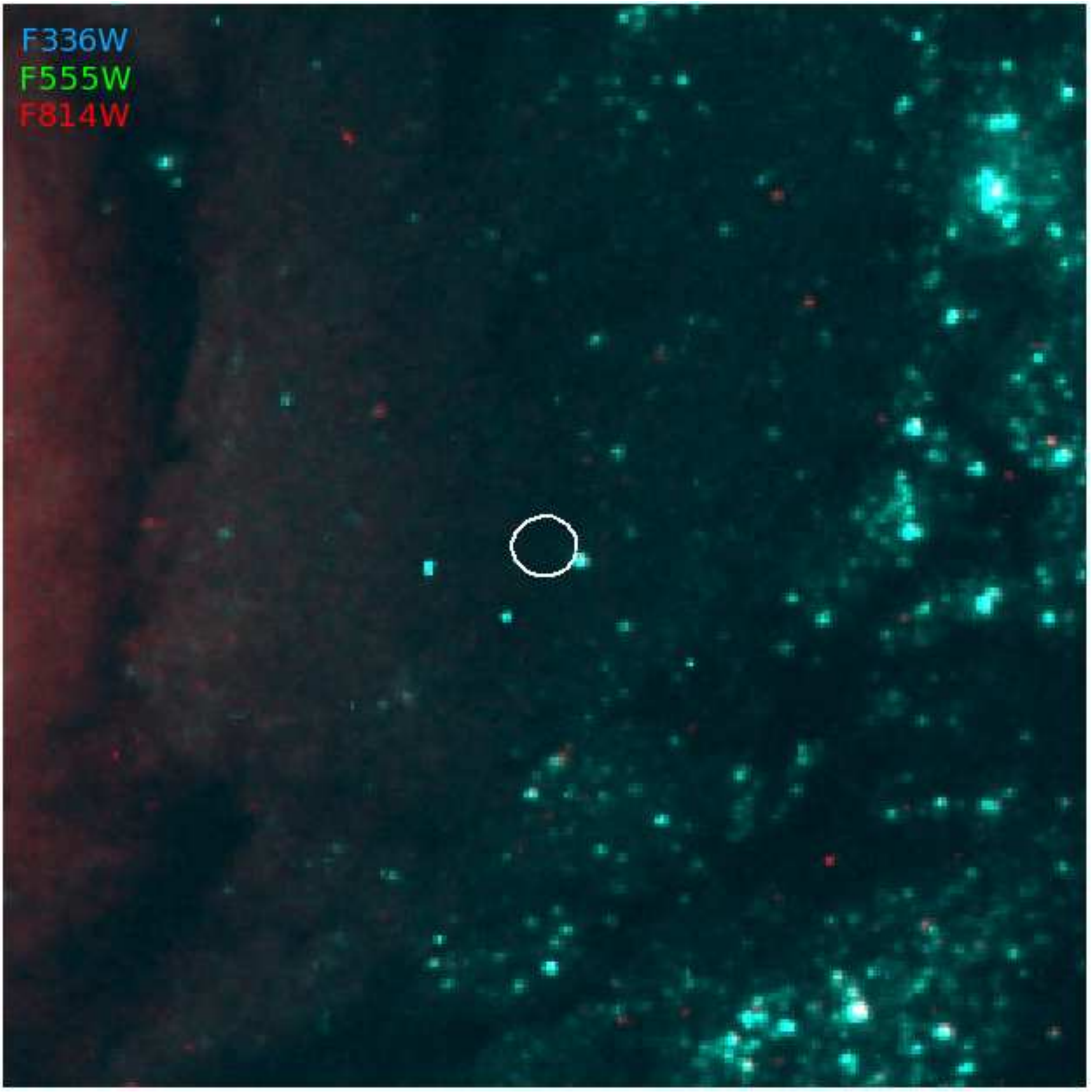} \hspace*{0.4cm}
\includegraphics[height=64mm, angle=0]{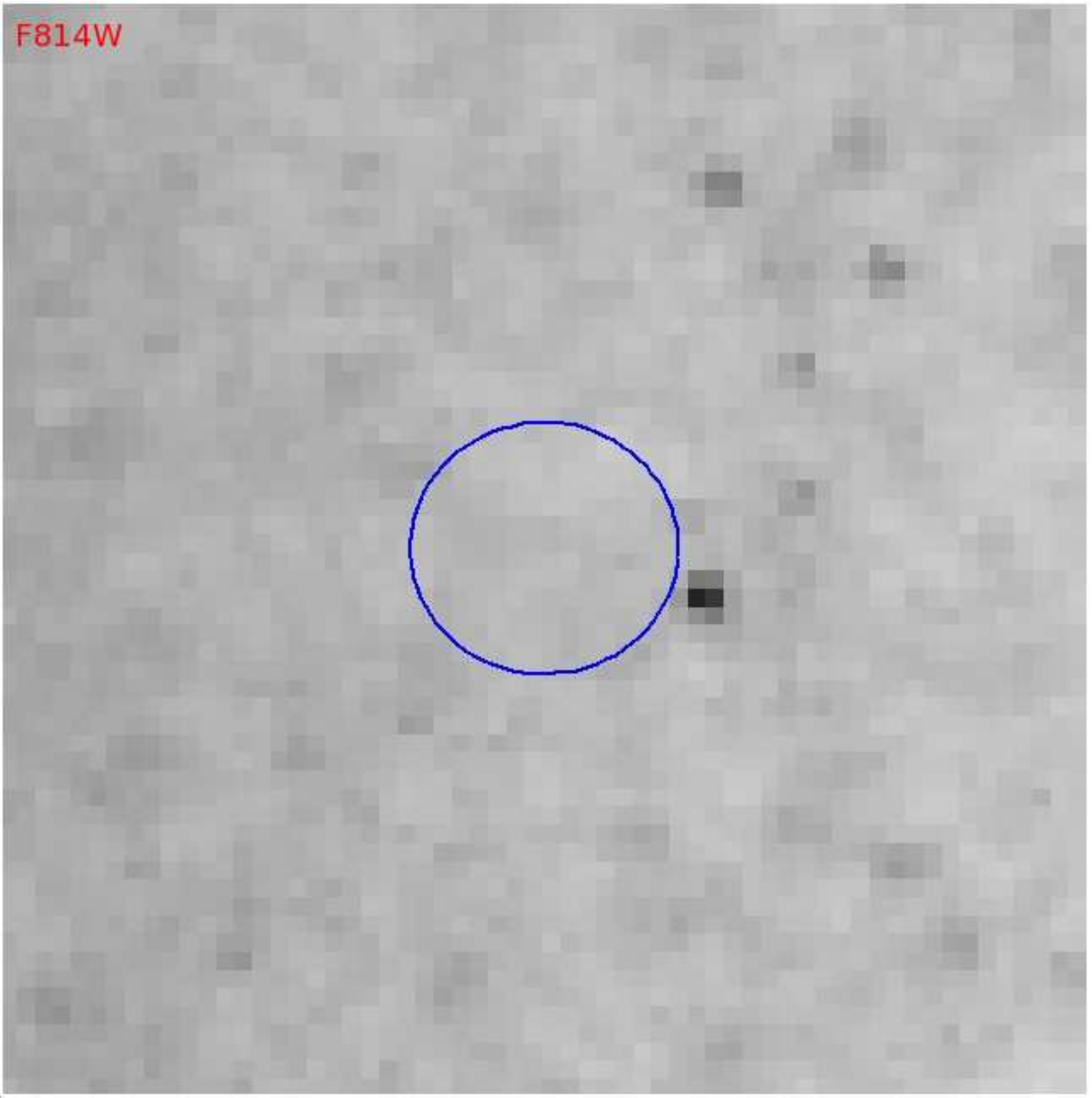} 
\vspace*{0.3cm} 

\includegraphics[height=64mm, angle=0]{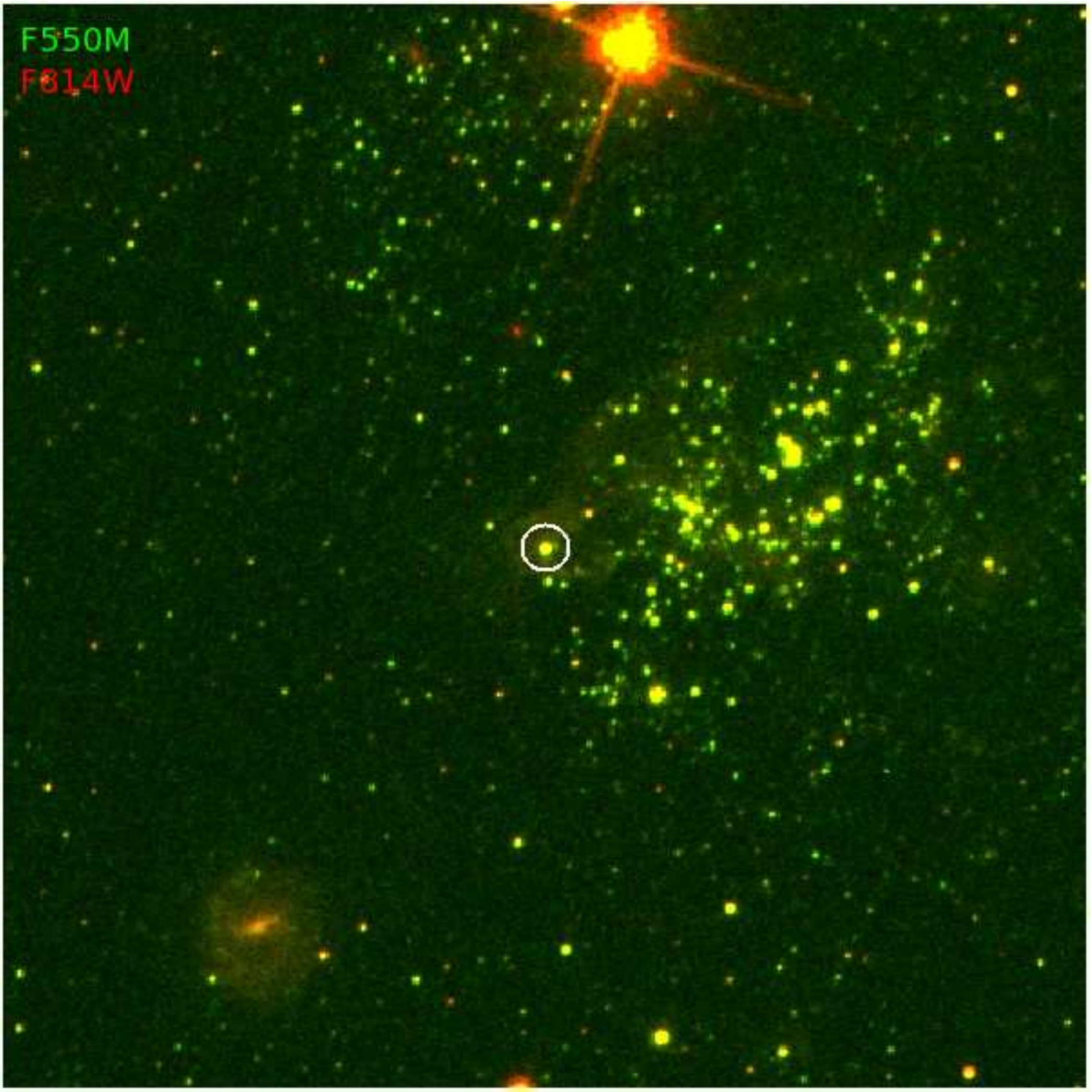} \hspace*{0.4cm}
\includegraphics[height=64mm, angle=0]{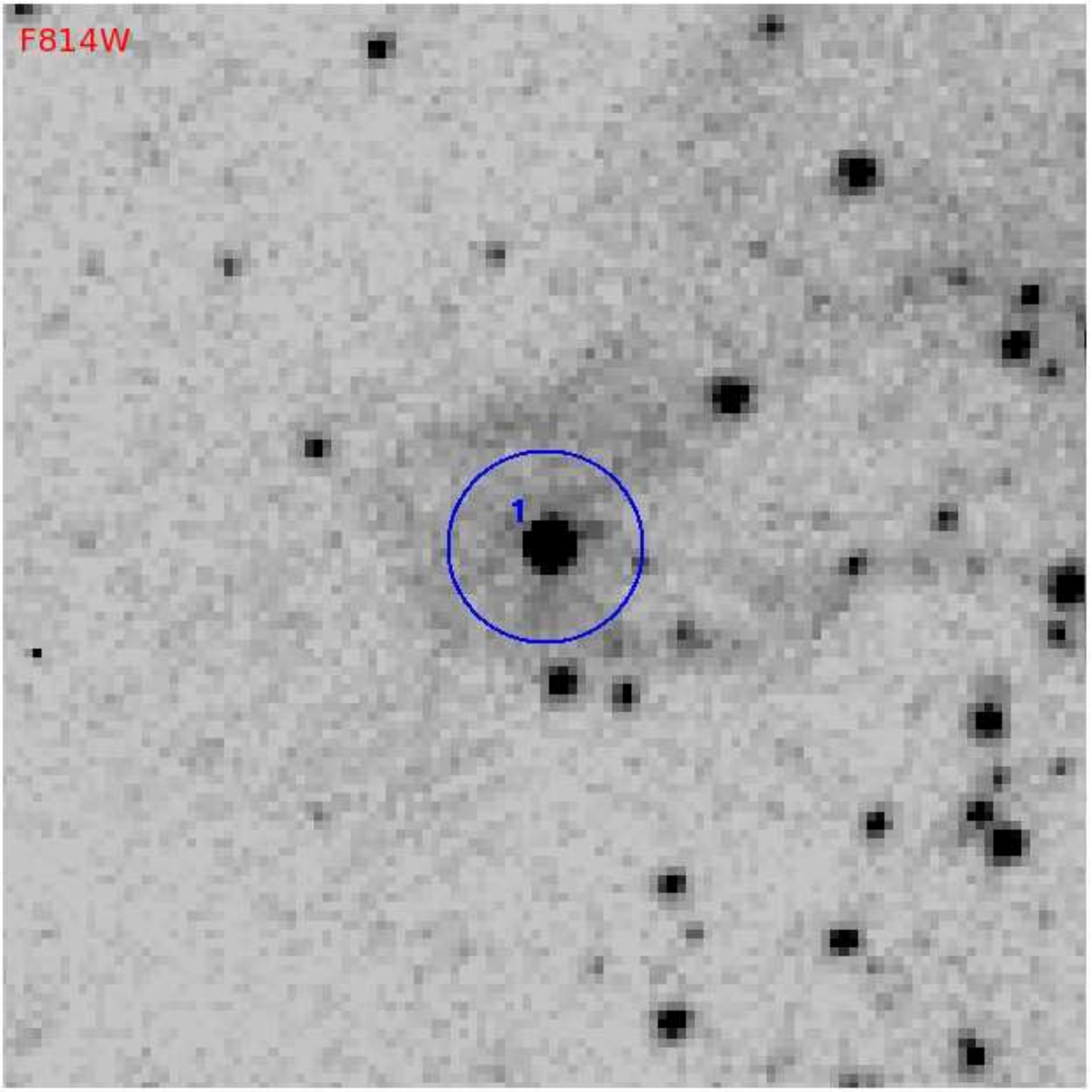} 
\vspace*{0.3cm} 

\includegraphics[height=64mm, angle=0]{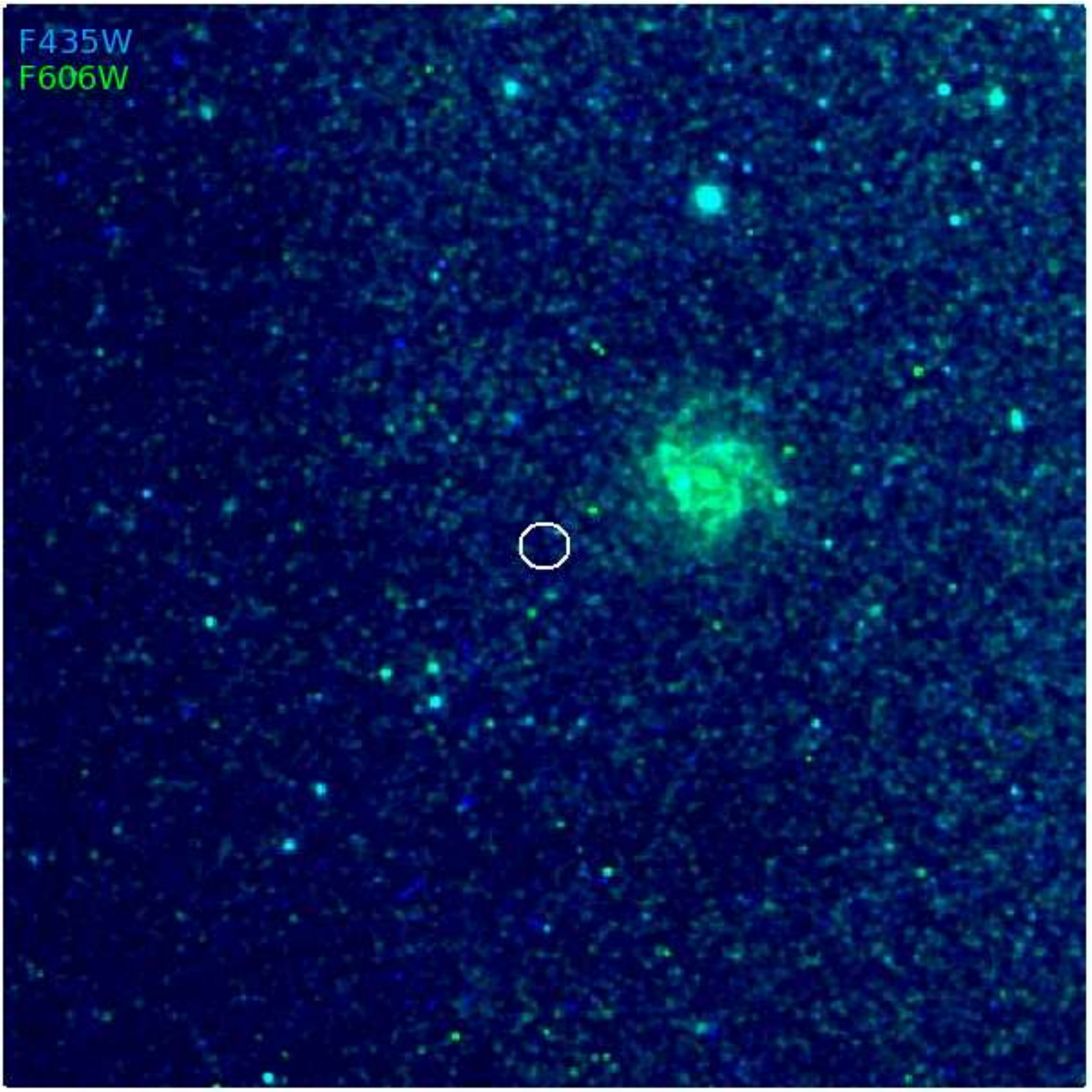} \hspace*{0.4cm}
\includegraphics[height=64mm, angle=0]{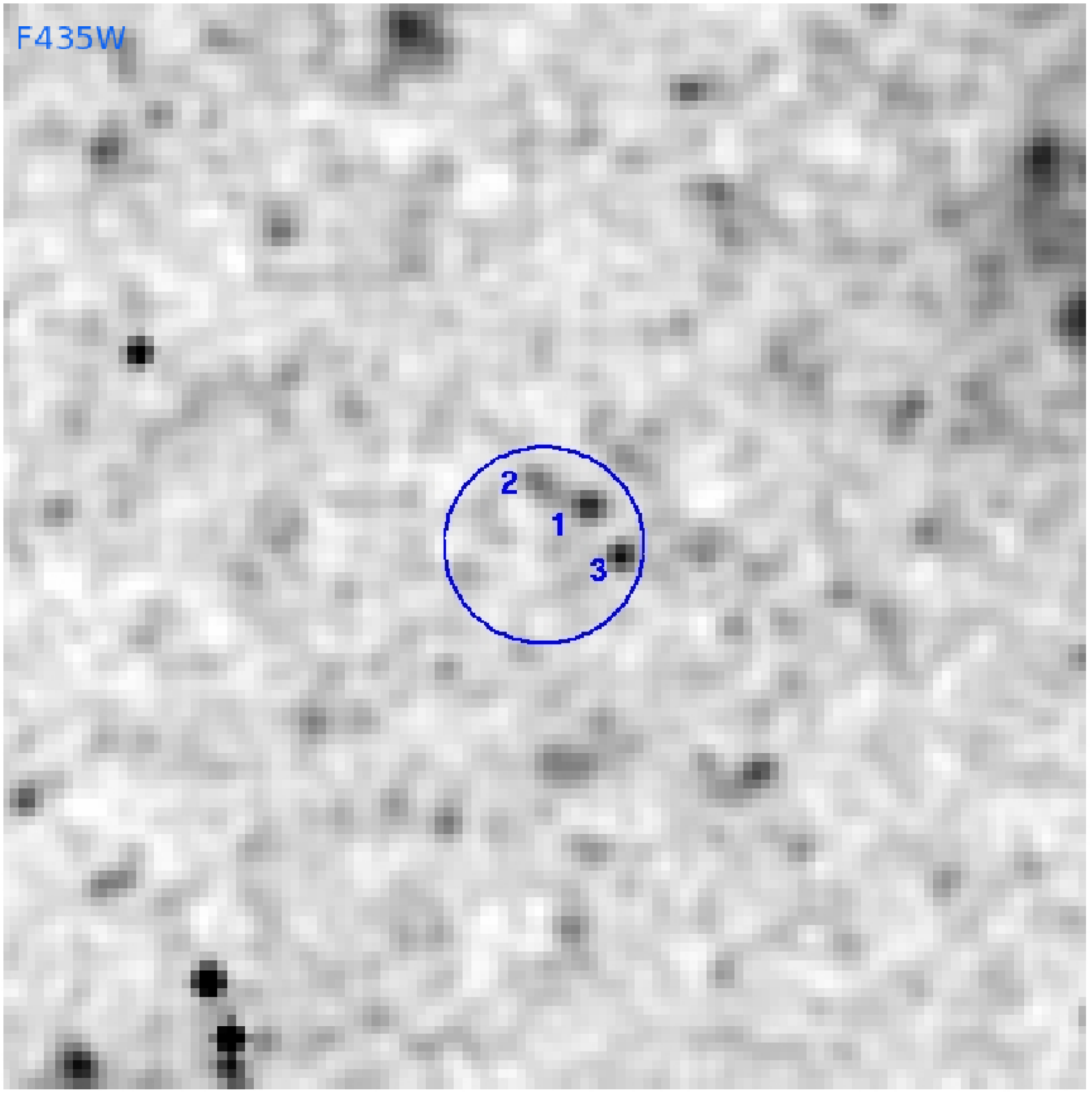} 

\end{center}
\addtocounter{figure}{-1}
\caption{\small{\emph{continued, pg. 9} -  Specific notes: displayed ULX regions are, from top to bottom, NGC 4736 X-1 (F814W), Holmberg II X-1 (F814W) \& M83 XMM1 (F435W).  Potential counterparts 2 \& 3 to M83 XMM1 are ruled out as they reside outside the error region in other bands. }}
\label{fig:pictures}
\end{figure*}

\begin{figure*}
\leavevmode
\begin{center}

\includegraphics[height=64mm, angle=0]{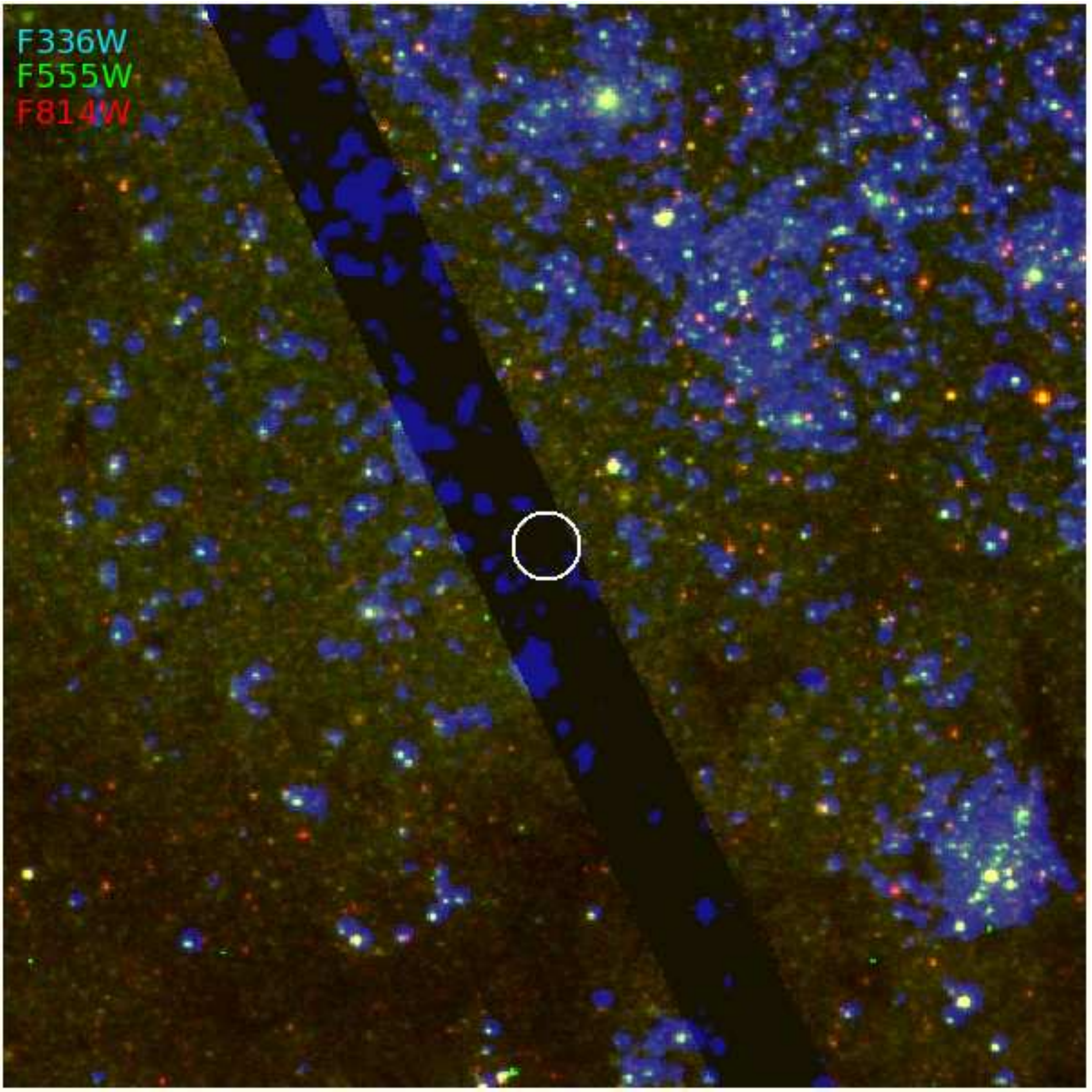} \hspace*{0.4cm}
\includegraphics[height=64mm, angle=0]{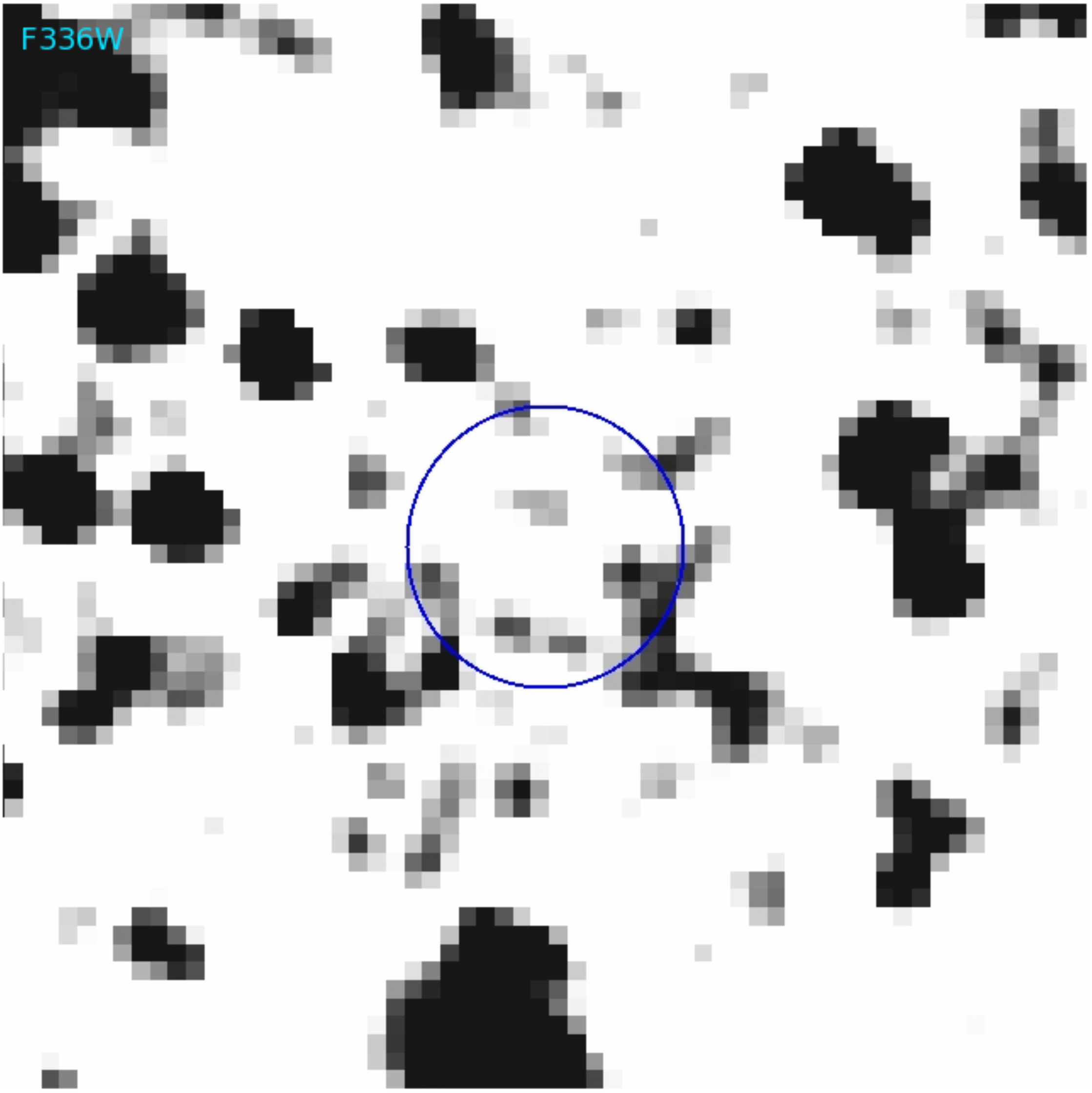} 
\vspace*{0.3cm} 

\includegraphics[height=64mm, angle=0]{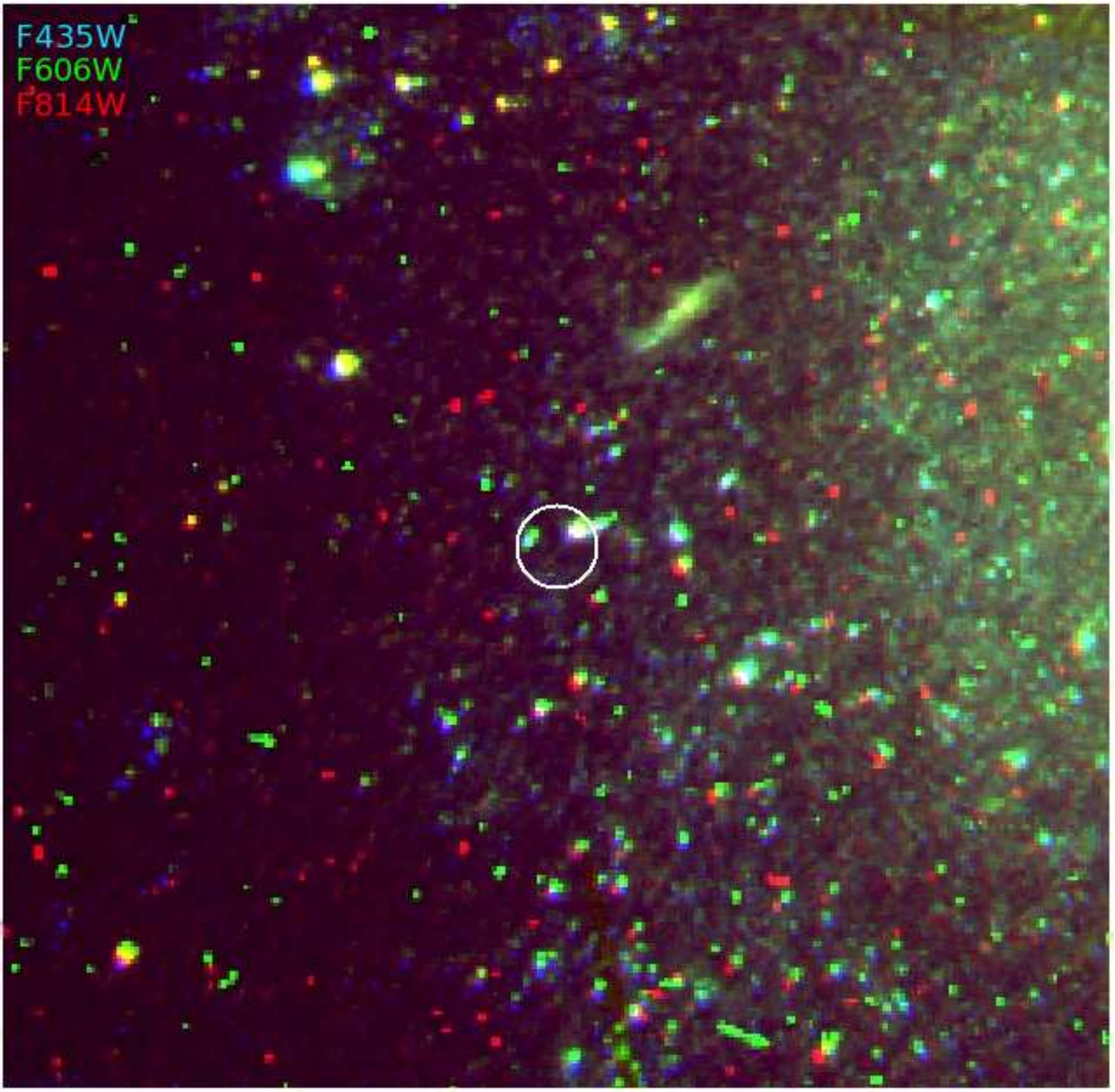} \hspace*{0.4cm}
\includegraphics[height=64mm, angle=0]{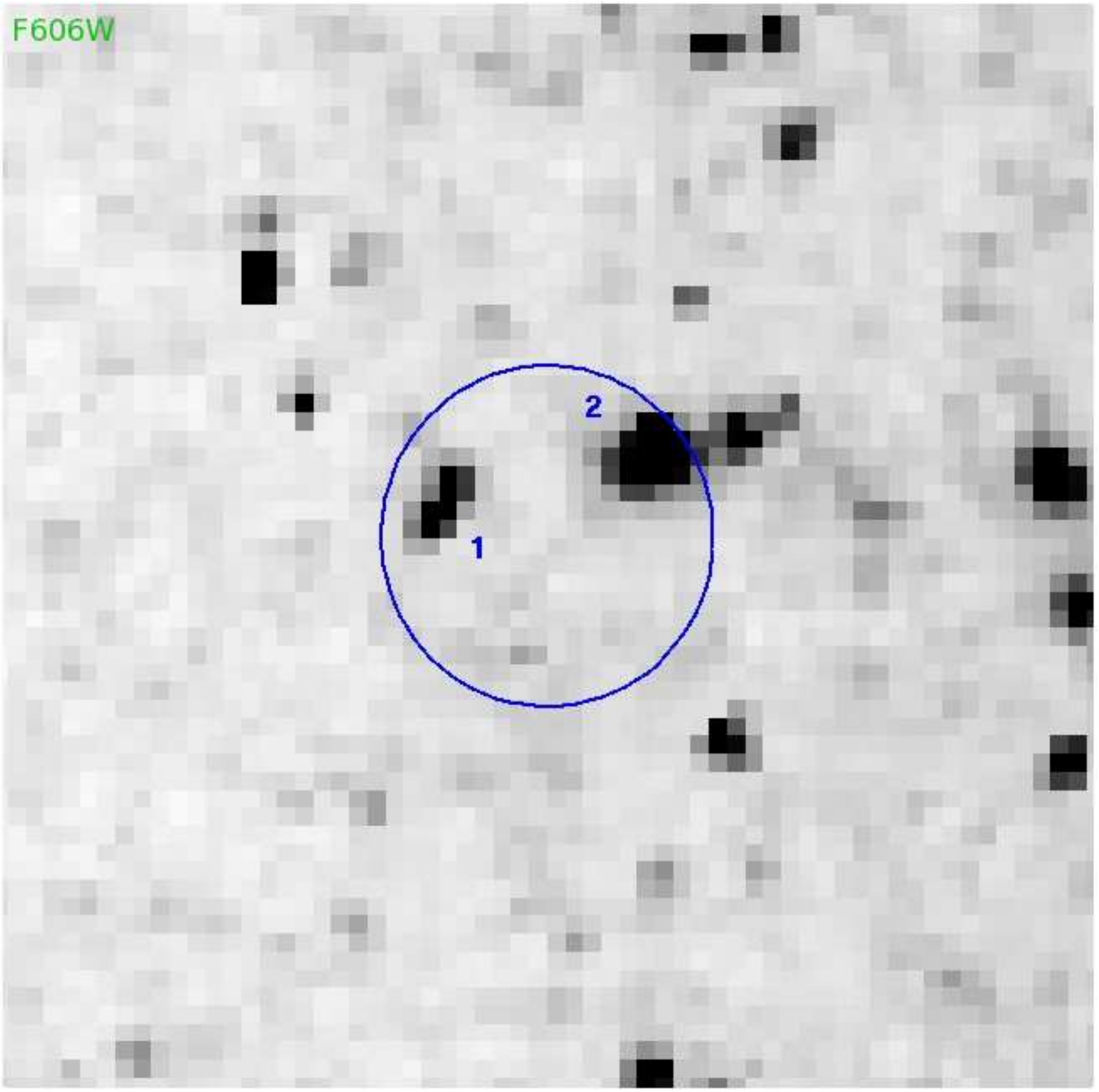} 
\vspace*{0.3cm} 

\includegraphics[height=64mm, angle=0]{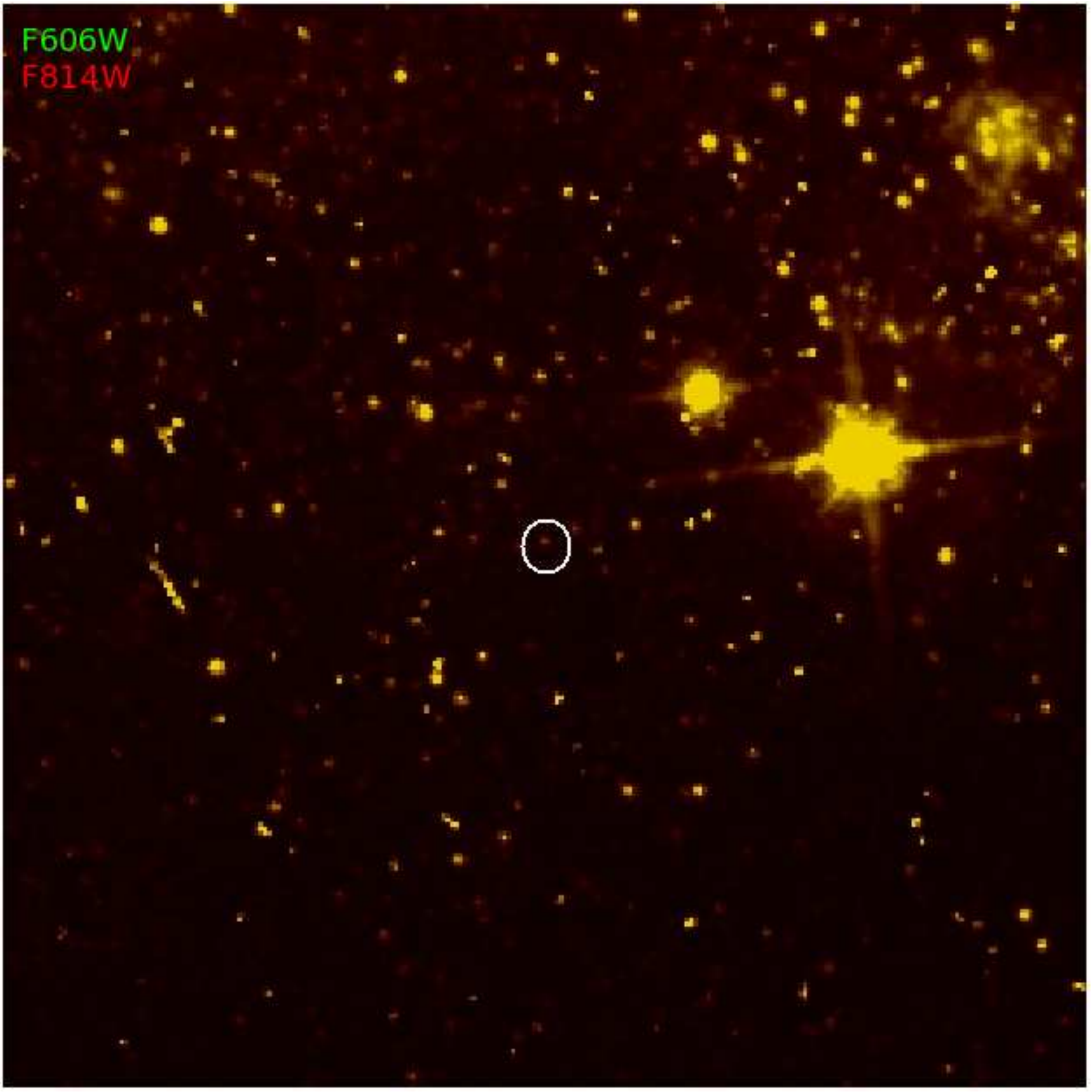} \hspace*{0.4cm}
\includegraphics[height=64mm, angle=0]{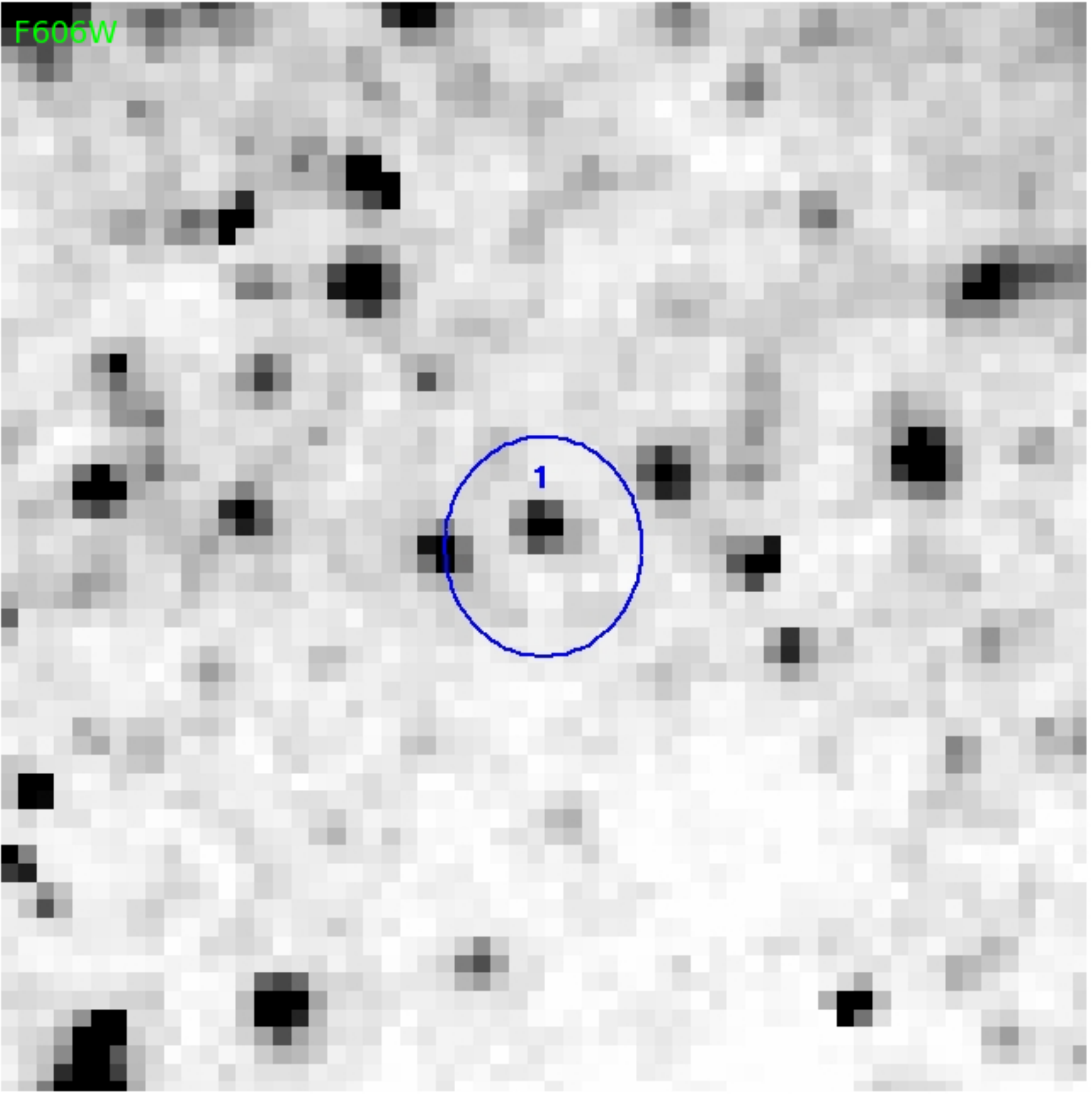} 

\end{center}
\addtocounter{figure}{-1}
\caption{\small{\emph{continued, pg. 10} -  Specific notes: displayed ULX regions are, from top to bottom, M83 XMM2 (F336W), NGC 5204 X-1 (F606W) \& NGC 5408 X-1 (F606W). }}
\label{fig:pictures}
\end{figure*}

\begin{figure*}
\leavevmode
\begin{center}

\includegraphics[height=64mm, angle=0]{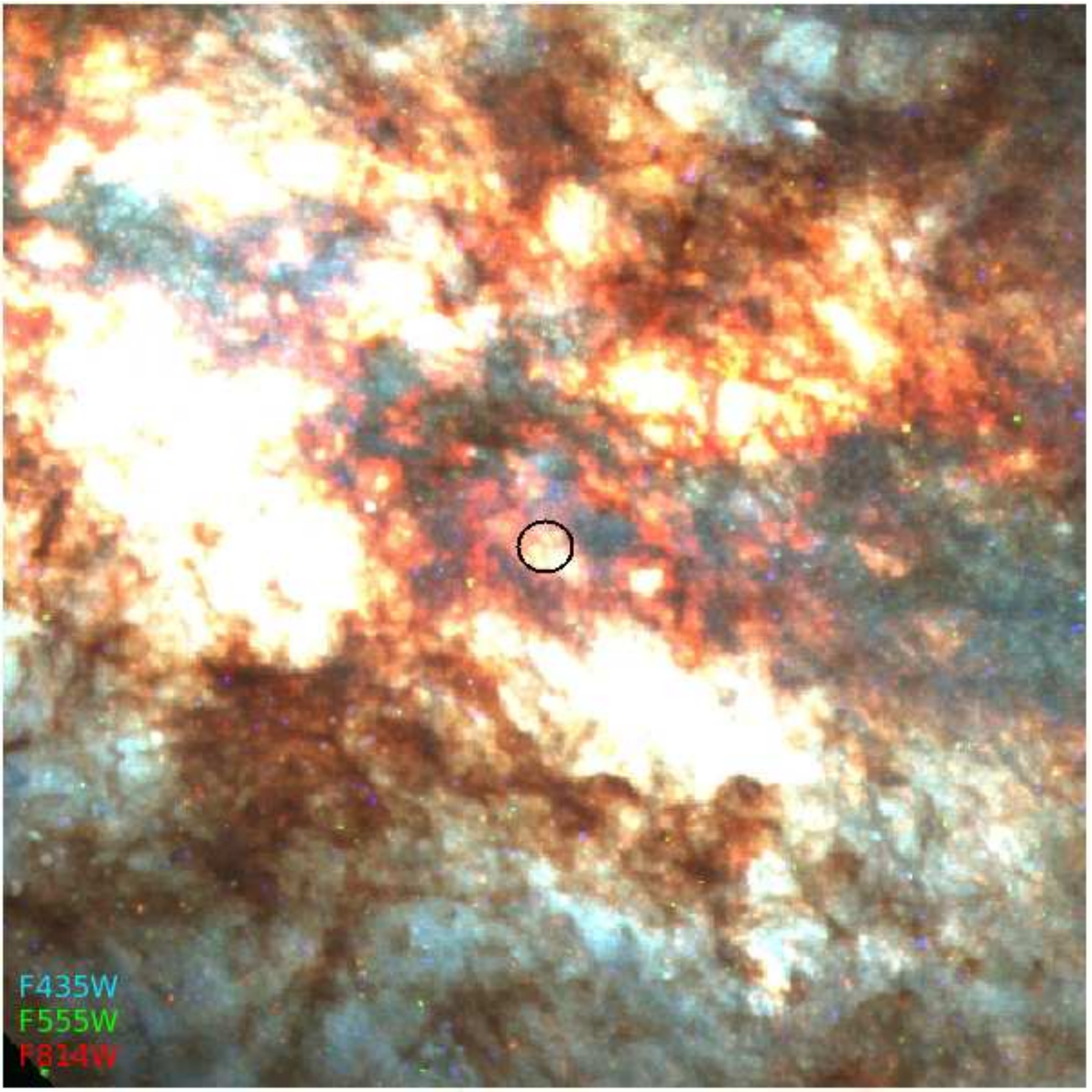}  \hspace*{0.4cm}
\includegraphics[height=64mm, angle=0]{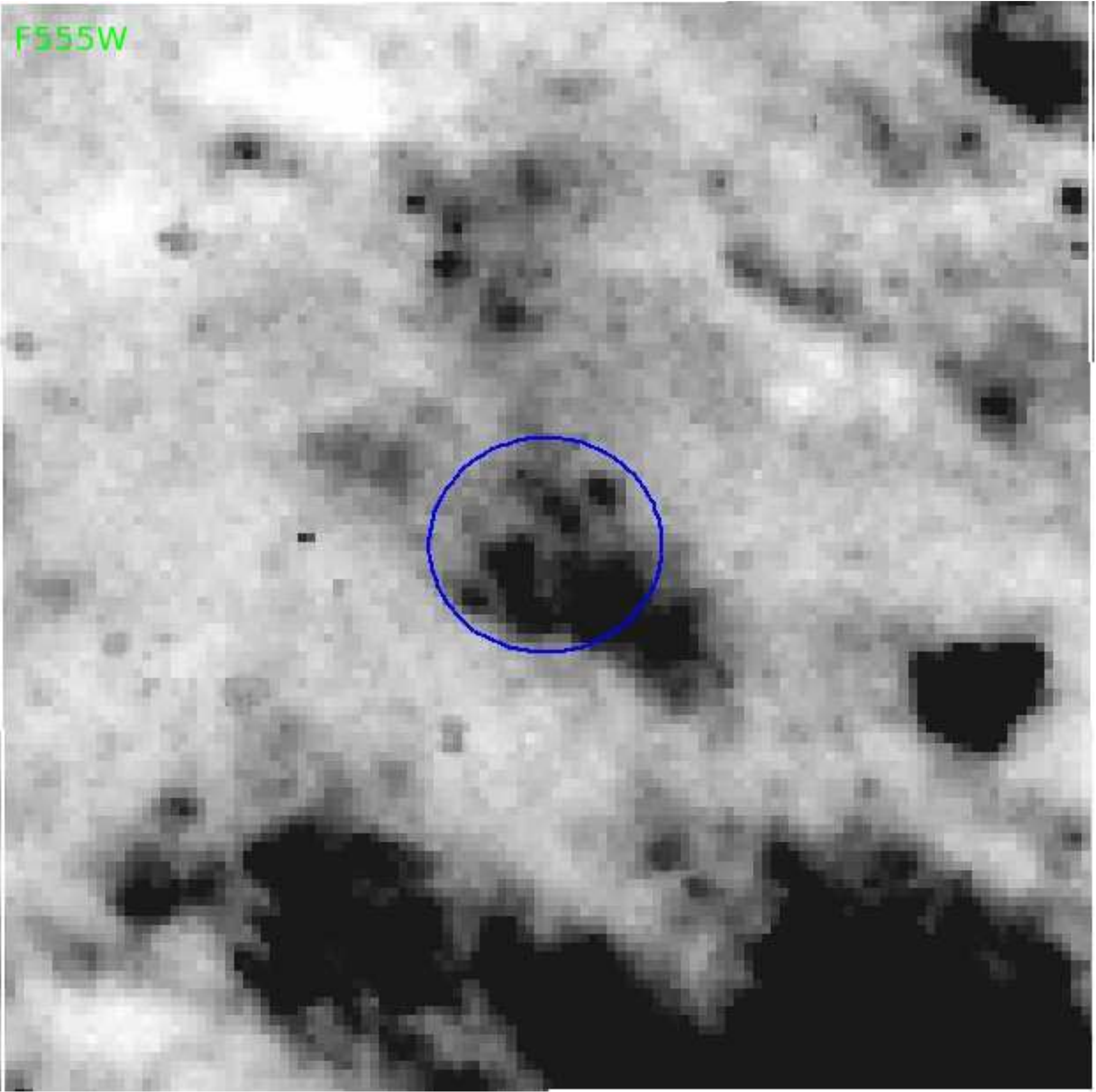}
\vspace*{0.3cm} 

\includegraphics[height=64mm, angle=0]{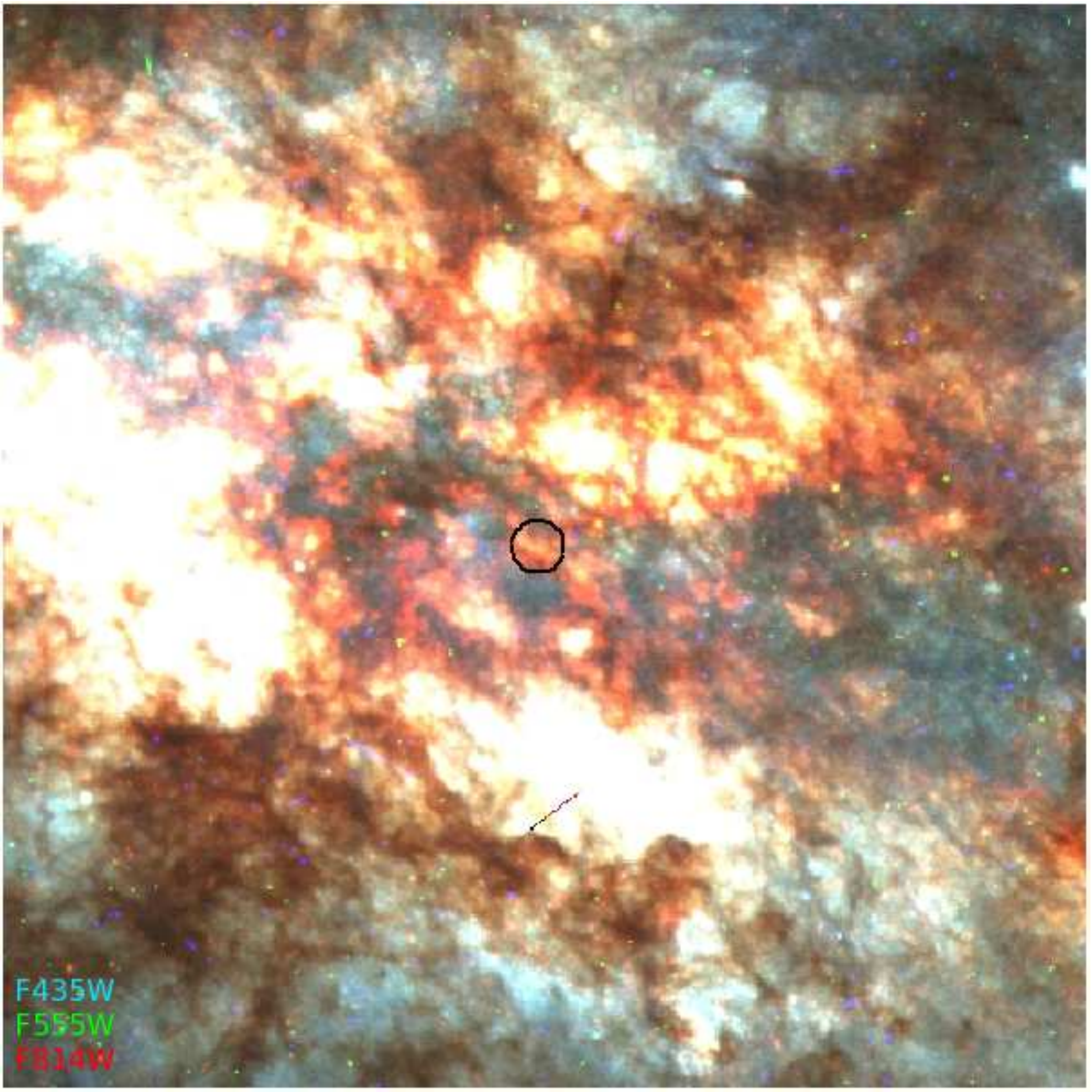}  \hspace*{0.4cm}
\includegraphics[height=64mm, angle=0]{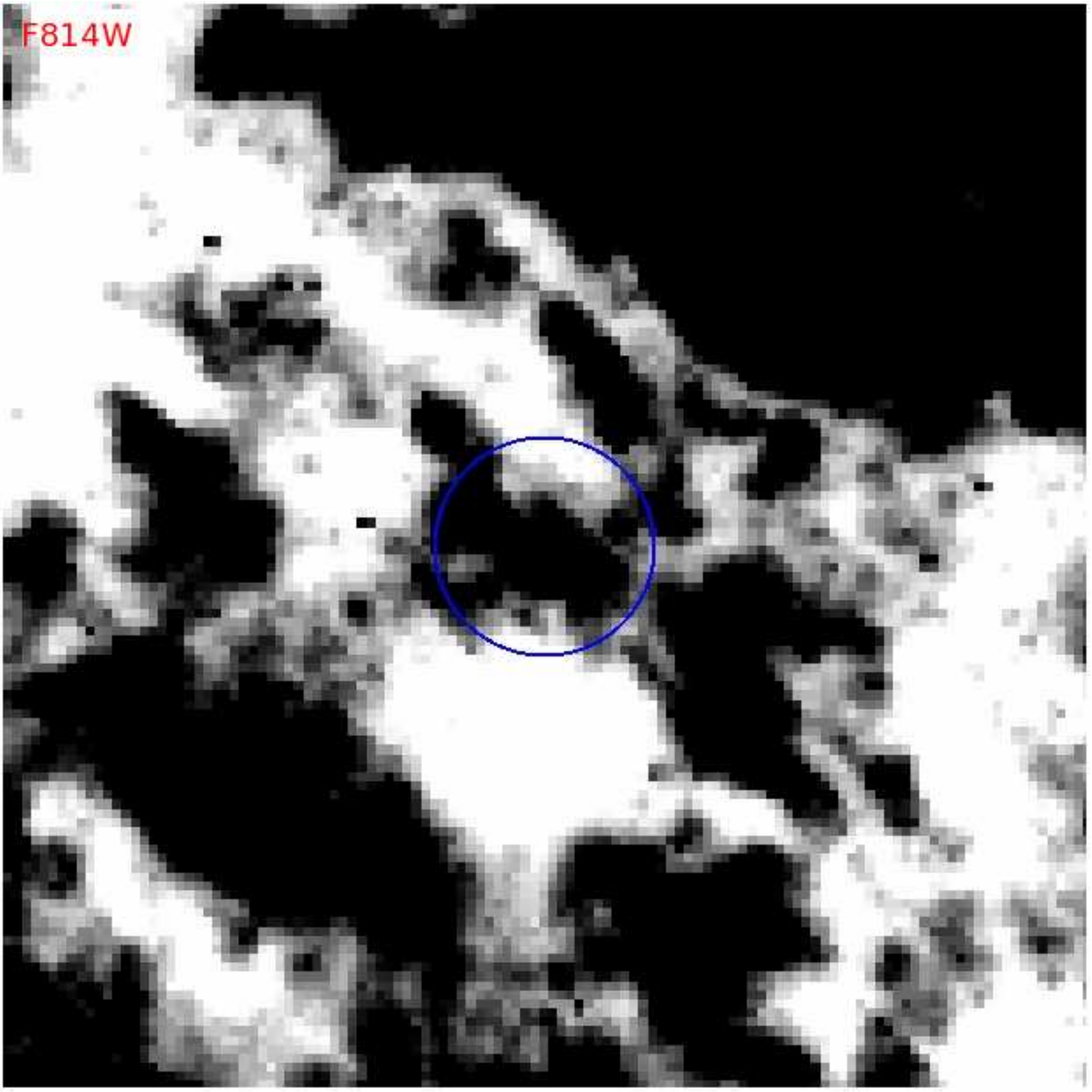}
\vspace*{0.3cm} 

\includegraphics[height=64mm, angle=0]{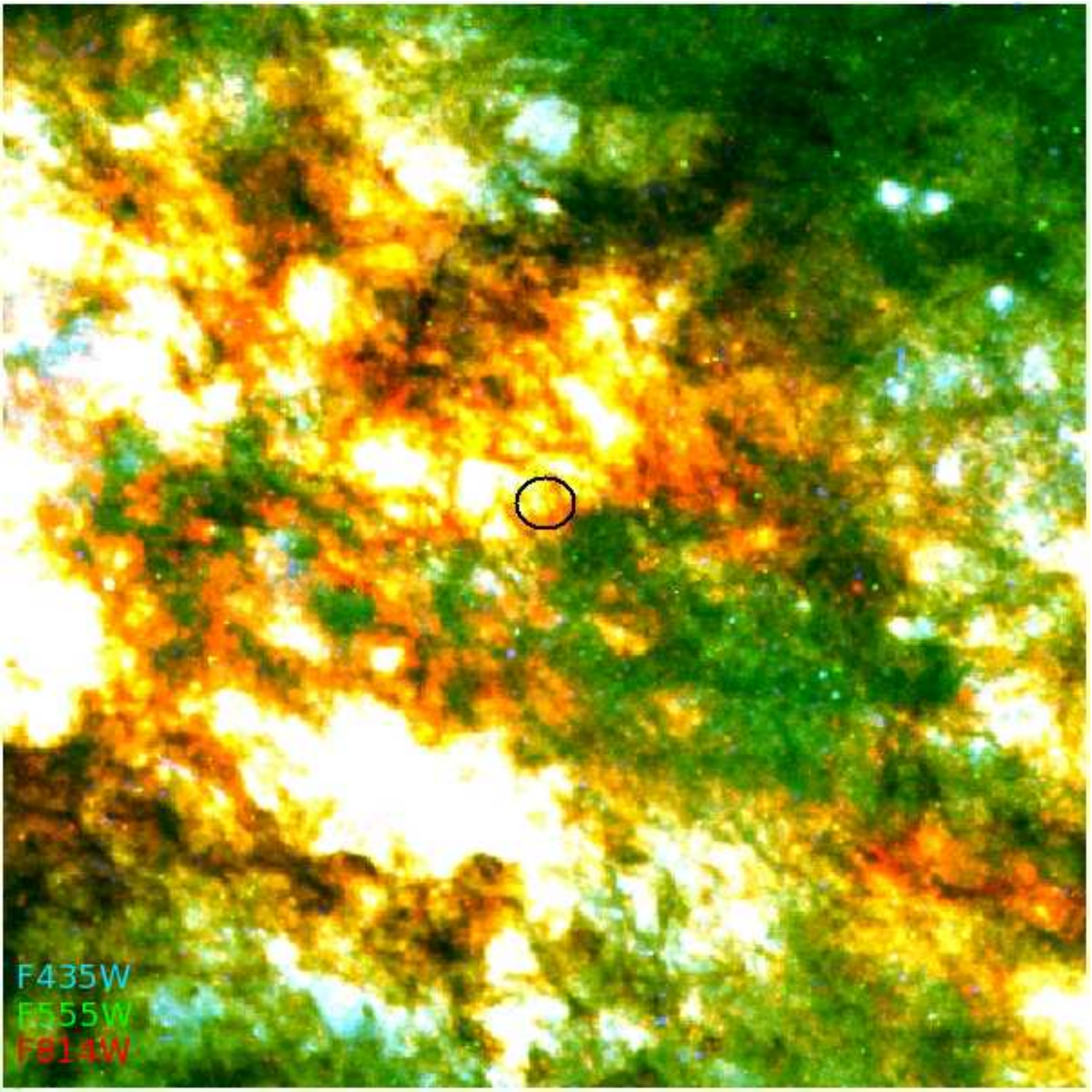} \hspace*{0.4cm}
\includegraphics[height=64mm, angle=0]{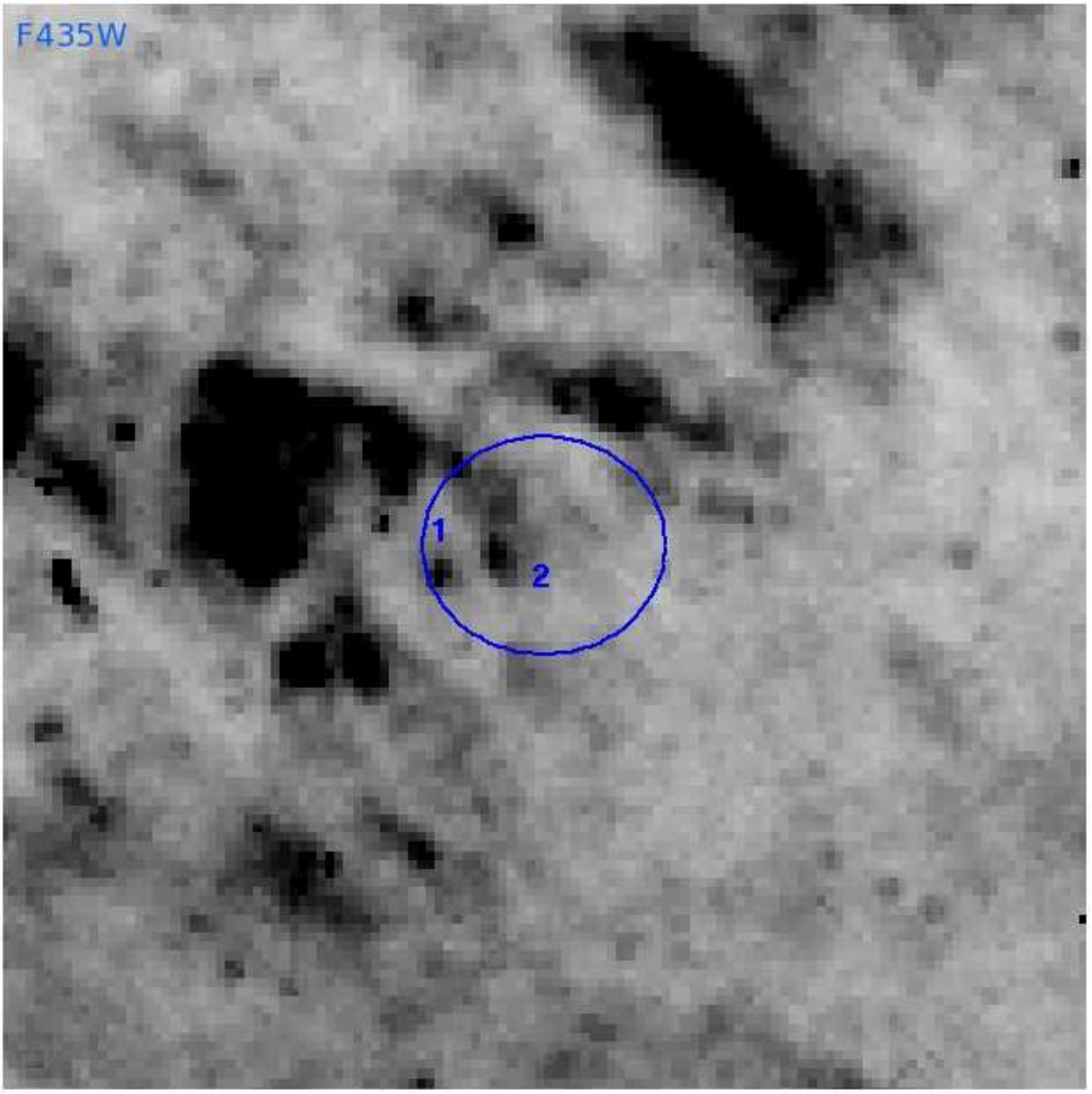} 

\end{center}
\addtocounter{figure}{-1}
\caption{\small{\emph{continued, pg. 11} -  Specific notes: displayed ULX regions are, from top to bottom, NGC 3034 ULX3 (F555W), NGC 3034 ULX4 (F814W) \& NGC 3034 ULX5 (F435W).}}
\label{fig:pictures}
\end{figure*}

\begin{figure*}
\leavevmode
\begin{center}

\includegraphics[height=64mm, angle=0]{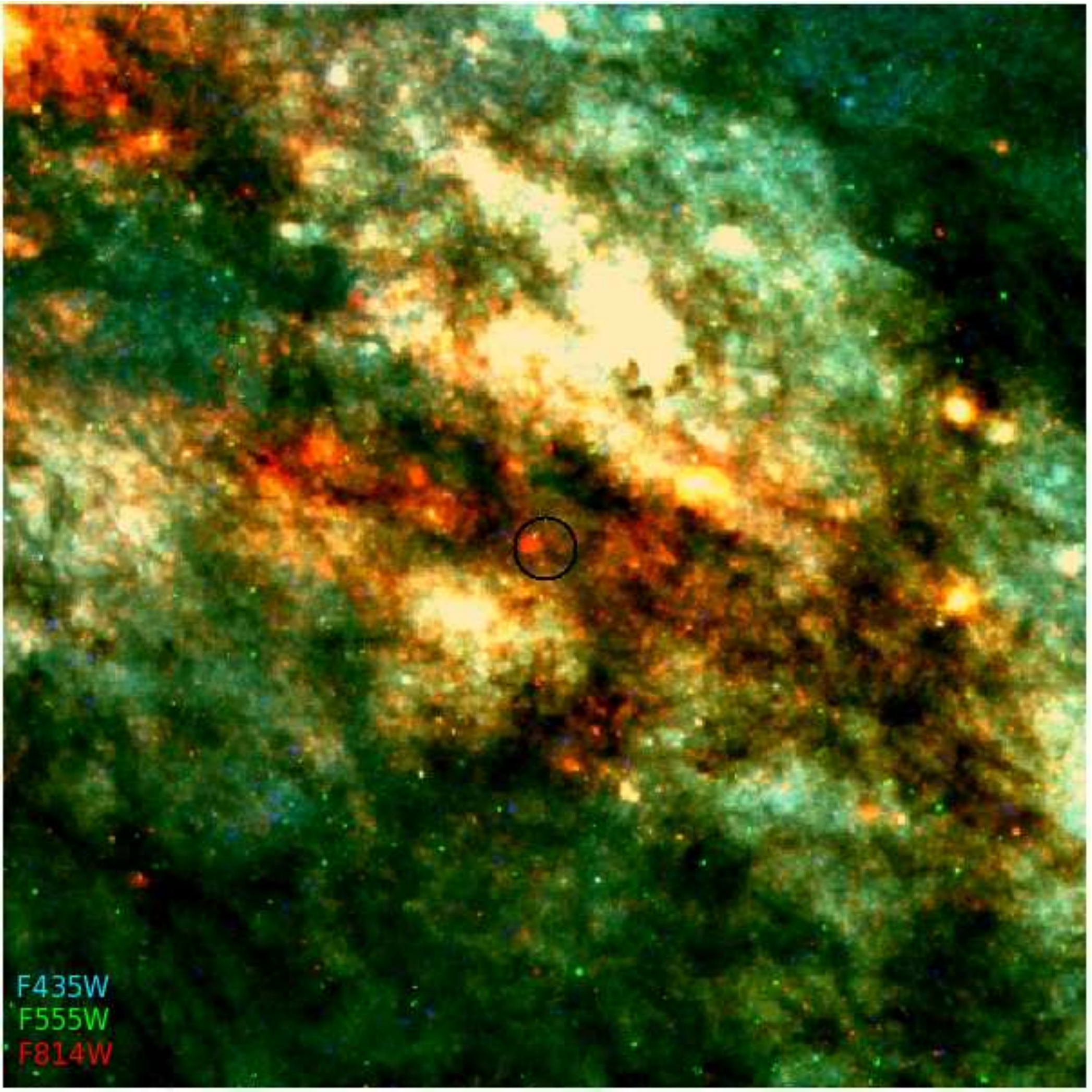} \hspace*{0.4cm}
\includegraphics[height=64mm, angle=0]{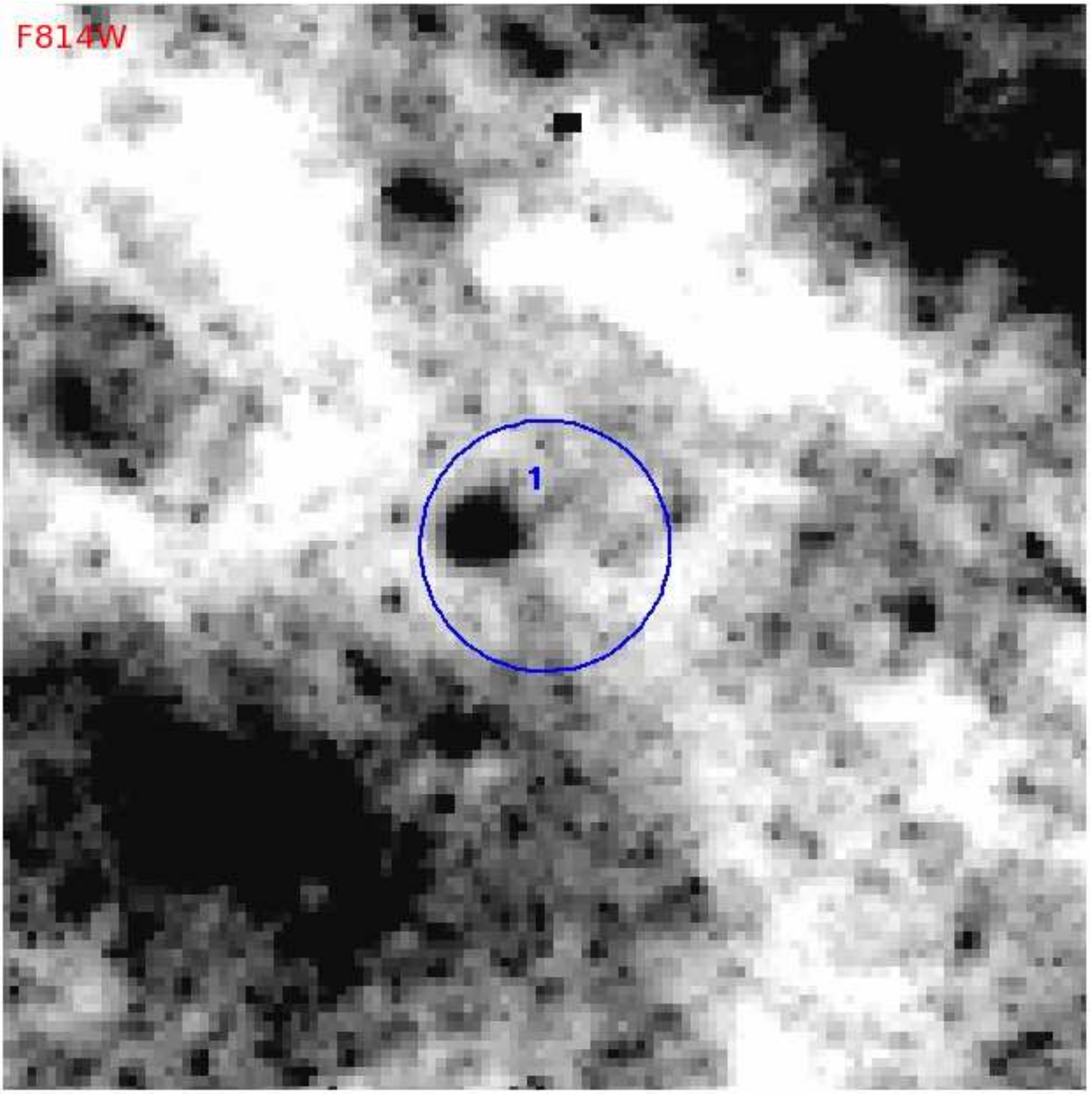} 
\vspace*{0.3cm} 

\includegraphics[height=64mm, angle=0]{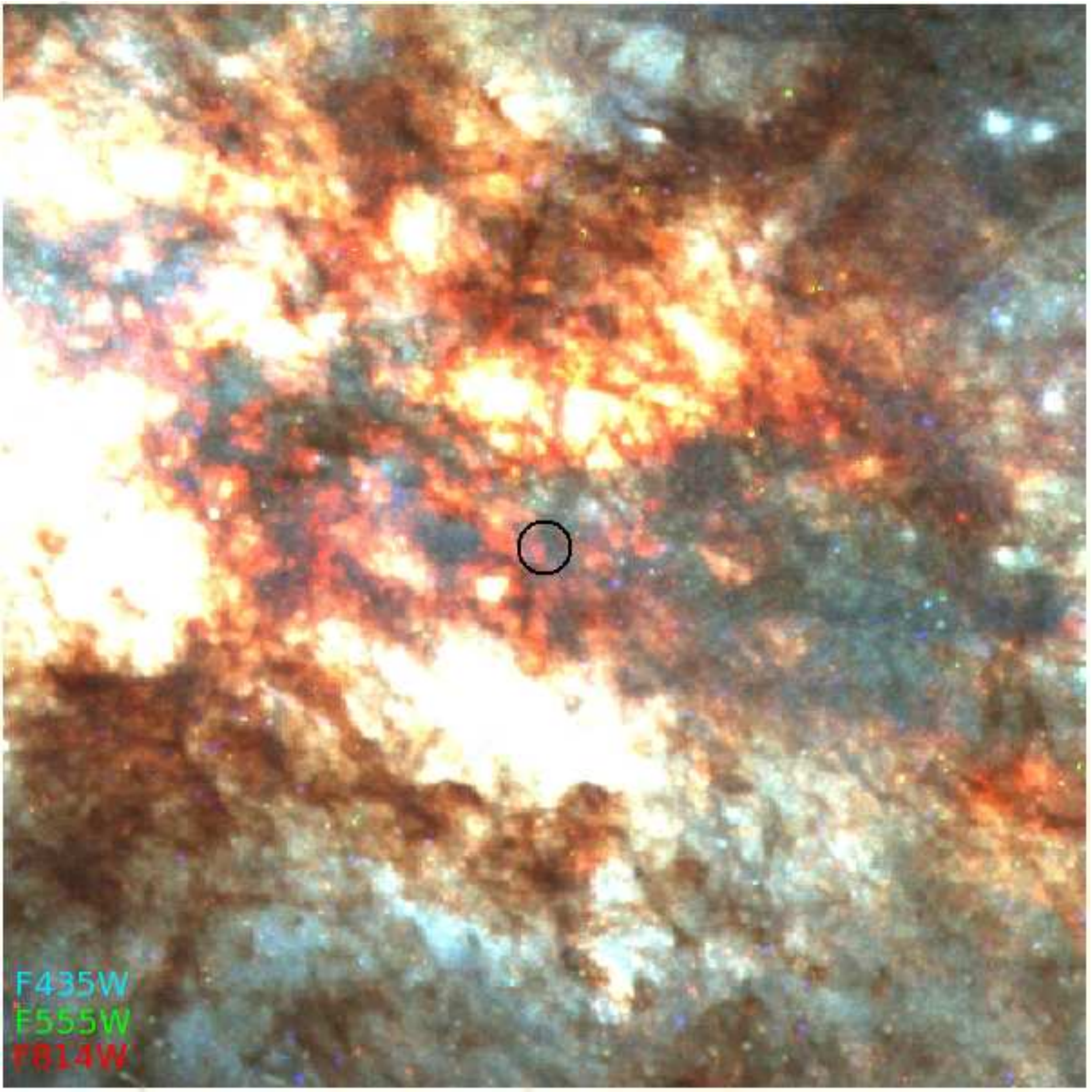} \hspace*{0.4cm}
\includegraphics[height=64mm, angle=0]{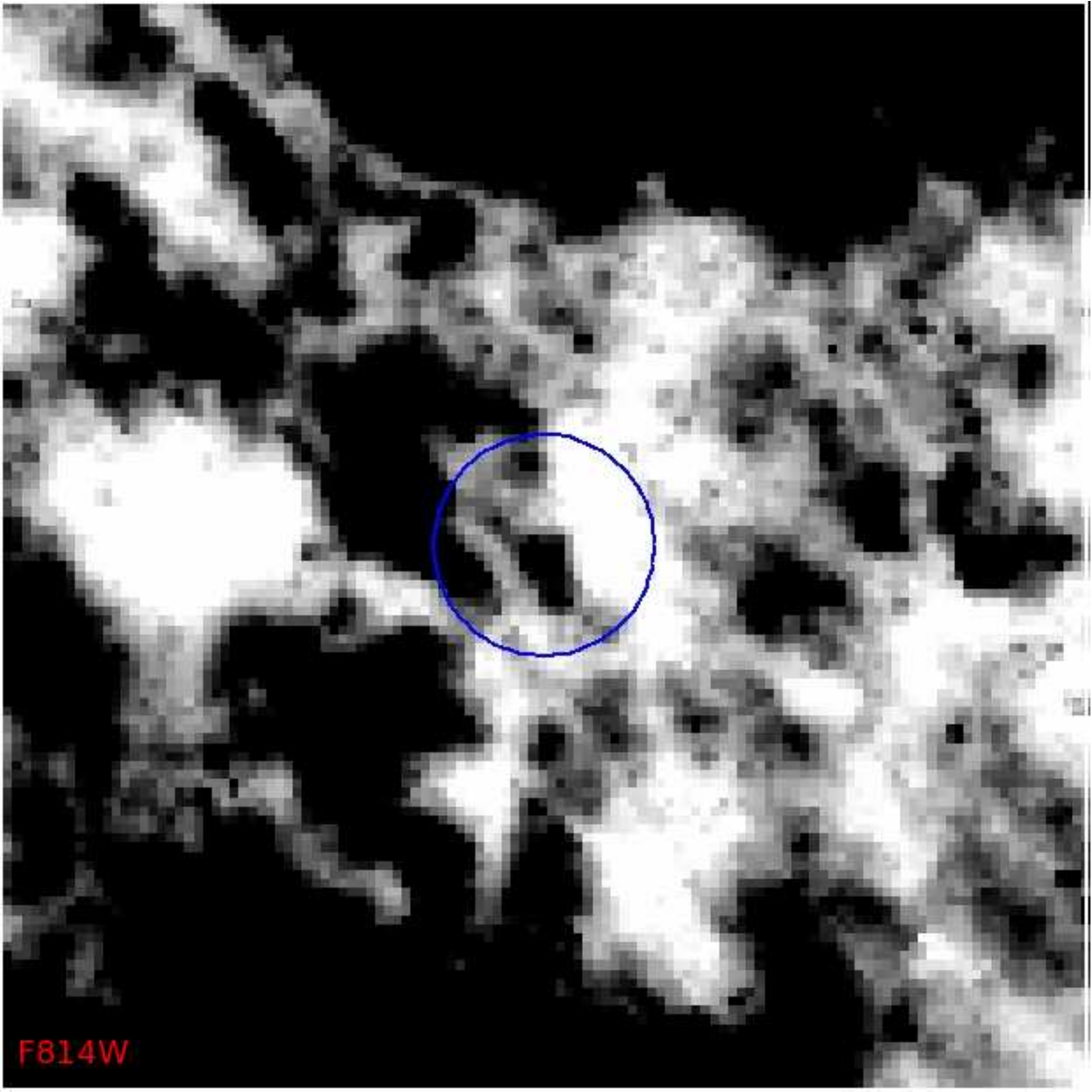} 

\end{center}
\addtocounter{figure}{-1}
\caption{\small{\emph{continued, pg. 12} -  Specific notes: displayed ULX regions are, from top to bottom, NGC 3034 ULX6 (F814W) \& CXOU J095550.6+694044 (F814W).}}
\label{fig:pictures}
\end{figure*}


\begin{figure*}
\leavevmode
\begin{center}

\includegraphics[height=135mm, angle=0]{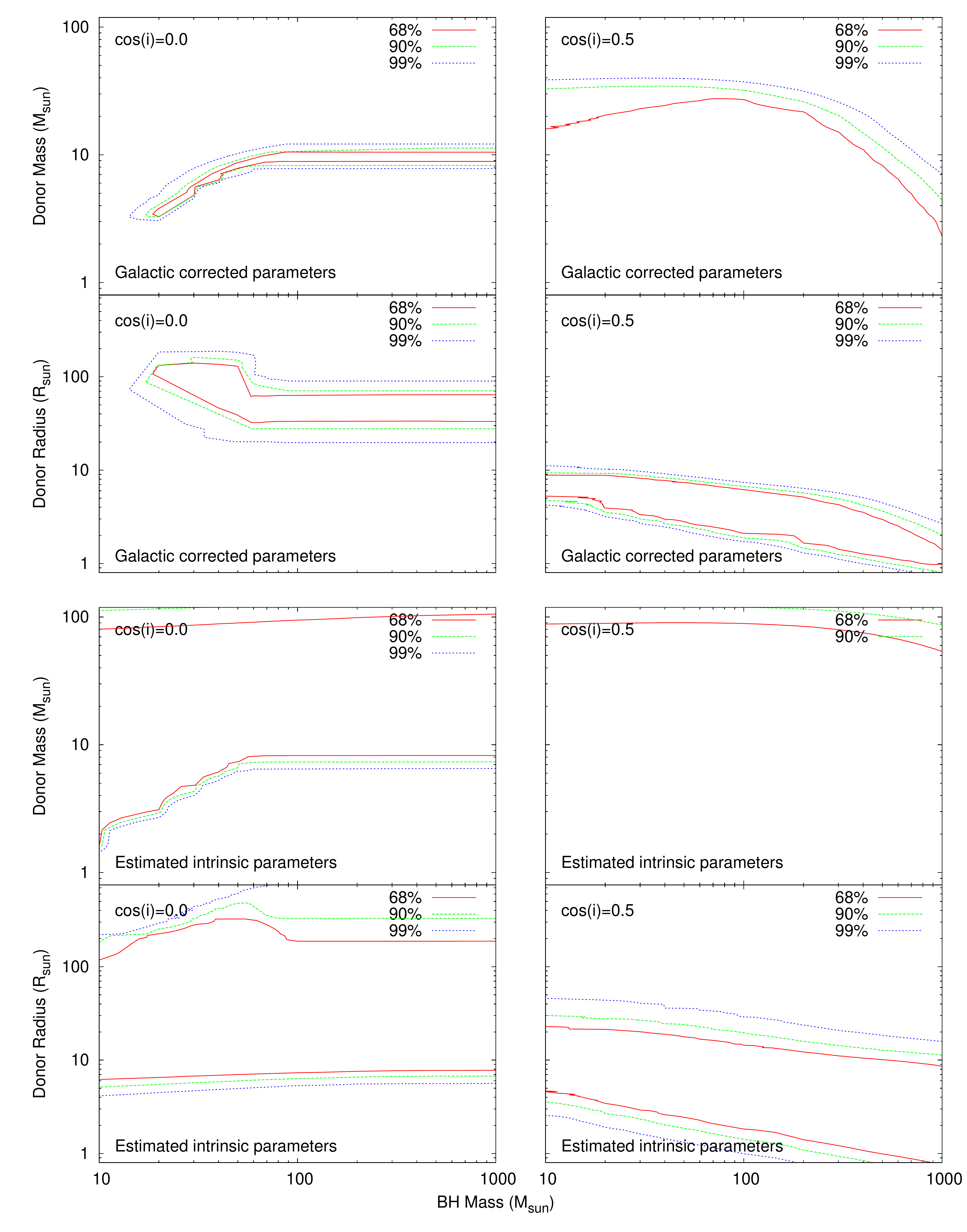} 

\end{center}
\addtocounter{figure}{+2}
\caption{
\small{
{\it continued, pg. 2} -  Specific notes: displayed above are fits for Holmberg IX X-1 (1), that was suggested by Liu et al. (2004) as the more likely candidate. We obtain black hole mass constraints for Galactic extinction/absorption correction data in superior conjunction with cos({\it i}) $=$ 0.0. We find that $M_{BH} > 18.6 M_\odot$ (assuming 1$\sigma$ constraints, see Table \ref{tab:copperwheat1}). We find that stellar mass and radius constraints are also quite confined, only allowing for an early F-type supergiant (Zombeck 1990), remarkably different from the B giant or supergiant suggested by our SED fitting (see Section \ref{subsection:SED}). The alternative is that the system is at low inclination, allowing us to see blue optical emission from the disc. Here, the stellar mass and radius constraints rule out the possibility of a supergiant, but still allow for a range of giants (A to mid G) and main sequence (B to mid G) companions. The black hole mass constraint is lost when we switch to intrinsic magnitudes, with fits ruling out the possibility of this system containing a main sequence star for all types except B, and the later-type supergiants (G or later), for the case of cos({\it i}) $=$ 0.0. At cos({\it i}) $=$ 0.5, we see that the F-type supergiants are also ruled out, but that the entire main sequence is available, depending on the assumed black hole mass.}}
\end{figure*}

\begin{figure*}
\leavevmode
\begin{center}

\includegraphics[height=135mm, angle=0]{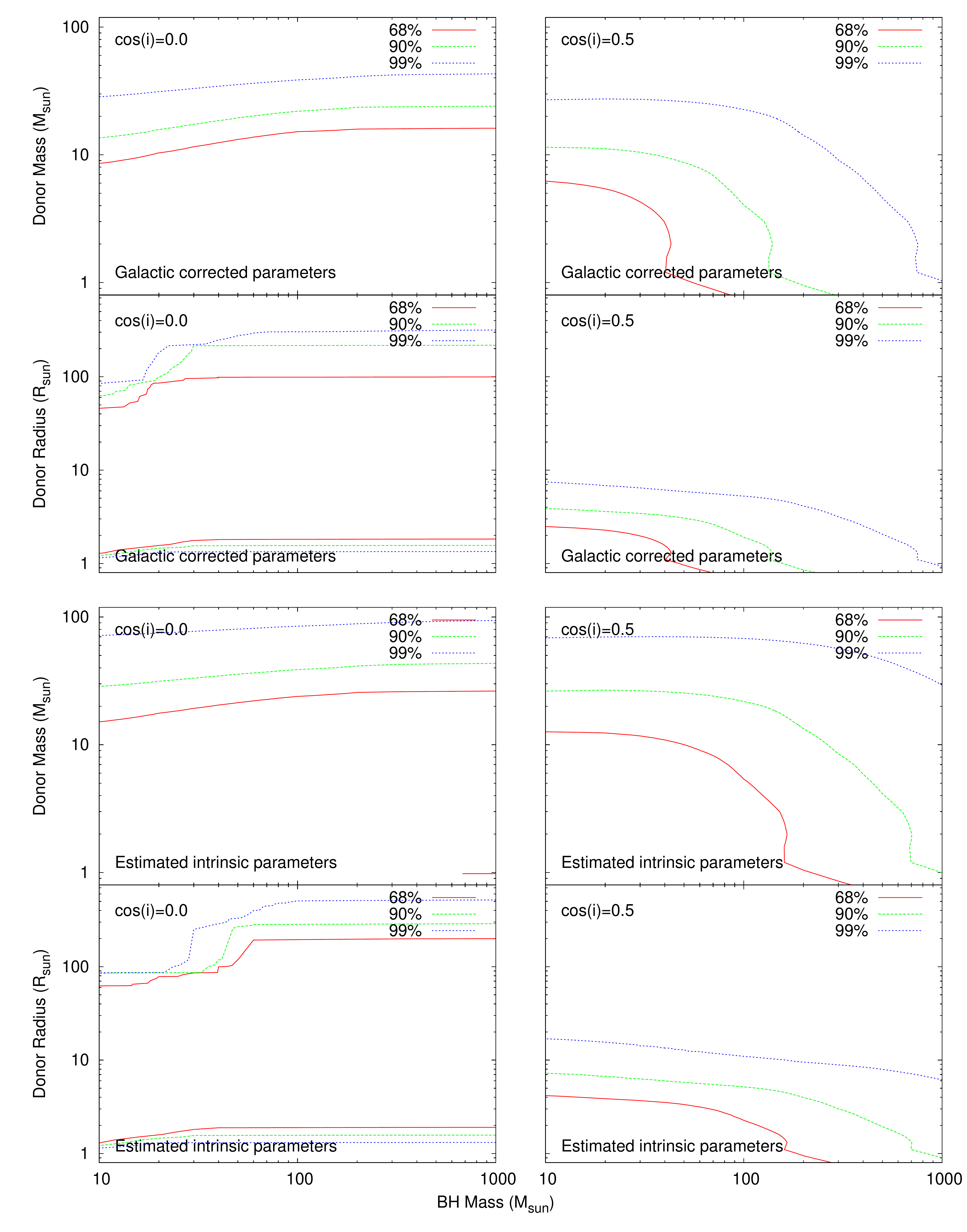} 

\end{center}
\addtocounter{figure}{-1}
\caption{
\small{
{\it continued, pg. 3} -  Specific notes: displayed above are fits for Holmberg IX X-1 (3). Here we see that the star's radius allows us to place constraints on the black hole mass in all cases, with upper limits ranging from 85 -- 350 $M_\odot$ (at 1 $\sigma$ level; see Table \ref{tab:copperwheat1} \& \ref{tab:copperwheat2}). If we assume that the top four panels are correct (Galactic extinction correction only), we rule out IMBHs, and the possibility of an HMXB. This is because the stellar constraints are $M_* < 6.2 M_\odot$ and $R_* < 2.5 R_\odot$ (for $M_{BH} = 10 M_\odot$). When we switch to the intrinsic scenario (no shielding of the star), the black hole mass constraints relax so that both MsBHs \& IMBHs are possible, but the star's constraints are still kept to $M_* < 12.6 M_\odot$ and $R_* < 4.2 R_\odot$. The companion could be a mid-B or later main sequence star, or a mid-class ($\sim$ F) giant. }}
\end{figure*}

\begin{figure*}
\leavevmode
\begin{center}

\includegraphics[height=135mm, angle=0]{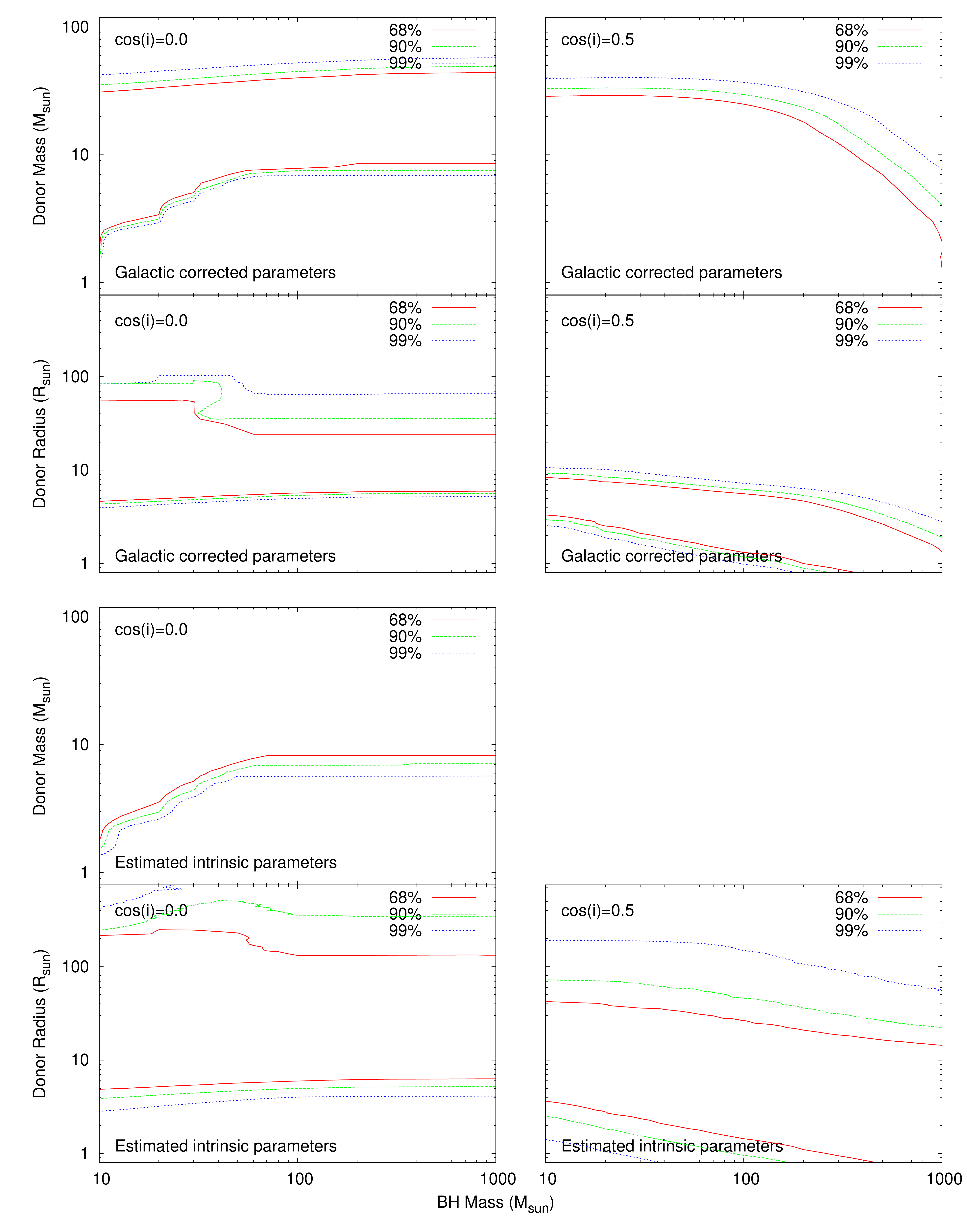}

\end{center}
\addtocounter{figure}{-1}
\caption{
\small{
{\it continued, pg. 4} -  Specific notes: fits for NGC 1313 X-2 (1). No constraints are found for the black hole mass in this ULX, but some constraints are achieved for the companion. Also, if we incorporate findings from studies of the surrounding stellar population, we can use an upper mass limit of $\sim 12 M_\odot$ (Gris{\'e} et al. 2008). This suggests a companion mass range of $1.7 \la M_{*} \la 12 M_\odot$ and a radius range of $2.6 \la R_{*} \la 325 R_\odot$, when using galactic extinction / absorption corrections and cos({\it i}) $=$ 0.0. This allows for early B to F0 type main sequence stars, along with A to mid-F giants (III) (Zombeck 1990). If we switch to the alternative inclination, we find that the range of stellar radii is reduced while lower limits on the mass range are lost. This implies that early-type main sequence stars are permissible, along with A to early G-like giants, while all supergiants are ruled out. 
Intrinsic data suggests only a lower-mass limit for the companion in the case of cos({\it i}) $=$ 0.0, while the radius range is large. Combining these with Gris{\'e} et al. (2008), we rule out all main sequence stars and all but F type supergiants, however, any giants are acceptable (using Zombeck 1990). If we switch to the inclination of cos({\it i}) $=$ 0.5, we lose stellar mass constraints but retain constraints on the star's radius, allowing any main sequence stars, all giants, but only OB-type super giants. Again, combining this with the mass limit 12 $M_\odot$,  this rules out O and early B stars of all classes. }}
\end{figure*}

\begin{figure*}
\leavevmode
\begin{center}

\includegraphics[height=135mm, angle=0]{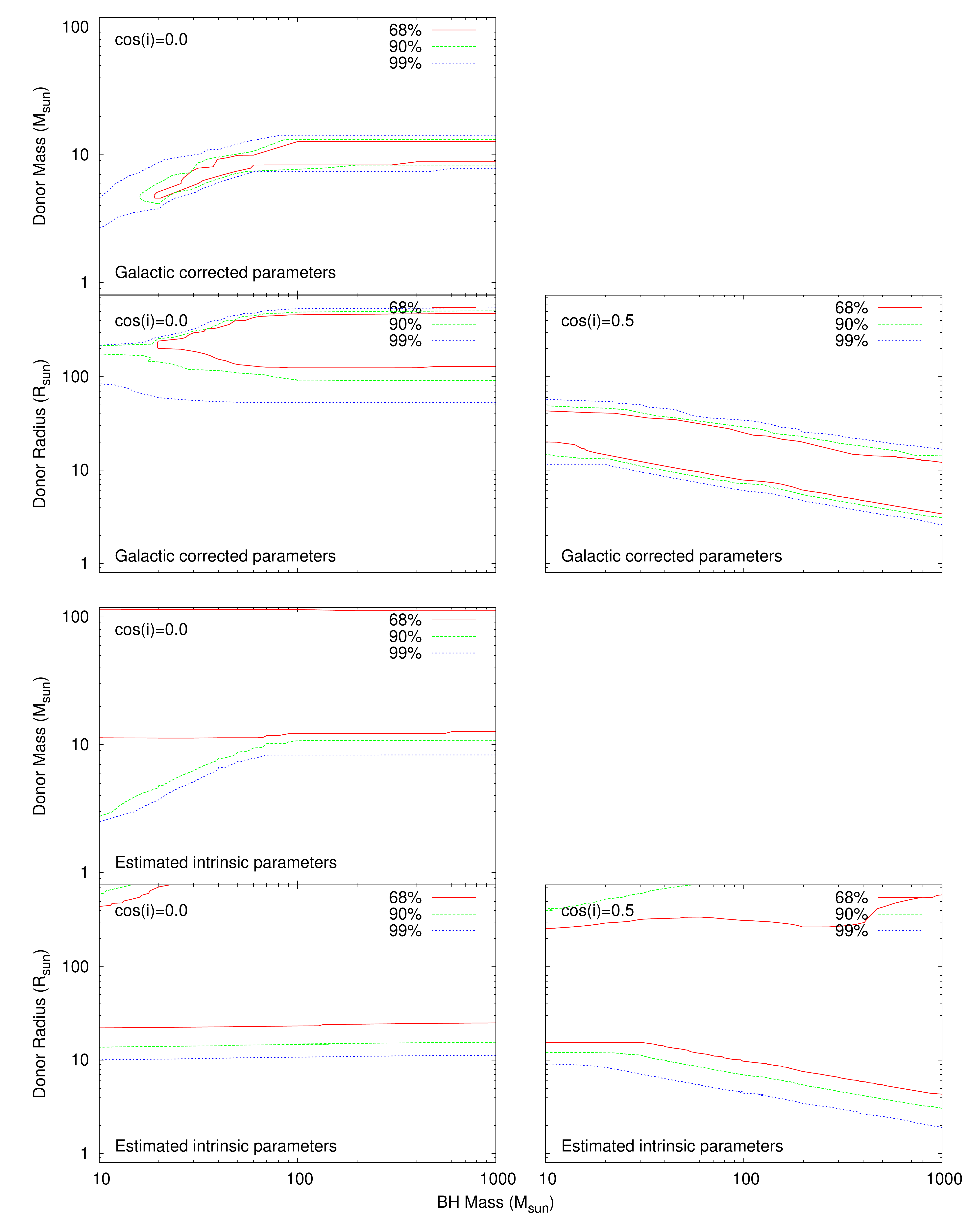} 

\end{center}
\addtocounter{figure}{-1}
\caption{
\small{
{\it continued, pg. 5} -  Specific notes: displayed above are fits for IC 342 X-1 (1). Fitting provides a lower limit on the black holes mass at the 1 $\sigma$ level in the upper two left hand plots (galactic corrected with inclination of cos({\it i}) $=$ 0.0. This limit indicates a lower-mass of $M_{BH} \ga 18.6 M_\odot$. The star mass and radius ranges, provided in the top four panels, are also narrow enough that they can also place good constraints on the companion star type. It indicates that we are most likely observing a late G or early K supergiant when cos({\it i}) $=$ 0.0 and a sliding scale for cos({\it i}) $=$ 0.5, covering a wide range of stellar types. For sMBHs the donor must be a high-mass companion that is either giant or supergiant in class. However, an $M_{555} = -5.7$ (from Table \ref{tab:mags}), rules out type Ia stars, although many 1b's are still allowable. If $M_{BH} = 100 M_\odot$ the companion can be an O or early B main sequence, or any giant (class III) star. Finally, if we are observing an IMBH ($M_{BH} = 1000 M_\odot$), the companion can be an OB main sequence, or a mid to late B to G III. Now we turn to the intrinsic magnitude for this source at an inclination of cos({\it i}) $=$ 0.5, where we are unable to obtain mass constraints once again. We also find that the radius range generally increases with increasing black hole mass, allowing for O \& early B main sequence stars, any III, or M0 and younger I, with the smallest range narrowing this to O-like main sequence, B or K III, K0 or younger supergiants.}}
\end{figure*}

\begin{figure*}
\leavevmode
\begin{center}

\includegraphics[height=135mm, angle=0]{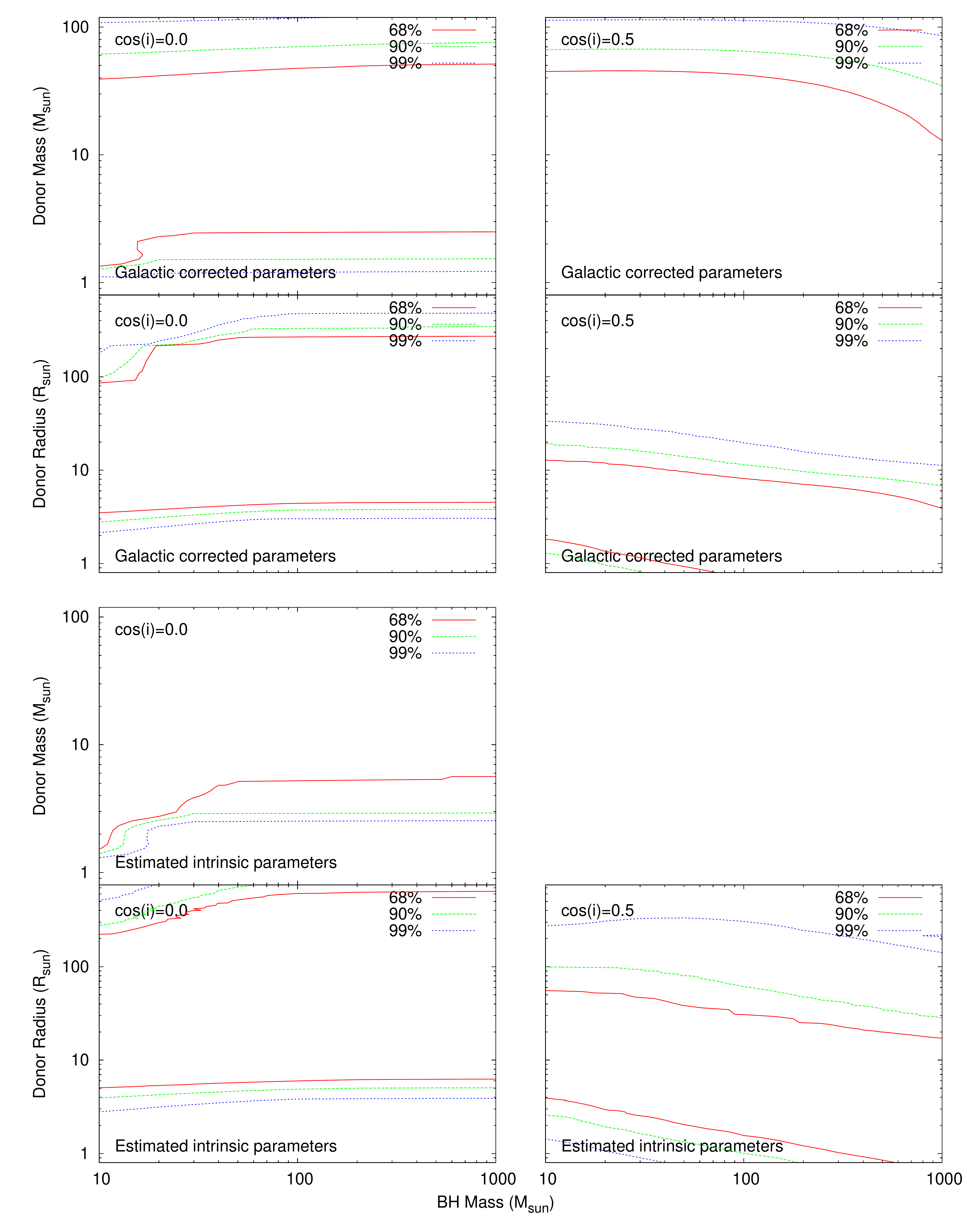} 

\end{center}
\addtocounter{figure}{-1}
\caption{
\small{
{\it continued, pg. 6} -  Specific notes: displayed above are fits for IC 342 X-1 (2). By combining these plots with the values from Tables \ref{tab:copperwheat1} \& \ref{tab:copperwheat2}, we obtain stellar mass constraints of 1.3 $\la M_{*} \la$ 51.3 $M_\odot$ for Galactic corrected magnitudes with radius constraints 3.5 $\la R_{*} \la$ 271.8 $R_\odot$ for an edge on system, which drops to $M_{*} \la 1.5 M_\odot$ and 5.1 $\la R_{*} \la$ 641.6 $R_\odot$ when intrinsic magnitudes are used. This allows for OB main sequence, any III, or B to mid K supergiant stars in the first instance, and O or early B main sequence, any III, or all but the reddest supergiant stars in the second instance. If we instead switch to cos({\it i}) $=$ 0.5, the star's mass and radius constraints provide us with a high upper limit on the mass of $M_{*} \la 44.9 M_\odot$ with a lower radius limit of $R_{*} \la 12.9 R_\odot$, with the radius decreasing with increasing black hole mass. This combination rules out all O classifications, all supergiants, and K or early B III. The intrinsic magnitudes provide us with even less constraints, obtaining only an upper limit on the companion star's radius ($R_{*} \la 55.5 R_\odot$) allowing for any main sequence star, any giant, or mid A or younger supergiants.}}
\end{figure*}

\begin{figure*}
\leavevmode
\begin{center}

\includegraphics[height=135mm, angle=0]{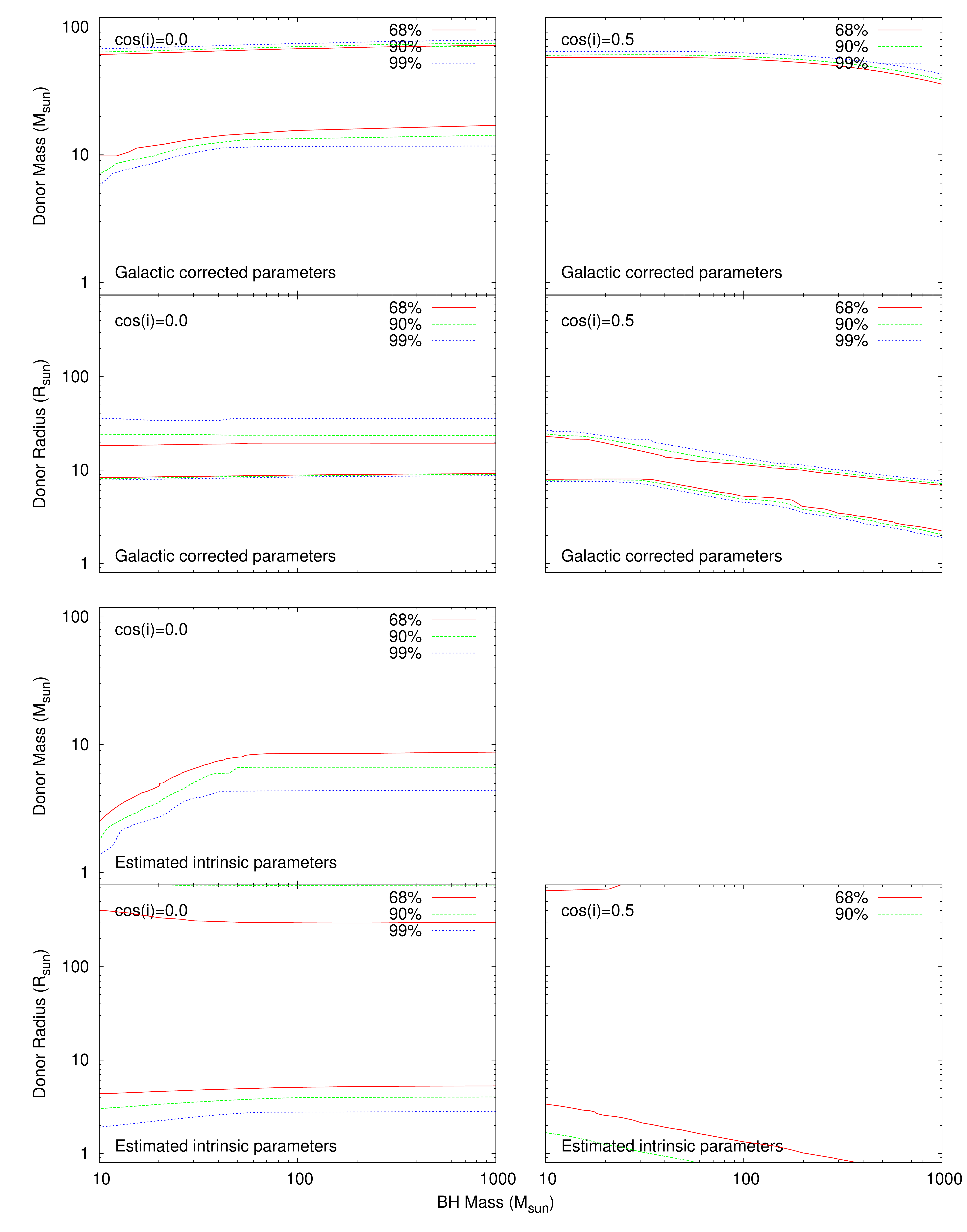} \hspace*{0.1cm}

\end{center}
\addtocounter{figure}{-1}
\caption{
\small{
{\it continued, pg. 7} -  Specific notes: displayed above are fits for NGC 5204 X-1 (1). We obtain mass and radius constraints for three of the four scenarios considered, losing mass constraints only in the intrinsic inclined (cos({\it i}) $=$ 0.5) case. For cos({\it i}) $=$ 0.0, in the galactic extinction corrected case the mass and radius constraints listed in Table \ref{tab:copperwheat1} constrain the allowable companion star types to O V, OB III or early B I, while intrinsic magnitudes extend the range to O or early B main sequence stars, any giants, or early K or younger supergiants. When we switch to an inclined system, we once again obtain a sliding radius scale for the companion star, which decreases with increasing black hole mass. For lower-mass black holes, we find strong constraints on the possible companion star, only allowing for early B supergiants. As the black hole mass increases, however, we find that none are comparable with the mean values (Zombeck 1990). This would indicate that either the the companion is not of a mean classification or that this is not the correct scenario or counterpart for this ULX. This could also be due to the fact that the assumption of galactic corrected magnitudes is incorrect in this case. The intrinsic case, however, allows little to no constraints, meaning that any companion type is possible.}}
\end{figure*}

\clearpage
\LongTables


\begin{deluxetable}{llcccccccc}
\tabletypesize{\small}
\tablecolumns{10}
\tablecaption{\textbf{The ULX sample listed by distance.}}
\tablehead{
\colhead{Source}   				& \colhead{Alternative names} 		   & \colhead{RA}    	& \colhead{Dec} & \colhead{$d$ $^a$} & \colhead{$N_{\rm H}$ $^b$} & \colhead{E(B-V)$^c$} & \colhead{$N_{\rm H}$ $^d$} & \multicolumn{2}{c}{$L_{\rm X}$ $^e$}              \\
                   				&                                                             			& (J2000)          &  (J2000)          & (Mpc)        		& 				  &                        	& 					& \multicolumn{2}{c}{(10$^{39}$ erg s$^{-1}$)} \\
}
\startdata
NGC 598  ULX1$^{({\it 1})}$		& Source 3$^{({\it 7})}$                                 	& 01 33 50.9  	& +30 39 37       & 0.9$^{({\it 8})}$	& 4.59	       			& 0.042	     	& 3.6 E$^{({\it 9})}$		& 2.5 	& $\alpha$XI$^{({\it 7})}$     \\  
                                       			& Source 7$^{({\it 10})}$ 				&                       	&                          &                      		&                       			&                    	&      					&  		& \\
                                         			& M33-X-8$^{({\it 11})}$                           	&                        	&                          &                      		&                       			&                   	&      					&       	&  \\                                        
                                         			& CXOUJ013351.0+303937$^{({\it 12})}$	&                        	&                          &                      		&                       			&                    	&       				&                      \\
						& \\
NGC 55 ULX1$^{({\it 2})}$     		& NGC 55 6$^{({\it 13})}$                                 	& 00 15 28.9    	& -39 13 19       & 1.74$^{({\it 2})}$    	& 1.71            			&   0.014        	& 23.9 E$^{({\it 9})}$		& 1.3 	& $\alpha$X$^{({\it 2})}$    \\ 
                                        			& XMMU J001528.9-391319$^{({\it 14})}$&                         	&                          &                      		&                       			&                    	&       \\ 
						& \\
\enddata
\end{deluxetable}
\clearpage

\begin{deluxetable}{lccccccc}
\tabletypesize{\small}
\tablecolumns{8}
\tablecaption{\textbf{Details of observations used to locate possible optical counterparts to our ULX sample.}}
\tablehead{
\colhead{Source}	&\colhead{$Chandra$} 	& \colhead{Expos.(ks)} 	& \colhead{HLA} 	& \colhead{Date of}	&\colhead{Instrument}	& \colhead{Filter} 	& \colhead{Expos.(s)}	\\
				& Obs ID					& Obs  ID	       			& Instr.         		& \emph{HST} obs 	&					&	     			&	 \\
}
\startdata
NGC 598 ULX1	& 6376	& 93			& 06038\_02	& 1995-10-02 		& WFPC2		& F170W	&  1800 \\
				& 		& ACIS-I		& 06038\_02	& 1995-10-02 	& WFPC2		& F336W	 & 1800 \\
				&		&			& 06038\_02	& 1995-10-02 	& WFPC2		& F439W	& 600 \\
				& 		&			& 06038\_02	& 1995-10-02 	& WFPC2		& F555W	& 160 \\
				& 		&			& 05464\_05	& 1994-09-26	& WFPC2		& F814W	  & 1280 \\
				& \\
NGC 55 ULX1		& 2255 	& 59			& 09765\_03	& 2003-09-23	&ACS$-$WFC	& F606W	& 400 \\
				& 	     	& ACIS-I		& 09765\_03	& 2003-09-23	& ACS$-$WFC	& F814W	& 676 \\
\enddata
\label{tab:observations}
\end{deluxetable}
\clearpage

\begin{deluxetable}{lccccccc}
\tabletypesize{\small}
\tablecolumns{8}
\tablecaption{\textbf{Details of observations used to locate possible optical counterparts to our ULX sample.}}
\tablehead{
\colhead{Source}	&\colhead{No. of}	& \colhead{Matches}	&\multicolumn{2}{c}{ULX Position$^c$}	& \colhead{Filter$^d$}	& \multicolumn{2}{c}{Positional error}		\\
				& matches$^a$		& used?$^b$		& RA				& DEC			&			&  RA$^{e(i)}$		& Dec$^{e(ii)}$			\\ }
\startdata
NGC 598 ULX1	& 0				& N				& 01 33 50.90	 	& +30 39 36.76 	& F170W		& 0.79   			& 0.81   	 	\\
				& 				&				& 				&				& F336W		& 0.81   			& 0.88   		 \\
				& 				&				& 				&				& F439W		& 0.91   			& 0.91   	  	\\
				& 				&				& 				&				& F555W		& 0.78    			& 0.78   		   	\\
				& 				&				& 				&				& F814W		& 0.68    			& 0.71   		 \\
				\\
NGC 55 ULX1		& 1				& N$^{({\it 1})}$		& 00:15:28.90 		& --39:13:18.77		& F606W		& 0.71   			& 0.71    	 \\
				& 				&				& 				&				& F814W		& 0.71   			& 0.66	   	 \\
\enddata
\end{deluxetable}
\clearpage

\begin{deluxetable}{lcccccc}
\tabletypesize{\small}
\tablecolumns{7}
\tablecaption{\textbf{Potential optical counterparts for each ULX, with their corrected magnitudes.}}
\tablehead{
\colhead{Source}	&\colhead{Filter}	& \colhead{C/P}		& \colhead{Previous }		& \colhead{Observed}			& \colhead{Galaxy corrected} 		& \colhead{Estimated intrinsic}		 \\
				&		&  ID$^a$ 		& ID$^b$			& Vega mag$^c$		& Vega mag$^d$		& Vega mag$^e$		\\
}
\startdata
NGC 598 ULX1         & F170W	& 0			& Y$^{({\it 1})}$		& -- 					& --					& --     				\\
				& F336W	&   			& 				& -- 					& -- 					& --  					 \\
				& F439W	&   			& 				&-- 					& --					& --  					\\
				& F555W	&			& 				&-- 					& --					& --  	 				\\
				& F814W	& 			& 				&-- 					& --					& --  					 \\
				\\
NGC 55 ULX1		& F606W  & 1			& N				& 23.2 $\pm$ 0.3		& 23.2 $\pm$ 0.3		& 22 $\pm$ 2		 \\
				& F814W	& 			&				& 23.2 $\pm$ 0.5		& 23.1 $\pm$ 0.5		& 22 $\pm$ 2		 \\
\enddata
\tablecomments{
$^a$Candidate counterpart identification number, 0 is listed if no counterpart is available 
$^b$Previous identification as the candidate optical counterpart to this source, references as follows:
$^{({\it 1})}$Dubus et al. (2004),
$^{({\it 2})}$Liu et al. (2002), 
$^{({\it 2})}$Ramsey et al. (2006), 
$^{({\it 3})}$Gris{\'e} et al. (2005),
$^{({\it 4})}$Yang et al. (2011),
$^{({\it 5})}$Liu et al. (2007),
$^{({\it 6})}$Feng \& Kaaret (2008),
$^{({\it 7})}$Roberts et al. (2008),
$^{({\it 8})}$Weisskopf et al. (2004),
$^{({\it 9})}$Ghosh et al. (2006),
$^{({\it 10})}$Kaaret et al. (2004),
$^{({\it 11})}$Liu et al. (2004), 
$^{({\it 12})}$Goad et al. (2002), 
$^{({\it 13})}$Lang et al. (2007).
$^c$Aperture corrected observed optical magnitude of each potential counterpart. CG denotes cases where error region resides in a chip gap. $\sim m_V$ limits are provided in those cases where no counterpart was observed. M83 XMM2 is located in the chip gap in the  $\sim$ V band, so a limit is obtained for the only band where this is not the case - F336W.
$^d$Galactic extinction corrected Vega magnitudes. 
$^e$Estimated intrinsic Vega-mag for each potential counterpart, calculated using published values of $N_H$ (provided in Table \ref{tab:full list}), assuming the Milky Way value of $R_v$.
}
\label{tab:c/ps}
\end{deluxetable}
\clearpage
\label{tab:mags}

\begin{deluxetable}{lccccccccccc}
\tabletypesize{\small}
\tablecolumns{7}
\tablecaption{\textbf{Typing of potential ULX counterparts.}}
\tablehead{
\colhead{Source}	& \colhead{C/P}  	& \colhead{Type$^b$}	& \colhead{Type} 	& \colhead{Distance} 	& \colhead{$m_{555}$$^e$} 	& \colhead{offset$^f$}	& \colhead{$\Delta m_{555}$$^g$}	& \colhead{Types from}		& \colhead{$M_{555}$$^i$}	& \colhead{Types from} 			& \colhead{Previous} 	\\
				& ID$^a$			&			& $M_V$$^c$			& modulus$^d$	&				&			&				&  colour$^h$			&				& $M_{555}$$^j$			& ID$^k$	   \\
}
\startdata
NGC 598 ULX1	& nucleus			& --			& --					& 24.771 			& --				& --			& --			& --					& --				& --					&nucleus$^{({\it 3})}$		\\
				\\
NGC 55 ULX1		& 1 				& A1 V		& +0.7$^*$$^{({\it 1})}$	& 26.203			& 26.898			& 23.179		& -3.719		& O, B, A, 				& -3.0			& late B V, 			& --	\\
				&				&			&					&				&				&			&			& early F				&				& IV, III, A  II			&	\\
				& 2				& M0V		& +9.0$^{({\it 1})}$		& 26.203			& 35.208			& 24.704		& -10.499		& late K, M			& -1.499			& B V, IV,   			& --	\\
				&				&			&					&				&				&			&			& early F				&				& \& late III, 				&	\\
				&				&			&					&				&				&			&			& early F				&				& F-K II				&	\\
				\\
\enddata
\tablecomments{
$^a$Candidate counterpart ID taken from Table \ref{tab:c/ps}. We remove those ULXs for which there are no candidate counterparts detected from the list. The only exception to this is NGC 5128 ULX1, as a candidate counterpart has been published elsewhere. $^b$Stellar type derived from $\chi^2$ fitting of {\it HST} band stellar templates, created using \textsc{synphot}. $^c$Absolute magnitude of identified stellar type in $V$ band (taken from similar type if particular $M_V$ is unavailable, sources are indicated by $^*$). $^d$Calculated distance modulus using values in Table \ref{tab:full list}. $^e$Derived Vega magnitude in F555W band, 
assuming stellar type classification is correct. $^f$offset required to fit observed data to stellar types, this can also be considered to be the apparent magnitude in the F555W band ($\sim$ $m_{555}$). This is found by $\chi^2$ fitting. $^g$Value $\Delta m_{555} = {\rm offset} - m_{555}$. $^h$All possible types allowed  from SED fitting, within errors. $^i$Galactic extinction corrected absolute magnitude in {\it HST} band F555W, calculated using offset value. $^j$All possible types allowed from absolute magnitude fitting, within errors. $^k$Stellar types collated from work of other authors, where Y refers to identification without typing. 
Figures shown in brackets relate to following references:
$^{({\it 1})}$Zombeck (1990), 
$^{({\it 2})}$Wegner (2006), 
$^{({\it 3})}$Dubus et al. (2004), 
$^{({\it 4})}$Liu et al. (2002), 
$^{({\it 5})}$Ramsey et al. (2006), 
$^{({\it 6})}$Gris{\'e} et al. (2005), 
$^{({\it 7})}$Yang et al. (2011), 
$^{({\it 8})}$Liu et al. (2007), 
$^{({\it 9})}$Feng \& Kaaret (2008), 
$^{({\it 10})}$Roberts et al. (2008),
$^{({\it 11})}$Weisskopf et al. (2004), 
$^{({\it 12})}$Ghosh et al. (2006), 
$^{({\it 13})}$Kaaret et al. (2004), 
$^{({\it 14})}$Liu et al. (2004), 
$^{({\it 15})}$Goad et al. (2002), 
$^{({\it 16})}$Lang et al. (2007).
}
\label{tab:mags}
\end{deluxetable}
\clearpage

\begin{deluxetable}{lccccc}
\tabletypesize{\small}
\tablecolumns{6}
\tablecaption{\textbf{Models, flux ratios and X-ray luminosities for each ULX that has a counterpart detected in multiple \emph{HST} bands.}}
\tablehead{
\colhead{Source}	& \colhead{Model$^{a}$ }						& \multicolumn{2}{c}{Flux ratio$^{b}$}	&  \multicolumn{2}{c}{$L_{X} ^c$ }	\\
				&										&				&				& \multicolumn{2}{c}{($10^{39}$ erg s$^{-1}$)} \\
				&										& Galactic			& Intrinsic			& Galactic			& Intrinsic		\\	
}
\startdata
NGC 55 ULX1		& DKBBFTH$^{(\it{1})}$						& 2.60			& 0.873			& 1.1				& 2.0		\\
NGC 4190 X-1		& DISKPBB$^{(\it{2})}$						& 6.80			& 3.17			& 4.0				& 4.7		\\
NGC 253 	ULX1	& PILEUP$\times$PEGPWRLW$^{(\it{*})}$		& 104			& 1.29			& 0.64			& 1.6		\\	
NGC 253 ULX2	& DISKBB$^{(\it{3})}$						& 13.1			& 6.62			& 1.5				& 1.9		\\
NGC 253 ULX3	& POWERLAW$^{(\it{*})}$						& 33.1			& 7.10			& 0.19			& 0.23	\\
NGC 253 	XMM6	& PILEUP$\times$(DISKBB+COMPTT)$^{(\it{*})}$	& 15.3			& 2.85			& 0.93			& 1.5		\\
M81 X-6			& DKBBFTH$^{(\it{1})}$						& 9.33			& 3.93			& 2.3				& 2.7		\\
Hol IX X-1			& DKBBFTH$^{(\it{1})}$						& 8.08			& 4.25			& 7.7				& 8.7		\\
\enddata
\tablecomments{
$^{a}$Published model taken from references denoted by superscript number, although not stated, all models are absorbed. $^{b}$Galactic corrected and intrinsic flux ratio (1.0 -- 10.0 / 0.3 -- 1.0 keV), derived within \textsc{xspec} using listed models. 
$^{c}$Galactic absorption corrected and intrinsic X-ray luminosity in the 0.3 --10.0 keV bandpass. These luminosities are taken from references given below whenever the bandpass matches. If this was not available, we scale using model fits within \textsc{xspec}.
References: 
$^{(\it{1})}$Gladstone et al. (2009), 
$^{(\it{2})}$Roberts et al. (\emph{in prep}), 
$^{(\it{3})}$Kajava et al. (2009), 
$^{(\it{4})}$Mak et al. (2011), 
$^{(\it{5})}$Stobbart et al. (2006), 
$^{(\it{6})}$Kaaret et al. (2006).
$^{(\it{*})}$Values not available from published results, so simple fitting performed on data using listed models.}
\label{tab:flux rat}
\end{deluxetable}
\clearpage

\begin{deluxetable}{lcccccccc}
\tabletypesize{\small}
\tablecolumns{9}
\tablecaption{\textbf{$1\sigma$ binary parameter constraints obtained for each potential ULX optical counterpart derived from Galactic absorption corrected magnitudes.}}
\tablehead{
				&             &               &\multicolumn{2}{c}{$M_{BH}=10 M_\odot$}  	&\multicolumn{2}{c}{$M_{BH}=100 M_\odot$} 	&\multicolumn{2}{c}{$M_{BH}=1000 M_\odot$}\\
Source          		& C/P ID$^a$    & cos({\it i})   	&$M_{*}$        	&$R_{*}$                &$M_{*}$        	&$R_{*}$                &$M_{*}$        	&$R_{*}$\\
                			&               &    		&($M_\odot$)    	&($R_\odot$)      	&($M_\odot$)    	&($R_\odot$)            &($M_\odot$)    	&($R_\odot$)\\
}
\startdata
NGC~55~ULX1     	&$1$            &$0.0$          &$1.4$ -- $15.4$ 	&$3.4$ -- $64.1$        &$3.8$ -- $19.7$  	&$4.1$ -- $59.1$        &$3.8$ -- $20.4$  	&$4.1$ -- $58.9$\\         
                			&               &$0.5$          &$<13.9$        	&$2.2$ -- $5.1$         &$<8.9$         	&$0.9$ -- $3.1$        	&\multicolumn{2}{c}{$M_{BH}<700M_\odot$}\\
				&$2$            &$0.0$          &$1.4$ -- $1.5$ 	&$50.9$ -- $52.3$       &$1.4$ -- $4.6$		&$29.3$ -- $55.4$       &$1.4$ -- $4.6$		&$29.3$ -- $55.5$\\
                			&               &$0.5$          &$<15.1$        	&$3.6$ -- $5.4$         &$<10.0$         	&$1.4$ -- $3.3$        	&\multicolumn{2}{c}{$M_{BH}<840M_\odot$}\\
				& \\
NGC~4190~X-1    	&$1$            &$0.0$          &$1.3$ -- $48.8$       	&$2.0$ -- $85.9$        &$2.9$ -- $58.2$	&$3.0$ -- $439.9$      	&$2.9$ -- $63.3$	&$3.0$ -- $447.6$\\ 
                			&               &$0.5$          &$<45.9$	  	&$<13.0$	       	&$<43.6$	  	&$<8.3$		       	&$<13.4$ 		&$<4.0$\\ 
				& \\
\enddata
\tablecomments{
$^{a}$Candidate counterpart IDs, as listed in previous tables. Irradiation model output parameters as derived from fits using Galactic extinction/absorption corrected magnitudes for each candidate counterpart. All stellar constraints are given in units of solar mass and radius. 
}
\label{tab:copperwheat1}
\end{deluxetable}

\begin{deluxetable}{lcccccccc}
\tabletypesize{\small}
\tablecolumns{9}
\tablecaption{\textbf{$1\sigma$ binary parameter constraints derived from intrinsic magnitudes for each potential ULX optical counterpart.}}
\tablehead{
				&             &               &\multicolumn{2}{c}{$M_{BH}=10 M_\odot$}  	&\multicolumn{2}{c}{$M_{BH}=100 M_\odot$} 	&\multicolumn{2}{c}{$M_{BH}=1000 M_\odot$}\\
Source          		& C/P ID$^a$    & cos({\it i})   	&$M_{*}$        	&$R_{*}$                &$M_{*}$        	&$R_{*}$                &$M_{*}$        	&$R_{*}$\\
                			&               &    		&($M_\odot$)    	&($R_\odot$)      	&($M_\odot$)    	&($R_\odot$)            &($M_\odot$)    	&($R_\odot$)\\
}
\startdata
NGC~55~ULX1     	&$1$            &$0.0$          &No constraint 		&$2.7$ -- $181.2$       &No constraint	 	&$3.4$ -- $294.7$       &No constraint  	&$3.4$ -- $299.9$\\            
                			&               &$0.5$          &No constraint        	&$1.4$ -- $33.4$        &No constraint         	&$<14.8$         	&No constraint		&$<9.6$\\
				&$2$            &$0.0$          &$<61.4$ 		&$2.3$ -- $117.8$       &$<66.7$  		&$2.9$ -- $261.3$       &$<70.3$  		&$2.9$ -- $266.1$\\            
                			&               &$0.5$          &$<63.1$        	&$1.0$ -- $23.4$        &$<60.1$         	&$<11.8$        	&$<39.0$		&$<7.2$\\
				& \\
NGC~4190~X-1   	&$1$            &$0.0$          &$>1.7$         	&$1.4$ -- $517.4$       &$>3.3$        		&$>2.2$               	&$>3.4$        		&$>2.3$\\ 
                			&               &$0.5$          &No constraint  	&No constraint       	&No constraint  	&No constraint       	&No constraint 		&No constraint\\ 
				& \\
\enddata
\tablecomments{
As with Table \ref{tab:copperwheat1}, $^{a}$Candidate counterpart ID, as listed in previous tables. All stellar constraints are given in units of solar mass and radius, with values derived from irradiation models fitting intrinsic X-ray and optical emission.}
\label{tab:copperwheat2}
\end{deluxetable}

\begin{deluxetable}{lcccccc}
\tabletypesize{\small}
\tablecolumns{7}
\tablecaption{\textbf{Black hole mass, and binary parameter constraints obtained during X-ray irradiation model fitting (1 $\sigma$ constraints).}}
\tablehead{
                			&               		& 			& 			& Black hole	& \multicolumn{2}{c}{Constraints obtained for}  	\\
Source          		& C/P ID$^a$  	& Fitting$^b$	& cos({\it i}) 	& mass     		& \multicolumn{2}{c}{companion star} 		 \\
                			&               		&    			& 			& constraints 	&  $M_{*}$        		&$R_{*}$  	  			\\
				&			&			&			& ($M_\odot$)  & ($M_\odot$)    	&($R_\odot$) 			\\
}
\startdata
NGC~55~ULX1     	& $1$            	& G			& $0.5$          	& $<700$		& $<13.9$        		& $0.9$ -- $5.1$        		\\
				& $2$            	& G			& $0.5$          	& $<840$		& $<15.1$        		& $1.4$ -- $5.4$        		\\
				& \\
NGC~253~ULX1    	& $3$            	& G			& $0.5$          	& $<550$	& $<12.2$ 		& $<4.6$       			\\
                			& \\
NGC~253~ULX2    	& $1$            	& G			& $0.0$          	& $>37.7$		& $11.8$ -- $14.8$	& $413.1$ -- $629.8$	\\ 
                			&               		& G			& $0.5$          	& $<590$		& $<6.8$		  	& $30.4$ -- $85.7$		\\
				& \\
\enddata
\tablecomments{
Notes: $^{a}$Candidate counterpart ID, as listed in previous tables. $^b$The type of magnitudes used in fitting X-ray irradiation models: G - Galactic extinction/absorption corrected values used, I intrinsic values used for fitting. All stellar constraints are given in units of solar mass and radius, with values derived from irradiation models fitting intrinsic X-ray and optical emission.}
\label{tab:BH mass}
\end{deluxetable}

\end{document}